%% file: PHD_THESIS.tex
\documentclass[twoside,11pt]{report}

\usepackage{guthesis}
\usepackage{amsmath}
\usepackage{graphicx}
\usepackage{longtable}
\usepackage{epstopdf}

\numberwithin{equation}{chapter}
\numberwithin{figure}{chapter}
\numberwithin{table}{chapter}

\begin{document}
\title{Retrodictive Quantum State Engineering}
\author{Kenneth Lyell Pregnell}
\school{Science}

\beforepreface 
\include{Acknowl}
\include{Pubs}
\include{Abstract}
\include{Symbols}
\include{Acronyms}

\afterpreface 


\newcommand{\smallfrac}[2]{\mbox{$\frac{#1}{#2}$}}
\newcommand{\half}{\smallfrac{1}{2}}
\newcommand{\tr}[1]{\textrm{Tr}[#1]}
\newcommand{\Tr}[1]{\textrm{Tr}\left[#1\right]}
\newcommand{\trs}[2]{\textrm{Tr}_{#1}[#2]}
\newcommand{\Trs}[2]{\textrm{Tr}_{#1}\left[#2\right] }
\newcommand{\pr}{\textrm{Pr}}

\include{Intro}
\include{Chap2}
\include{Chap3}
\include{Chap4}
\include{Chap5}
\include{Conc}

\appendix
\include{AppendixA}
\include{AppendixB}
\include{AppendixC}
\include{AppendixD}


\bibliographystyle{PHDThesis}
\bibliography{Bibliography}

\end{document}

%% file: Acknowl.tex
\prefacesection{Acknowledgments}
Several people deserve mention for their direct, and indirect,
contribution towards this thesis.

In particular, Prof. David Pegg has shown much patience in his
supervision of this work. His intuitive and unique understanding
of nature and of the physical processes which contribute to it has
greatly broadened my outlook towards physics. He has an
extraordinary ability to reduce seemingly complex problems to the
essentials, which he successfully applies to physics as well as
everyday life.

The depth of knowledge of Ass. Prof. Howard Wiseman is remarkable.
At times it provided a great resource in offering the `correct'
opinion when debates between us students could not be settled. I
am yet to find a physics problem into which Howard cannot offer
some physical insight.

I am quite privileged to have kept such good company throughout my
years at Griffith. In my first year as an undergraduate, some
eight years ago, I formed what has become a long standing
friendship with Jay Gambetta, Laura Thomsen and Jonathan Ashmore.
The time has come when we are all to go our separate ways. I wish
them luck. May our friendship continue.

On a more personal note, I would like to thank Jayne Bullock. We
have been together for a large part of my candidature, and,
together, we have endured. Her commitment to me, especially in the
writing stages, has contributed enormously to the completion of
this thesis.

Finally, I would like to mention the support of my parents and
family, not just throughout my candidature, but throughout my
existence. Their love and commitment to me are the only true
constants in my life, without which, I would not be the person I
am today.

%% file: Pubs.tex
\prefacesection{List of Publications}

In this chronological list of my refereed publications, those
which are published in a conference proceedings are marked with an
asterisks.
\begin{enumerate}

\item K.~L. Pregnell and D.~T. Pegg, ``{Measuring the phase variance of light,}''
  \textit{J. Mod. Opt.} \textbf{48}, 1293-1302 (2001).

\item K.~L. Pregnell and D.~T. Pegg, ``{Quantum phase distribution by operator
  synthesis,}'' \textit{J. Mod. Opt.} \textbf{49}, 1135-1146 (2002).

\item K.~L. Pregnell and D.~T. Pegg, ``{Measuring the elements of the optical
  density matrix,}'' \textit{Phys. Rev. A} \textbf{66}, 013810 (2002); (quant-ph/0209132).

\item K.~L. Pregnell and D.~T. Pegg, ``{Single-shot measurement of quantum
  optical phase,}'' \textit{Phys. Rev. Lett.} \textbf{89}, 173601 (2002).

\item * K.~L. Pregnell and D.~T. Pegg, ``{Measuring the individual elements of
  the optical density matrix,}'' in ``{Proceedings of the sixth
  International Conference on Quantum Communication, Measurement and
  Computing,}'' , edited by J.~H. Shapiro and O.~Hirota, pp. 325--328 (Rinton
  Press, New Jersey, 2003).

\item K.~L. Pregnell and D.~T. Pegg, ``{Binomial states and the phase
  distribution measurement of weak optical fields,}'' \textit{Phys. Rev. A}
  \textbf{67}, 063814 (2003).
  
\item  K.~L. Pregnell and D.~T. Pegg, ``Retrodictive quantum state engineering," {\em J. Mod. Opt}. {\bf 51}, 1613 (2004).

\end{enumerate}

%% file: Abstract.tex
\prefacesection{Abstract}

This thesis is concerned with retrodiction and measurement in quantum optics. The latter of these two
concepts is studied in particular form with a general optical multiport device, consisting of an arbitrary
array of beam-splitters and phase-shifters. I show how such an apparatus generalizes the original projection
synthesis technique, introduced as an in principle technique to measure the canonical phase distribution.
Just as for the original projection synthesis, it is found that such a generalised device can synthesize any
general projection onto a state in a finite dimensional Hilbert space. One of the important findings of this
thesis is that, unlike the original projection synthesis technique, the general apparatus described here only
requires a classical, that is a coherent, reference field at the input of the device. Such an apparatus lends
itself much more readily to practical implementation and would find applications in measurement and
predictive state engineering.

If we relax the above condition to allow for just a single non-classical reference field, we show that the
apparatus is capable of producing a single-shot measure of canonical phase. That is, the apparatus can
project onto any one of an arbitrarily large subset of phase eigenstates, with a probability proportional to
the overlap of the phase state and the input field. Unlike the original projection synthesis proposal, this
proposal requires a binomial reference state as opposed to a reciprocal binomial state. We find that such a
reference state can be obtained, to an excellent approximation, from a suitably squeezed state.

The analysis of these measurement apparatuses is performed in the less usual, but completely rigorous,
retrodictive formalism of quantum mechanics.

%% file: Symbols.tex
\prefacesection{List of Symbols}

The following is a listing of the most frequently used symbols in this thesis.

\begin{tabular}{ll}

  $\hat a$ $(\hat a^\dag)$      & an annihilation (creation) operator for an optical mode \\
  $n,|n\rangle$                 & a photon number, eigenstate\\
  \emph{P}($\cdots$)                   & probability density of $\cdots$ \\
  Pr($\cdots$)                  & probability of $\cdots$ \\
  Pr($i|j$)                     & Bayesian probability: probability of $i$ given $j$ \\
  $\textbf{R}(N+1)$             & a unitary rotation matrix of $N+1$ dimensions\\
  $r$                           & transmission coefficient of a beam-splitter \\
  $\hat S$ $(\hat S^\dag)$      & the forward (backward) time unitary evolution operator of a linear optical device\\
  $t$                           & reflection coefficient of a beam-splitter \\
  Tr[$\cdots$]                  & trace of $\cdots$ \\
  $\hat U$ $(\hat U^\dag)$      & a forward (backward) time unitary evolution operator\\
  $\textbf{U}(N+1)$             & a unitary transformation matrix of $N+1$ dimensions\\
  $U_{nm}$                      & matrix elements of $\textbf{U}(N+1)$\\

  $|\alpha\rangle$              & a coherent state\\
  $|\theta\rangle$              & a phase state\\
  $|\theta_n\rangle$            & a truncated phase state\\
  $\delta\theta$                    & $=2\pi/(N+1)$, a discrete phase increment \\
  $|\tilde\psi\rangle$          & an unnormalised state\\
  $\hat\Gamma_j$                & a measurement device operator\\
  $\eta$                        & an efficiency \\
  $\hat\Lambda_i$               & a preparation device operator\\

\end{tabular}

\begin{tabular}{ll}
  $\hat \Xi_i$                  & a (preparation) POM element \\
  $\hat \Pi_j$                  & a (measurement) POM element \\
  $\hat\rho$                    & a predictive density matrix\\
  $\rho_{n,m}$                  & $=\langle n|\hat\rho |m\rangle$, density matrix elements of $\hat\rho$ in the photon number basis\\
  $\hat\rho^{\textrm{ret}}$     & a retrodictive density matrix\\
  $\mathbf{\Omega}(N+1)$        & a discrete Fourier transformation matrix of $N+1$ dimensions\\
  $\Omega_{nm}$                 & $=(N+1)^{-1/2}\omega^{nm}$, the matrix elements of $\mathbf{\Omega}(N+1)$\\
  $\omega$                      & $=\exp[i2\pi/(N+1)]$, the $(N+1)^\textrm{th}$ root of unity\\

  ${n\choose m}$                & a binomial coefficient\\
  $\hat 1$                      & unit operator of the infinite dimensional Hilbert space\\
  $\hat 1_{N}$                      & unit operator of an $N+1$ dimensional Hilbert space\\
  $\langle\cdots\rangle$        & expectation value of $\cdots$\\
  $\hat{}$                      & indicates a Hilbert space operator\\
\end{tabular}

%% file: Acronyms.tex
\prefacesection{List of Abbreviations}


\begin{tabular}{ll}
  BS        & Beam-splitter\\
  DFT       & Discrete Fourier transformation\\
  MDO       & Measurement device operator. (For an unbiased measuring\\
            & device this becomes the element of a POM.) \\
  PDO       & Preparation device operator. (This includes information about\\
            & the prepared state and its preparation probability. It can be \\
            & normalised by dividing by its trace to become a density operator.) \\
  POM       & Probability operator measure \\
  PS        & Phase-shifter\\
\end{tabular}

%% file: Intro.tex
\chapter{Introduction}\label{intro}

Retrodictive state engineering concerns the generation of states that evolve backwards in time. At first
glance this may appear to be in strong violation of our notion of causality. In the usual, that is
predictive, formalism of quantum mechanics a state is produced by a preparation event and then evolves
forwards in time. Retrodictive states, on the other hand, evolve backwards in time from the measurement
event. As we shall see in this thesis, however, causality is not manifest in the time direction of
evolution of the states. Rather it is encapsulated, and guaranteed by, an asymmetry in the normalisation of
the operators associated with the preparation and measurement events.

It is the purpose of this chapter to introduce two intimately related concepts in the theory of quantum
mechanics: retrodiction and the quantum theory of measurement. Retrodiction, in contrast to prediction, is a
technique whereby information about some previous event is gained from knowledge of some related event
occurring at a later time. If only partial knowledge is known about the later event, then retrodicting
probabilities must suffice. Consider, for example, the event to be the outcome of some horse race, say the
Melbourne cup. Typically a punter will try to \emph{predict} the outcome of the race \emph{before} it has
happened. If he is an experienced punter, then he may favour a horse more if it has been in good form leading
up to the race. That is to say he may increase the chances of the horse to win the race, based upon the
results of the horse in previous races. In complete symmetry, the punter could \emph{retrodict} the outcome
of the race \emph{after} is has happened. Take the unlikely situation where the punter was not privileged to
know the outcome of the Melbourne cup, but does know how a particular horse performed in races \emph{after}
the Cup. If the horse is in good form after the race, he may decide to favour the chance that the horse had
won the Melbourne cup. The punter would then increase the chance he assigns to the horse having won the race,
based upon the results of the horse in subsequent races.

In this situation, retrodiction is generally a redundant exercise since the results of the race can usually
be found, after the race, in the public domain, for example in the newspaper or on the internet. For this
reason bookmakers generally do not pay favourable odds to punters after the race has been run. However, such
a technique does find useful applications when the past results cannot easily be found in the public domain.
The rapidly growing field of secure communications is one such example where enormous efforts are placed to
avoid information entering into the public domain. Unlike the punter in the above example, retrodiction is
then a valuable tool for an eavesdropper attempting to acquire some useful information about a message, sent
some time in the past, that he was not privileged to. Not only may information not enter into the public
domain through the concerted efforts of a select few, but also it is possible that some part of the intended
information may just become lost. Again, this is a real problem when information is sent down a noisy channel
resulting in a distorted message being received by the receiver. In such cases the receiver can retrodict the
intended message from the distorted one \cite{Shannon63}.

Despite the internal symmetry between the techniques of prediction and retrodiction there is, however, an
undeniable asymmetry in the frequency in which we use prediction over retrodiction. This is because there is
an inherent asymmetry in nature in that we remember the past and not the future. So although the techniques
are time symmetric, because we can never be completely certain of the future, but at times can be completely
sure of the past, prediction is more often than not the most used technique.

\section{Retrodiction in quantum mechanics}\label{sec1.1}

\subsection{Arrow of time}\label{sec1.1.1}

The perennial question as to why do we remember the events of the past as opposed to the events of the future
is one of the great unresolved questions of theoretical physics. The observed distinction between past and
future allows a direction, or an ``arrow", of time to be defined. Such an arrow is evidence of a fundamental
asymmetry inherent in nature, which to date, is void of an orthodox physical explanation. In fact, there are
at least two other widely-believed versions of the arrow of time \cite{Hawking88} aside from the
psychological arrow just mentioned. The first of these is the cosmological arrow of time. This is the
direction of time in which the universe is expanding rather than contracting. Such a result is the conclusion
of the general theory of gravity. The second is the thermodynamical arrow and is encapsulated by the second
law of thermodynamics which states that the disorder or entropy of an open system will not decrease over
time. To illustrate, consider the frequently observed process of combustion. If we were to take a match and
set it to a piece of paper we would not be alarmed if the piece of paper caught fire and after a short period
of time turned to ash and smoke. However, if instead we were to observe the air above a pile of ash suddenly
begin to go smokey and, for no apparent reason, begin to take fire we would think something strange was going
on. Furthermore, if at the end of the process when the flames where finally extinguished there remained an
unburnt piece of paper in place of the ash, we would be truly astonished. Both situations described are,
obviously, the time reversal of each other and are therefore equivalent physical processes, however, there is
an undeniable asymmetry in the frequency at which we observe one process over the other. The direction in
which processes of this kind can proceed defines an arrow of time. A rather interesting argument was put
forward by Hawking \cite{Hawking88} to suggest that all three arrows necessarily point in the same direction.

In an attempt to provide a physical explanation for the thermodynamical arrow of time, it was suggested by
Bohm \cite{Bohm51}, and von Neumann \cite{vonNeumann55}, that the ``reduction of the wavefunction'', implicit
in the orthodox interpretation of quantum measurement theory, introduces into the foundations of quantum
physics a time asymmetric element, which in turn leads to the thermodynamical arrow of time. If this
reduction were a real physical process, we would have a fundamental basis for the arrow of time.

It is the orthodox view that there are two distinct types of evolution present in the theory of quantum
mechanics. There is the deterministic or reversible evolution, postulated by the Schr\"{o}dinger or
Heisenberg equations of motion, which is time symmetric. Such an evolution is represented in the theory by a
unitary operator and, generally speaking, describes the evolution of a quantum system either not interacting,
or interacting with only ``small'' systems. Then there is the non-deterministic and irreversible evolution
resulting from the interaction of the system with a measuring device. Such an interaction is responsible for
the ``reduction of the wavefunction'' and is represented in the theory by a non-unitary operator. The
question naturally arising from such a bifurcation is then: at what point does the evolution of a system
cease to be deterministic and begin to be non-deterministic? That is, when do we consider a system to be
``measured''? Such a question lies at the very heart of what is referred to as the ``measurement problem''.

The idea of Bohm and von Neumann \cite{Bohm51, vonNeumann55} that such an irreversible process necessarily
introduces into the theory a time asymmetric process, suggests that state reduction associated with the act
of measurement is a physical process responsible, at least in part, for the thermodynamical arrow of time in
the quantum domain. In response to these suggestions it was later shown by Aharonov, Bergmann and Lebowitz
\cite{Aharonov64}, following the work of Watanabe \cite{Watanabe55}, that, although the reduction of the
wavefunction is an irreversible process, it is not time asymmetric. They proceeded to demonstrate this by
taking the standard probability expression of quantum mechanics, which implicitly deals with ensembles that
have been ``preselected'' on the basis of some initial observation, and deriving a probability expression
that favours neither the past nor the future. From such a time symmetric probability expression, they
demonstrated how the conventional predictive, as well as a ``retrodictive'', probability expressions can be
obtained. It was their findings that an additional postulate was necessary to regain from the time symmetric
expressions the conventional predictive probability expression. This led them to conclude;
\begin{quote}
``If, as we believe, the validity of this postulate and the falsity of its time reverse result from the
macroscopic irreversibility of our universe as a whole, then the basic laws of quantum physics, including
those referring to measurements, are as completely symmetric as those of classical physics.''
\end{quote}

It was shown, ``as a by-product'' of their analysis, that a quantum state could be assigned based on the
outcome of a future measurement event with as much justification as one, ordinarily, assigns the state base
on the outcome of an initial preparation event. Since then the formalism of retrodiction in quantum mechanics
has been further developed \cite{Barnett00a,Barnett00b,Pegg02,Pegg02b} in addition to finding practical
applications \cite{Pegg99b,Jeffers02a,Jeffers02b}. As we will see in Chapter~\ref{chap2}, in the retrodictive
formalism it is the preparation apparatus, as opposed to the measurement apparatus, that is responsible for
the non-unitary evolution resulting in the reduction of the wavefunction. With the predictive and
retrodictive formalisms being equally valid, it is apparent that the reduction of the wavefunction is a time
symmetric process, while still remaining an irreversible process in agreement with the conclusions of
\cite{Aharonov64}.

\section{Structure of this thesis}

Just as a preparation device is said to be responsible for the production of a predictive quantum state, the
foundational work of \cite{Aharonov64} tells us that, with the same justification, a measurement device is
responsible for the production of a retrodictive quantum state. It was the findings of \cite{Aharonov64} that
selecting ensembles of systems favouring either preselection or postselection leads to the predictive or
retrodictive formalism respectively. Traditionally, the conventional formalism of quantum mechanics has been
the predictive formalism, presumably for the same reason as why a bookmaker will not pay favourable odds
after a horse race. The retrodictive formalism is then usually derived from the conventional formalism using
Bayes' theorem \cite{Aharonov64, Barnett00a}. I begin this thesis from a more general approach. Following the
work of Pegg, Barnett and Jeffers \cite{Pegg02}, I discuss in Chapter~\ref{chap2} a formalism of quantum
mechanics which, similar to that of Aharonov \emph{et al.} \cite{Aharonov64}, is time symmetric. Such a
formalism begins by postulating a probability measure in which both the preparation and measurement
procedures are represented by the same class of operators. Along similar lines to the original paper
\cite{Pegg02}, I demonstrate how the predictive and retrodictive formalism can be derived from this.

With both the preparation and measurement processes represented by the same class of operators, it is then
possible to formally consider a hybrid process involving combinations of both preparation and measurement
devices. I show that such a combined device can always be reduced to either a preparation device or a
measurement device. This allows one to redefine the boundaries between preparation and measurement for a
joint system under study.

Utilising this, I consider in this thesis linear optical multiport devices with equally many input ports as
output ports. By considering everything except one of the input ports as a measuring device, I am able to
effectively synthesize novel measuring devices. By changing the preparation device at the input port we can
change the form of the synthesized measuring device. Such a technique was first introduced into quantum
optics by Pegg, Barnett and Phillips \cite{Pegg97,Phillips98} as a way to measure the phase distribution of
light in the quantum regime. Although this technique was novel, it was largely an in principle demonstration
as the reference states necessary to synthesise the measurement requires, in general, a non-classical
reference state.

It is the findings of Chapter~\ref{chap3} that the non-classical reference state in the original projection
synthesis can be replaced by an easily prepared coherent reference state, provided we extend the
beam-splitter to a linear optical multiport with as many input ports as the maximum number of photons in the
retrodictive state we wish to engineer. Such a finding demonstrates how any observable can be measured using
only linear optics, coherent reference states and photo-detection. We find, in keeping with an emphasis on
realistic measurement, that only photodetectors that can discriminate between zero, one and many photons are
required.

Such a finding is important to the field of quantum optics because it demonstrates that any quantum optical
observable can be measured using only three basic elements; linear optics, coherent states and single photon
detection. As all these elements are experimentally available, this results offers an experimentally
achievable way to measure any quantum optical observable.

We find, quite remarkably, that although any general retrodictive state can be produced by this method, it is
difficult to physically implement the time reversed situation (that is swap the measured output with the
prepared input and vice versa) to generate any general predictive state. The problem arises because of the
difficulty associated with simultaneously producing multiple single photon states. By conditioning involving
simultaneous photo-detector outcomes it is not so difficult to produce the retrodictive equivalent of this.
So although the range of predictive states which can be generated in the laboratory is relatively small, we
find that the range of retrodictive states is, remarkably, quite large. With this in mind we investigate in
Chapter~\ref{chap4} two experimental proposals which are proving difficult to implement due to a need for a
non-classical predictive reference state. We consider redesigning the proposed experiments in such a way to
replace the predictive reference state with a more readily prepared retrodictive state in order to achieve
the same result. The experiments on the whole are much simpler, involving only linear optics, a coherent
reference state and photo-detection.

Although the general experiment introduced in Chapter~\ref{chap3} can measure the probability distribution of
any quantum optical observable, it does not provide a ``single-shot'' measurement of that observable. A
single-shot measurement is defined as a measurement which can project onto any eigenstate in the set of
eigenstates representing the observable of interest with the correct probability. That is to say that the
measurement device can produce the complete set of retrodictive states defined by the observable. Such an
example is a perfect photon number measurement where the measurement device could project onto any one of the
infinite photon number eigenstates $|n\rangle$. Although such a measurement is not currently available, it is
assumed that such a measurement could be performed in principle if we could eliminate all physical sources of
imperfection inherent in the devices existing today. Conversely, an observable for which there does not exist
even an imperfect single-shot measurement device is the observable canonically conjugate to photon number.
Such an observable, commonly referred to as canonical phase, has had an interesting history in the field of
quantum optics\footnote{For a review see, for example, \cite{Leonhardt95} and for a chronological listing of
most of the relevant papers on phase see the detailed bibliography of \cite{Pegg97b}.}. Despite these
challenges, there nevertheless does exist a well-defined physical observable without, until now, a technique
capable of providing a single-shot measurement. In Chapter~\ref{chap5} I present the first proposal capable
of producing a single-shot measure of canonical phase. The technique is similar to the general measuring
device introduced in Chapter~\ref{chap3}, with one of the coherent reference states replaced with a binomial
reference state. Such a tradeoff is the price paid to obtain a single-shot measuring device.

It is perhaps a little ironic then that the apparatus used to measure ``measured phase'', which was
introduced by Noh, Foug\`{e}res and Mandel \cite{Noh91,Noh92a,Noh92b,Noh93a,Fougeres94} as a way to
circumvent the difficulties associated with measuring canonical phase, can in fact be used to measure the
canonical phase distribution for the range of fields considered by the same authors. I show in
Chapter~\ref{chap5} that the only alteration to the apparatus needed is to either suitably squeeze one of the
coherent reference states at the input of the optical multiport device or, even more simply, set the
amplitudes of the three coherent reference states to a predetermined value. I conclude this thesis with some
final remarks on the work presented within this thesis.

%% file: Chap2.tex
\chapter[Preparation and measurement]{Quantum theory of preparation and measurement}\label{chap2}

Traditionally, the theory of quantum mechanics is regarded as a predictive theory. Take for example the
situation were Alice prepares some system in a particular state which she then sends off to Bob who performs
a measurement on it. We typically assign a quantum state $\hat\rho_i$ to the system Alice prepared based on
the outcome $i$ of her preparation device and ask the question: with what probability will Bob observe the
outcome $j$ at his measurement device given that Alice prepared the system in the state $\hat\rho_i$? It
would also be perfectly reasonable, albeit less common, to ask this question in reverse. That is, given that
Bob has observed the outcome $j$ at his measurement device, with what probability did Alice prepare the
system in the state $\hat\rho_i$? Provided the answer to the first of these two questions is known then it is
possible to infer an answer to the latter using Bayes' theorem \cite{Bayes}. Alternatively, using the less
usual but completely rigorous formalism of retrodiction
\cite{Watanabe55,Aharonov64,Penfield66,Aharonov84a,Aharonov84b,Aharonov91} it is possible to answer questions
of this kind more directly. The theory assigns a \emph{retrodictive} quantum state
$\hat\rho^{\textrm{ret}}_j$ to the system based on the outcome $j$ of Bob's measurement device. As opposed to
the predictive state $\hat\rho_i$ that Alice assigns to the system which evolves forward in time,
interestingly, the retrodictive state is said to evolve backwards in time, away from the measurement event.

To introduce both the predictive and retrodictive formalisms of quantum mechanics, it is perhaps simplest to
begin with a theory which treats preparation and measurement on an equal footing. Following the work of Pegg
\emph{et. al.} \cite{Pegg02} such a theory is introduced in Sections~\ref{sec2.1} and \ref{sec2.2}. To show
that the theory is consistent with all the prediction of quantum mechanics, the authors derive the
conventional predictive formalism of quantum mechanics from their symmetric theory which is presented here in
Section~\ref{sec2.3}. From within this framework it is then a straightforward matter, as was originally done
in \cite{Pegg02}, to extend this derivation to obtain the retrodiction formalism of quantum mechanics. This
to is presented in Section~\ref{sec2.3}. From there the work becomes original as I address in
Section~\ref{sec2.5} the seemingly paradoxical issue associated with the very existence of a retrodictive
sate. We resolve this issue and explain why a predictive state can be used by Alice to send information to
Bob some time into the future, but why Bob cannot use a retrodictive state to send useful information about
the future backwards in time to Alice. This however does not render the concept of a retrodictive state as
useless. Indeed, in Section~\ref{sec2.6} it is shown how, with the aid of post selection, retrodictive states
can be conditionally generated. This is an important result to this thesis as the remaining chapters are
focused on using post selection to engineer specific tailored retrodictive states to provide arbitrary
measurements. We find that not only are retrodictive states practically useful, they also provide a deeper
conceptual understanding of quantum mechanics where in Section~\ref{sec2.5} we briefly examine the
measurement problem and highlight the ambiguity in defining when the state collapse `really' occurs.

\section{Preparation and measurement devices}\label{sec2.1}

Following Pegg \emph{et al.} \cite{Pegg02}, consider the situation where Alice prepares a quantum system
which is then automatically sent to Bob who performs a measurement upon it. Alice is capable of preparing the
system in any one of a number of, not necessarily orthogonal, states labelled by $i=1,2,\dots$. When the
desired system is prepared she sends the label $i$ associated with her successful preparation event to a
computer for recording. The combined process of preparation and recording of a particular event by Alice is
described mathematically by a preparation device operator (PDO) $\hat\Lambda_i$. The PDO acts on the state
space of the system and represents the successfully prepared state, including any biasing that might arise
from the preparation device not being able to produce certain states or from Alice choosing not to record
states of a particular kind. The set of all such operators $\{\hat\Lambda_i\}$ is sufficient to describe
mathematically the preparation procedure dictated by Alice.

The measuring device has a readout mechanism which indicates the outcome of a measurement event labelled
$j=1,2,\dots$. To each possible measurement event $j$ a measurement device operator (MDO) $\hat\Gamma_j$ is
associated. This operator also acts in the state space of the system and, among other things, represents the
measured state of the system. At Bob's discretion, the label $j$ of the actual outcome is sent to the
computer for recording. If the computer receives a label from both Alice and Bob then it records the combined
event $(i,j)$. If however the computer does not receive both a value of $i$ and $j$ then the single event
recorded by either Alice or Bob is discarded by the computer automatically. Any biasing on behalf of Bob or
of the measuring apparatus to record an event is also incorporated in the MDO. As an example, the MDO for a
perfect von Neumann type measurement \cite{vonNeumann55} which is faithfully recorded by Bob would be the
projector formed by the eigenstate corresponding to the detection event $j$. The combined process of
detection and recording by Bob is called the measurement procedure and is completely described mathematically
by the set of all MDOs $\{\hat\Gamma_j\}$. For now no restrictions are imposed on the operators
$\hat\Lambda_i$ and $\hat\Gamma_j$ other than they represent their associated preparation and measurement
procedures. In the following section a single restriction upon these operators is postulated and from this
postulate standard measurement theory \cite{Helstrom76} is derived.

By repeating the experiment many times, with both Alice and Bob independently performing their own
preparation and measurement procedures, a list of combined events $(i,j)$ is compiled on the computer from
which particular occurrence frequencies can be found.

\section{Fundamental postulate}\label{sec2.2}

A sample space of mutually exclusive outcomes can be constructed from the combined recorded events $(i,j)$,
such that each point in the space represents a particular event. To each identical event we assign the same
point in the sample space. To this space a probability measure can be introduced that assigns probabilities
between zero and one to each point so that the sum of the probabilities over the whole space is unity. This
measure is such that the probability is proportional to the number of recorded events $(i,j)$ associated with
each point, that is, the probability is proportional to the occurrence frequency of the recorded event
$(i,j)$. The fundamental \emph{postulate} of \cite{Pegg02} is that the probability associated with each point
in this sample space is given by
\begin{equation}\label{2.1}
\pr^{\Lambda\Gamma}(i,j)=\frac{\tr{\hat\Lambda_i\hat\Gamma_j}}{\tr{\hat\Lambda\hat\Gamma}},
\end{equation}
where the trace is over the state space of the system and
\begin{equation}\label{2.2}
\hat\Lambda=\sum_i\hat\Lambda_i,
\end{equation}
\begin{equation}\label{2.3}
\hat\Gamma=\sum_j\hat\Gamma_j.
\end{equation}
That is to say that if a recorded event is chosen at random from the computer list, then Equation (\ref{2.1}) is the
probability that this event is $(i,j)$.

The postulate of (\ref{2.1}) imposes an implicit restriction upon the operators $\hat\Lambda_i$ and $\hat\Gamma_j$, in
that these operators must be non-negative or non-positive definite to ensure that no probabilities
$\pr^{\Lambda\Gamma}(i,j)$ are negative. Without loss of generality, we assume they are non-negative definite. The
denominator in (\ref{2.1}) is for normalisation purposes, that is to make certain that all probabilities assigned to
the sample space add to unity. As a result, each set of operators $\{\hat\Lambda_i\}$ and $\{\hat\Gamma_j\}$ need only
be specified up to an overall arbitrary constant, as this constant always cancels in the expression for the
probabilities (\ref{2.1}). This freedom will become useful later in defining a more convenient set of operators
$\{\hat\Lambda_i\}$ and $\{\hat\Gamma_j\}$ such that the operators $\hat\Lambda$ and $\hat\Gamma$ have unit trace.

From (\ref{2.1}), the following probabilities can be derived,
\begin{eqnarray}
\pr^{\Lambda\Gamma} (i)&=&\sum_j
\pr^{\Lambda\Gamma}(i,j)=\frac{\tr{\hat\Lambda_i\hat\Gamma}}{\tr{\hat\Lambda\hat\Gamma}}
\label{2.4}\\
\pr^{\Lambda\Gamma} (j)&=&\frac{\tr{\hat\Lambda\hat\Gamma_j}}{\tr{\hat\Lambda\hat\Gamma}}
\label{2.5}\\
\pr ^{\Lambda\Gamma}(j|i)&=&\frac{\pr^{\Lambda\Gamma}(i,j)}{\pr ^{\Lambda\Gamma}(i)}=\frac{\tr
{\hat\Lambda_i\hat\Gamma_j}}{\tr{\hat\Lambda_i\hat\Gamma}}
\label{2.6}\\
\pr ^{\Lambda\Gamma}(i|j)&=&\frac{\tr{\hat\Lambda_i\hat\Gamma_j}}{\tr{\hat\Lambda\hat\Gamma_j}} \label{2.7}.
\end{eqnarray}
and can be interpreted as such: If one event is chosen at random from the list of combined recorded events on
the computer, then (\ref{2.4}) is the probability that this combined event includes the preparation and
recording of the event $i$. Similarly, $\pr ^{\Lambda\Gamma}(j)$ is the probability that a randomly selected
combined event will involve the detection and recording of measurement event $j$. The expression (\ref{2.6}) is the
probability that if the preparation event $i$ was observed and recorded by Alice, then the corresponding
measurement event $j$ will be observed and recorded by Bob. The reverse situation may also be considered. That
is, if Bob observed and recorded the detection event $j$, then the probability that this detection event
coincided with the preparation event $i$ of Alice is given by (\ref{2.7}).

Expression (\ref{2.6}) may be used for prediction. If the PDO for Alice's preparation event is known, say
$\hat\Lambda_i$, then (\ref{2.6}) can be used to calculate the probability that Bob, some time in the future,
will detect and record the event $j$ represented by the MDO $\hat\Gamma_j$. To do this, the operator
$\hat\Gamma$ must be known. This requires having some information about the measurement procedure of Bob.
Expression (\ref{2.7}), in contrast, can be used for retrodiction. When the detection event $j$ is observed
and recorded by Bob, a corresponding MDO $\hat\Gamma_j$ can be assigned. With this knowledge, Bob can use
expression (\ref{2.7}) to calculate the probability that Alice, some time in the past, prepared and recorded
the state represented by the PDO $\hat\Lambda_i$. To do this he also needs to have some information about the
preparation procedure of Alice, in that the operator $\hat\Lambda$ needs to be known. For this reason,
(\ref{2.7}) can be viewed as the time reversal of Eqn~(\ref{2.6}) and vice versa.

\section{Density matrices, POMs and unbiased devices}\label{sec2.3}
The postulate of Eqn~(\ref{2.1}) treats preparation and measurement on an equal footing. This is evident in that the
class of operators representing each procedure are subject to the same constraints. In standard measurement theory
there is, however, an asymmetry in the class of operators representing the preparation and measurement procedures.
\emph{Density matrices}, with unit trace, are used to describe the system prepared by Alice while a \emph{Probability
Operator Measure} (POM) $\hat\Pi$, with elements $\hat\Pi_j$ which sum to unity, is used to describe the measurement
procedure of Bob. In what follows we relate the respective operators in these two formalisms and find an equivalence in
the predictive ability of both formalisms. That is we derive the conventional predictive formalism of quantum mechanics
from the postulate of Eqn~(\ref{2.1}). We find however, that the postulate of Eqn~(\ref{2.1}) allows for a wider class
of measurements procedures than standard measurement theory allows.

Interestingly, quantum mechanics allows for the class of operators representing each procedure to be
exchanged. In the standard theory where the class of operators is asymmetric, an unorthodox formalism arises.
Such a formalism, known as the retrodictive formalism, assigns a density matrix to the state of the system
measured by Bob, while a POM can be used in some circumstances to describe the preparation procedure of
Alice. In Section~\ref{sec2.3.2} we derive the retrodictive formalism from the postulate of Eqn~(\ref{2.1})
and, in doing so, find an expression relating the corresponding operators in each formalism.

\subsection{Predictive formalism of quantum mechanics}\label{sec2.3.1}

Following the formalism of Ref. \cite{Pegg02} it is understood that a PDO represents, among other things, the
state of the system prepared by Alice. A normalised version of this operator can be introduced as
\begin{equation}\label{2.8}
  \hat\rho_i=\frac{\hat\Lambda_i}{\tr{\hat\Lambda_i}}.
\end{equation}
Since this operator is positive definite and has unit trace it serves as valid density matrix representing
the state of the system prepared by Alice. Consider now the situation were Bob's measurement procedure is
such that $\hat\Gamma$, the sum of the MDOs, is proportional to the unit operator acting on the state space
of the system,
\begin{equation}\label{2.9}
  \hat\Gamma=\gamma\hat 1.
\end{equation}
Such a measurement procedure is referred to as an unbiased measurement procedure \cite{Pegg02}. For an unbiased
measurement procedure it is useful to introduce the set of operators
\begin{equation}\label{2.10}
  \hat\Pi_j=\gamma^{-1}\hat\Gamma_j.
\end{equation}
Summing the above equation over all measurement outcomes $j$ and comparing to Eqn~(\ref{2.9}) shows that the
elements $\hat\Pi_j$ sum to the identity. As these operators are non-negative definite it is appropriate to
regard the set of such operators $\{\hat\Pi_j\}$ as a POM. We note that the constant of proportionality
$\gamma$ always cancels in the expression for the probabilities (\ref{2.4})-(\ref{2.7}) so it is appropriate
to regard the POM elements as MDOs. Therefore it is understood that each POM element corresponds to the
measuring and recording of a particular outcome by Bob.

By way of Eqns~(\ref{2.8}), (\ref{2.9}) and (\ref{2.10}), it is straightforward then to recast (\ref{2.6})
as,
\begin{equation}\label{2.11}
  \pr ^{\Lambda 1}(j|i)=\Tr{\hat\rho_i\hat\Pi_j},
\end{equation}
thereby recovering the standard probability postulate of quantum mechanics \cite{Helstrom76}. In doing this
however, an asymmetry in the normalisation properties of the two operators $\hat\rho_i$ and $\hat\Pi_j$,
representing the preparation and detection events respectively, is introduced. This asymmetry is not of a
fundamental origin, but rather comes from the added restriction that the measurement process satisfy
Eqn~(\ref{2.9}). Under this restriction, the symmetric postulate of (\ref{2.1}) reduces to the asymmetric
form given above. Interestingly, we show in Section~\ref{sec2.5} that the measurement procedure for all
faithfully recording measurement devices must satisfy Eqn~(\ref{2.9}) in order to preserve causality. In this
case the above equation is sufficient to predict the probability of all future measurement events $j$, given
that the system was initially prepared in the state $\hat\rho_i$. In general, however, not all measurement
procedures are required to faithfully record all outcomes. For example, in the operational phase measurements
of Noh \emph{et al.} \cite{Noh91,Noh93a} certain photo-detector readings are not recorded because they do not
lead to meaningful values being measured. This is a specific example where the MDOs do not sum to unity so
the probabilities obtained from (\ref{2.11}) must be appropriately renormalised to be consistent with the the
statistics of the experiment. This is precisely what the more general measurement postulate leading to
(\ref{2.6}) does. Substituting Eqn~(\ref{2.8}) for $\hat\Lambda_i$ in (\ref{2.6}) we can express this in
terms of density matrices as
\begin{equation}\label{2.12}
  \pr ^{\Lambda\Gamma}(j|i)=\frac{\tr{\hat\rho_i\hat\Gamma_j}}{\tr{\hat\rho_i\hat\Gamma}}.
\end{equation}
which is a more general expression than Eqn~(\ref{2.11}) as it accounts for experiments where measurement results may
be discarded.

\subsection{Retrodictive formalism of quantum mechanics} \label{sec2.3.2}

The decision to express the PDO as a density matrix was arbitrary. For an unbiased measurement device it
gives a simple expression for the predictive formula $\pr (j|i)$ of standard measurement theory.
Alternatively, the MDO could be expressed in terms of a density matrix as
\begin{equation}\label{2.13}
  \hat\rho_j^{\textrm{ret}}=\hat\Gamma_j/\tr{\hat\Gamma_j}
\end{equation}
which reduces the retrodictive formula $\pr ^{\Lambda\Gamma}(i|j)$ of Eqn~(\ref{2.7}) to
\begin{equation}\label{2.14}
  \pr ^{\Lambda\Gamma}(i|j)=\frac{\tr{\hat\Lambda_i\hat\rho_j^{\textrm{ret}}}}{\tr{\hat\Lambda\hat\rho_j^{\textrm{ret}}}}.
\end{equation}
We say in general that the operation of the preparation device is unbiased if the PDOs are proportional to
$\hat\Xi_i$ where
\begin{equation}\label{2.15}
  \sum_i\hat\Xi_i=\hat 1,
\end{equation}
that is, if the operators $\hat\Xi_i$ form the elements of a preparation device POM. For such a preparation
procedure the retrodictive formula $\pr ^{\Lambda\Gamma}(i|j)$ of Eqn~(\ref{2.7}) simplifies to
\begin{equation}\label{2.16}
  \pr ^{1\Gamma}(i|j)=\Tr{\hat\Xi_i\hat\rho_j^{\textrm{ret}}},
\end{equation}
which is identical to the standard predictive formula of Eqn~(\ref{2.11}) with the roles of preparation and measurement
exchanged.

A specific example of a preparation device with an unbiased operation is where Alice prepares a spin-half
particle in either the \emph{up} state or the \emph{down} state with equal probability. For such a
preparation device, $\hat\Lambda$, the sum of the PDOs is proportional to the identity operator as the PDOs
$\hat\Lambda_\uparrow$ and $\hat\Lambda_\downarrow$ associated with the \emph{up} and \emph{down} states are
proportional to the projectors $|\!\!\uparrow\rangle\langle\uparrow\!\! |$ and
$|\!\!\uparrow\rangle\langle\uparrow\!\! |$ respectively. From Eqn~(\ref{2.7}) the retrodictive probability
that Alice prepared the system in the \emph{up} state given that Bob measured the system in the state
$\hat\rho_j^{\textrm{ret}}=\hat\Gamma_j/\tr{\hat\Gamma_j}$ can be calculated to be
\begin{equation}\label{2.17}
  \pr ^{\Lambda\Gamma}(i=\uparrow|j)=\Tr{|\!\uparrow\rangle\langle\uparrow\! |\hat\rho_j^{\textrm{ret}}}.
\end{equation}
This is consistent with $\hat\Xi_\uparrow=|\!\uparrow\rangle\langle\uparrow\! |$ in Eqn~(\ref{2.16}).

In general however, many preparation procedures are biased in their operation in which case Eqn~(\ref{2.16})
would not apply. For example, the preparation device for an optical field is limited in that it cannot
produce energies above some value. As such the sum of the PDOs is not the identity operator acting in the
entire state space of the system. In the case of the spin-half particle, it may be such that Alice only
produces states that are \emph{up} or in and equal superposition of the states \emph{up} and \emph{down}. For
such situations one must use the more general retrodictive formula of Eqn~(\ref{2.14}).

\section{Time evolution} \label{sec2.4}

In the conventional approach, when a system evolves between the time of preparation $t_p$ and the time of measurement
$t_m$, the density operator describing the state of the system $\hat\rho_i$ undergoes a unitary change. The final state
of the system at the time of the measurement $\hat\rho_i(t_m)=\hat U\hat\rho_i\hat U^\dag$, where $\hat U$ is a unitary
operator, is then substituted for $\hat\rho_i$ in the appropriate probability formulae. Akin to this, the operator
representing the state of the system in this approach, $\hat\Lambda_i$, is considered to undergo a unitary change and
is replaced by $\hat\Lambda_i(t_m)=\hat U\hat\Lambda_i\hat U^\dag$. Noting that $\tr{\hat U\hat\Lambda_i\hat
U^\dag}=\tr{\hat\Lambda_i}$, it is apparent from (\ref{2.8}) that this approach is consistent with the conventional
approach and yields the standard predictive formula (\ref{2.11}) with $\hat\rho_i$ replaced by $\hat\rho_i(t_m)$.

The time reversal of this is to consider the retrodictive picture, in which the general retrodictive probability
formula (\ref{2.14}) would read
\begin{equation}\label{2.18}
  \pr ^{\Lambda\Gamma}(i|j)=\frac{\tr{\hat U\hat\Lambda_i\hat U^\dag\hat\rho_j^{\textrm{ret}}}}
  {\tr{\hat U\hat\Lambda\hat U^\dag\hat\rho_j^{\textrm{ret}}}}.
\end{equation}
From the cyclic property of the trace this can be written as
\begin{equation}\label{2.19}
  \pr ^{\Lambda\Gamma}(i|j)=\frac{\tr{\hat\Lambda_i\hat\rho_j^{\textrm{ret}}(t_p)}}
  {\tr{\hat\Lambda \hat\rho_j^{\textrm{ret}}(t_p)}}
\end{equation}
where $\rho_j^{\textrm{ret}}(t_p)=\hat U^\dag\rho_j^{\textrm{ret}}\hat U$ is the retrodictive density
operator evolved backwards in time to the time of the preparation. From equation (\ref{2.19}) it could be
suggested that the state collapse occurs at the time of the preparation, $t_p$. This is not in conflict with
the conventional predictive formalism as can be seen by replacing $\hat\rho_i(t_m)$ by $\hat U\hat\rho_i\hat
U^\dag$ in the predictive probability formula (\ref{2.11}) and using the cyclic property of the trace to give
the probability as $\tr{\hat\rho_i\hat U^\dag\hat\Pi_j\hat U}$. In this case the set of operators $\{\hat
U^\dag\hat\Pi_j\hat U\}$ constitute a valid POM and can therefore be interpreted as a measuring device that
makes the measurement on the prepared state immediately after the the preparation time $t_p$. This highlights
the arbitrariness as to when the state collapse `really' occurs, and it is seen to be a direct result of the
ambiguity in defining the physical boundary of the measuring device.

\section{Unidirectional flow of information}\label{sec2.5}

In their formalism of the quantum theory of preparation and measurement, Pegg \emph{et al.} \cite{Pegg02}
defined an unbiased measuring device as one for which the sum of the MDOs are proportional to the unit
operator. In this section we show that the physical requirement of causality requires that all faithfully
recording measurement devices must be unbiased. Thus the arrow of time does not arise from quantum mechanics,
it is inserted by means of the conventional postulate that sets of MDOs are represented by POMs but sets of
PDOs are, in general, not. From this we are able to derive an expression representing the choices Alice makes
in preparing a quantum system. The expression relates the PDO to the probability that she chooses to prepare
the system in the state $\hat\rho_i$. We call such a probability the \emph{a priori} probability as we
explicitly show that Alice has sole choice in the states prepared by the preparation device. As a consequence
of causality we find, in contrast, that Bob cannot be assigned such an \emph{a priori} probability as we show
that he cannot choose the outcomes of his measurement device.

\subsection{Unbiased measuring devices}\label{sec2.5.1}
With the arrangement we have described and the associated derived probabilities, we ask the question: can Bob employ
retrodictive states to communicate with Alice at an earlier time by means of a series of a large number of experiments?
Let us assume, for example, that Alice and Bob are a light day apart and the series of experiments takes an hour for
Alice to prepare and equally as long for Bob to measure.

We take it that Bob has control over the choice of the measurement device and whether or not he records a measurement
event that the readout on this device shows has happened. Utilizing this, Bob could then send a message by using his
recording control to vary the probability that the preparation event $i$ is recorded on the computer list, the
expression for which is given by Eqn~(\ref{2.4}). However, Alice could only determine the probability for $i$ and
receive the message \emph{after} Bob has made his contributions to the list. Essentially this process is making use of
the list as a classical communication channel.

We could eliminate this classical means of communication as follows. If we ensure that Alice and Bob must
record every preparation and measurement event shown by the readouts on their devices, then Eqn~(\ref{2.1})
becomes equal to the probability that events $i$ and $j$ are shown on the readouts of the preparation and
measurement devices. We can thus eliminate the computer list entirely and apply the probability formulae to
refer to the readout events themselves. It follows that Eqn~(\ref{2.4}) is then the probability that event
$i$ will be shown on Alice's \emph{readout}, which she can access \emph{before} Bob receives his readout
signal. Bob's recording control is now useless and cannot help Bob play a part in determining Alice's readout
probability given by Eqn~(\ref{2.4}). Can Bob exert his free choice of \emph{measuring device} to alter
$\hat\Gamma$ at will and thus send a message to Alice by changing $\pr ^{\Lambda\Gamma}(i)$ in (\ref{2.4})? A
controllable physical operation on the device can result in a unitary transformation of the MDOs that will
alter $\hat\Gamma$ to $\hat\Gamma'=\hat U\hat\Gamma\hat U^\dag$. If $\hat\Gamma'\neq\hat\Gamma$ then Bob has
a means of communicating with Alice at an earlier time via the retrodictive state by altering the probability
of Eqn~(\ref{2.4}). In order to prevent this, and thus preserve causality, we need to ensure that
$\hat\Gamma'=\hat\Gamma$. The only way this can be done for all such transformations is for $\hat\Gamma$ to
be proportional to the unit operator as this is the only operator that commutes with all possible unitary
transformation operators. We note that Bob could also try to alter $\hat\Gamma$ by deciding not to make a
measurement at all. To ensure that this is impossible, the single `non-measurement' MDO must also be
proportional to the unit operator.

We see, therefore, that \emph{causality}, in the form of a unidirectional flow of information, demands that
the sum of the MDOs for \emph{any} measuring device for which the measurements are faithfully recorded are
proportional to the elements of a \emph{probability operator measure}. Causality thus ensures that any
faithfully recording measuring device provides an unbiased measurement. On the other hand, there is no such
restriction on the PDOs, because information \emph{is} allowed to be transferred via the predictive state into
the future. Faithfully recorded preparation devices therefore \emph{can} be biased. Thus the original time
symmetric equations of Pegg \emph{et al.}, when pertaining to controllable information flow, reduce to a
unidirectional form that encapsulates the essence of the arrow of time. In this form, the predictive and
retrodictive density operators in the absence of information about the actual preparation or measurement
events are, respectively,
\begin{eqnarray}
 \hat\rho &=&\frac{\hat\Lambda}{\tr {\hat\Lambda}}\label{2.20}\\
 \hat\rho^{\mathrm{ret}}&=&\frac{\hat 1}{\tr {\hat 1}},\label{2.21}
\end{eqnarray}
where Eqn~(\ref{2.20}) comes from Eqns~(\ref{2.5}) and (\ref{2.8}). The asymmetry in the normalisation of the
non-negative operators in the usual postulate
\begin{equation}\label{2.22}
  \pr (j|i)=\tr{\hat\rho_i\hat\Pi_j}
\end{equation}
for faithfully recording measurements can now be seen to result from causality.

The `no-information' retrodictive density operator (\ref{2.21}) is the state that is \emph{always} sent back
in time by a non-measurement. In this context it is totally controllable or deterministic. We wish to stress
that this is the \emph{only} retrodictive state that we can send back in time deterministically.

In this thesis, we shall show how to engineer a variety of retrodictive states that are not proportional to
the unit operator by \emph{non-deterministic} means. It is this lack of determinism that prevents such states
from being used to send messages into the past. This can be seen explicitly from Eqn~(\ref{2.7}) and also
(\ref{2.14}) where general retrodictive states are generated conditioned, however, on the outcome $j$ of
Bob's measurement. Causality is preserved in this case because Bob cannot reliably guarantee the outcome $j$
of the measurement, thus Alice must use Eqn~(\ref{2.4}) to try and infer a message from Bob which we
constrained previously to be consistent with causality.

\subsection{Independent \emph{a priori} probability}\label{sec2.5.2}

We postulated in the preceding section that all faithfully recording measuring devices must, in order to
preserve causality, be such that the sum of all the MDOs be proportional to the identity operator acting on
the state space of the system. With such a condition it is then impossible for Bob to communicate with Alice
through the outcomes of her preparation device, the probability of which is given by Eqn~(\ref{2.4}). This
can be seen from (\ref{2.4}) with $\hat\Gamma=\gamma\hat 1$
\begin{equation}\label{2.23}
  \pr ^{\Lambda 1}(i)=\frac{\tr{\hat\Lambda_i}}{\tr{\hat\Lambda}},
\end{equation}
as this probability is $\hat\Gamma_j$-independent. Since this probability depends only on the PDOs describing
Alice's preparation procedure, we attribute \emph{choice} to Alice and say that she can choose the
probability $i$ of an outcome. We can therefore refer to this probability as Alice's \emph{a priori}
probability $\pr^{\Lambda}(i)$,
\begin{equation}\label{2.24}
  \pr^{\Lambda}(i)=\frac{\tr{\hat\Lambda_i}}{\tr{\hat\Lambda}}.
\end{equation}
This \emph{a priori} probability relates to Alice's choice in the states that she decides to prepare, and is
independent of Bob and his subsequent measurement procedure. In contrast, the probability that Bob observes
the outcome $j$ at his measuring device, Eqn~(\ref{2.5}), \emph{does} depend on Alice and her preparation
procedure. This is a direct consequence resulting from the fact that Alice's preparation procedure can be
biased in its operation. That is $\hat\Lambda$ need not be proportional to the identity operator acting on
the space of the system. As such Bob's measurement outcomes are \emph{not} independent of Alice and her
preparation procedure and so Bob cannot be associated with an independent \emph{a priori} probability
$\pr^\Gamma(j)$.

We find that the condition of causality represented by Eqns~(\ref{2.20}) and (\ref{2.21}) allow Alice to be assigned an
independent \emph{a priori} probability representing the choices she makes, but forbids such an assignment for Bob.
Such an asymmetry, in that our choices can only influence the events of the future, again encapsulates the essence of
the arrow of time in the form of unidirectional flow of information.

With Alice's choice represented by the \emph{a priori} probability of Eqn~(\ref{2.24}) we find, from
Eqn~(\ref{2.8}), that we can write $\hat\Lambda_i$ as proportional to $\pr^{\Lambda}(i)\hat\rho_i$. The
constant of proportionality, $\tr{\hat\Lambda}$, always cancels in the expressions for the various
probabilities so there is no loss in generality in setting it to unity. Then we have
\begin{equation}\label{2.25}
  \hat\Lambda_i=\pr^{\Lambda}(i)\hat\rho_i.
\end{equation}
It is now apparent how the PDO $\hat\Lambda_i$, as well as the representing the state of the system, also contains
information about the biasing in its preparation. The biasing factor is just the \emph{a priori} preparation
probability. Because we have taken $\hat\Lambda$ to have unit trace, it too is a density operator given by
\begin{equation}\label{2.26}
   \hat\Lambda=\hat\rho=\sum_i\pr^\Lambda(i)\hat\rho_i,
\end{equation}
which is the sum of all possible states that Alice prepares weighted by their \emph{a priori} probabilities of being
prepared. Void of any knowledge of the preparation outcome $i$, this is the best possible description of the state that
Alice has prepared, when the only known information is the \emph{a priori} probabilities in which she prepares and
records the states.

\section{Conditional state generation}\label{sec2.6}
In principle it should be possible for Alice to generate any predictive state she so desires. Similarly, it
should be possible, at least in principle, for Bob to measure in any basis he desires provided, of course,
his measurement procedure is consistent with Eqn~(\ref{2.9}). Unfortunately, in practice, this is rarely the
case as the range of precise preparation and measurement apparatuses are often limited. It is therefore
always of considerable interest to find novel ways to extend the range of states, both predictive and
retrodictive, that can be generated.

A technique which has long been used to generate a wide range of predictive states is what is referred to
here as conditional preparation. It is a simple technique that begins with a entangled state at the initial
time and, at some time later, involves a measurement on part of the system. The description of the remaining
subsystem is then correlated to the outcome of the measurement event. In this section we show that the
remaining subsystem can indeed be described mathematically by a predictive density operator conditioned on
the outcome of the measurement event. By considering the measurement as part of the preparation procedure we
are able to derive, from the fundamental postulate of Eqn~(\ref{2.1}), a PDO associated with this event. From
this we then derive an expression for the \emph{a priori} probability in which this conditional state is
produced.

What is then interesting is to consider the time reversal of this situation. The technique, which could be
referred to as conditional measurement, has a joint measurement on both subsystems, similar in form to a
Bell-type measurement at the final time. In the retrodictive formalism a retrodictive state is assigned to
the outcome of this joint measurement. This state evolves backwards in time until one of the subsystems is
prepared. As a result of the correlations inherent in the retrodictive state, the remaining subsystem will be
correlated (at earlier times!) to the outcome of the preparation event. By regarding this preparation
procedure as part of the total measurement procedure we show that the remaining subsystem at earlier times
can be described by a retrodictive state conditioned on the outcome of the preparation event. The interesting
feature of a conditional measurement is that we have \emph{control} over the working of the preparation
device. Utilizing this control allows a wider range of retrodictive states to be engineered than those
produced by simple measurement devices alone. The importance of this technique is fundamental to the
remainder of this thesis as all following sections are devoted to the generation of retrodictive states in
this manner.

\subsection{Predictive entangled state}\label{sec2.6.1}
\begin{figure}
\begin{center}
\includegraphics{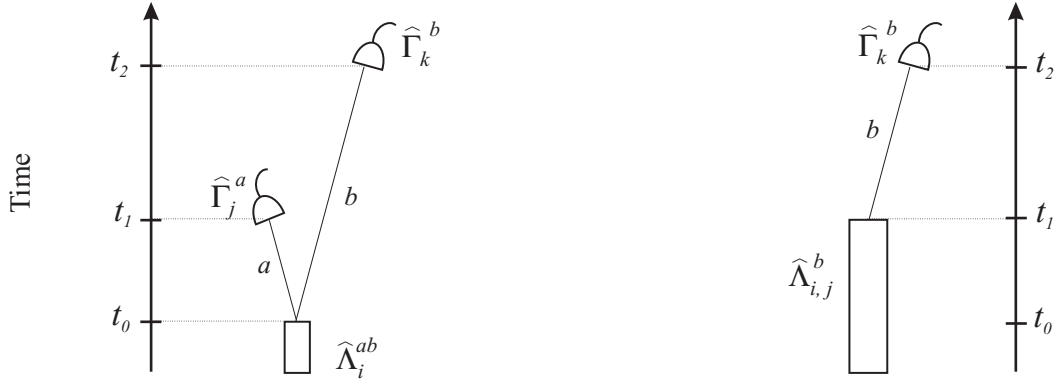}
\caption[Comparison of a bi-part and single system undergoing measurement]{On the left is a bi-part system
undergoing independent measurement of each sub-system at separate times. The square boxes represent
preparation events and the domes represent measurement events. On the right is a single system undergoing
measurement. By considering the measurement at time $t_1$ to be part of the preparation device, we show that
these two systems are physically indistinguishable.}\label{fig2.1}
\end{center}
\end{figure}
Consider the situation illustrated in Figure~\ref{fig2.1}. Depicted is a quantum system with subsystems $a$
and $b$ which are initially prepared in an entangled state described by the PDO $\hat\Lambda^{ab}_i(t_0)$ at
some time $t_0$ associated with a preparation event $i$. After preparation the systems do not interact but
evolve unitarily until a time $t_1$ where a measurement is performed on subsystem $a$. The evolved PDO of the
system just prior to this measurement can be written as $\hat\Lambda^{ab}_i\equiv\hat\Lambda^{ab}_i(t_1)$.
The measurement event is associated with an MDO $\hat\Gamma^a_j$, corresponding to the measurement outcome
$j$. At a later time $t_2$, subsystem $b$ is measured and the outcome $k$ is recorded. The later measurement
event is described by the MDO $\hat\Gamma^b_k$, corresponding to the measurement outcome $k$.

As discussed in Section~\ref{sec2.4} we can obtain the same measured probabilities by redefining the second measurement
to take place at an earlier time $t_1$ provided we replace the MDOs $\hat\Gamma^b_k$ with $\hat\Gamma'{}_k^b=\hat
U_b^\dag(t_2,t_1)\hat\Gamma^b_k\hat U_b(t_2,t_1)$ where $\hat U_b(t_2,t_1)$ is the unitary time displacement operator
acting in the state space of system $b$. Formally these MDOs represent a different measuring device, one in which the
measurement takes place at an earlier time $t_1$. The probability of the joint event $(i,j,k)$ occurring is then, from
Eqn~(\ref{2.1})
\begin{equation}\label{2.27}
  \pr(i,j,k)=\frac{\textrm{Tr}_{ab}[\hat\Lambda^{ab}_i\hat\Gamma_j^a\otimes\hat\Gamma'{}_k^b]}
  {\textrm{Tr}_{ab}[\hat\Lambda^{ab}\hat\Gamma^a\otimes\hat\Gamma'{}^b]},
\end{equation}
where $\hat\Lambda^{ab}=\sum_i\hat\Lambda^{ab}_i$, $\hat\Gamma^{a}=\sum_j\hat\Gamma^{a}_j$ and
$\hat\Gamma'{}^{b}=\sum_k\hat\Gamma'{}^{b}_k$ and the trace is over the state space of both subsystems. The probability
that the later measurement event is $k$ if the earlier one is $j$ is given by
\begin{equation}\label{2.28}
  \pr (k|i,j)=\frac{\pr(i,j,k)}{\pr(i,j)}=\frac{\pr(i,j,k)}{\sum_k\pr(i,j,k)}.
\end{equation}
Substituting from Eqn~(\ref{2.27}) gives
\begin{equation}\label{2.29}
\pr (k|i,j)=\frac{\textrm{Tr}_{b}[\hat\Omega^{b}_{ij}\hat\Gamma'{}_k^b]}{\textrm{Tr}_{b} [\hat\Omega^{b}_{ij}\hat\Gamma'{}^b]},
\end{equation}
where
\begin{equation}\label{2.30}
  \hat\Omega^{b}_{ij}\equiv\frac{\textrm{Tr}_{a}[\hat\Lambda^{ab}_i\hat\Gamma_j^a]}
  {\textrm{Tr}_{ab}[\hat\Lambda^{ab}\hat\Gamma^a]}.
\end{equation}
Using the definition of $\hat\Gamma'{}^b_j$ and the cyclic property of the trace we can rewrite this as
\begin{equation}\label{2.31}
  \pr (k|i,j)=\frac{\textrm{Tr}_{b}[\hat U_b(t_2,t_1)\hat\Omega^{b}_{ij}\hat U_b^\dag(t_2,t_1)\hat\Gamma_k^b]}
  {\textrm{Tr}_{b}[\hat U_b(t_2,t_1)\hat\Omega^{b}_{ij}\hat U_b^\dag(t_2,t_1)\hat\Gamma^b]}.
\end{equation}

It is clear from Eqn~(\ref{2.29}) that $\hat\Omega^{b}_{ij}$ is non-negative, so an alternative interpretation can be
offered to the above configuration. The probability given by Eqn~(\ref{2.31}) is precisely what we would obtain by
decomposing the entire dynamics into a single preparation event at time $t_1$ associated with the PDO
$\hat\Omega^{b}_{ij}$ and a measurement event associated with the MDO $\hat\Gamma_k^b$ at time $t_2$. The preparation
produces a state of the quantum subsystem $b$ that is conditioned on the preparation outcome $(i,j)$. This state
evolves unitarily until a later time when a measurement is performed on this subsystem. If we wish to describe this
state by a density operator instead of a PDO, we can define
\begin{equation}\label{2.32}
  \hat\rho_{ij}^b=\frac{\hat\Omega^{b}_{ij}}{\textrm{Tr}_b[\hat\Omega^{b}_{ij}]}
\end{equation}
at time $t_1$.

The \emph{a priori} probability for this state to be prepared is just the probability for the joint event
$(i,j)$ in the absence of any measurement information about the subsystem $b$. Note that we are now
considering the measurement event $j$ in the original interpretation as part of the joint preparation event.
To find this \emph{a priori} probability we put $\hat\Gamma_k^b=\hat 1_b$ where $\hat 1_b$ is the unit
operator in the state space of subsystem $b$, which is the 'non-measurement' MDO \cite{Pegg02}. From the
formula corresponding to Eqn~(\ref{2.1}), we find the \emph{a priori} probability for the state
$\hat\rho_{ij}^b$ to be prepared is
\begin{equation}\label{2.33}
  \frac{\textrm{Tr}_{ab}[\hat\Lambda^{ab}_i\hat\Gamma^a_j\otimes\hat 1_b]}{\textrm{Tr}_{ab}
  [\hat\Lambda^{ab}\hat\Gamma^a\otimes\hat 1_b]}
\end{equation}
This is precisely the result obtained by finding $\textrm{Tr}_b[\hat\Omega^{b}_{ij}]$, showing that
$\hat\Omega^{b}_{ij}$ is just the state $\hat\rho_{ij}^b$ multiplied by the \emph{a priori} probability that
this state is produced. We are thus justified in representing the state of the remaining subsystem, after the
measurement at $t_1$ is performed, by a predictive density operator prepared by a combined preparation device
comprising the original preparation device and the first measurement device. Using this density operator as
the state of the system immediately after $t_1$, we can calculate correctly the probabilities for any
subsequent measurement.

The result derived in this subsection is one which most physicists accustomed to the predictive formalism
would use intuitively. The purpose of deriving it formally is to establish a method for deriving the
corresponding time-reversed result in the next subsection, which is not so intuitive.

\subsection{Retrodictive entangled state}\label{sec2.6.2}
\begin{figure}
\begin{center}
\includegraphics{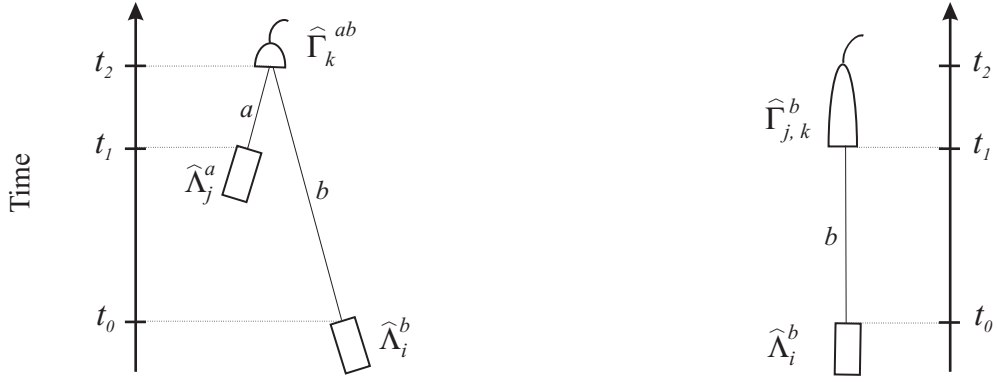}
\caption[Comparison of a bi-part and single system undergoing preparation]{On the left is a system, initially
prepared at two different times by independent preparation procedures, undergoing a joint measurement. On the
right is a single system undergoing measurement. By considering the preparation at time $t_1$ to be part of
the measurement device, we show that these two systems are physically indistinguishable.}\label{fig2.2}
\end{center}
\end{figure}
The retrodictive equivalent to the situation in the previous subsection is illustrated in
Figure~\ref{fig2.2}. We consider a quantum system consisting of two subsystems $a$ and $b$ which undergo a
joint measurement procedure at time $t_2$ which is characterised by the MDO $\hat\Gamma^{ab}_k(t_2)$
associated with the measurement outcome $k$. Before the measurement the systems do not interact but evolve
unitarily from separate preparation procedures, the latest of which occurs at a time $t_1$. As discussed in
Section~\ref{sec2.4} we can formally define the measurement procedure to occur at the earlier time $t_1$ and
associate with it a different MDO $\hat\Gamma^{ab}_k(t_1)\equiv\hat\Gamma^{ab}_k$. To each preparation event
we assign a PDO corresponding to the outcome of the preparation device. For the preparation of subsystem $a$,
occurring at time $t_1$, the PDO $\hat\Lambda_j^a$ is assigned based on the outcome $j$ of the preparation
device, whereas the preparation of subsystem $b$, occurring at the earliest time of $t_0$, is associated with
the PDO $\hat\Lambda^b_i$. Allowing for unitary evolution of the system between preparation events, we can
write the PDO of subsystem $b$ at the later time $t_1$ as $\hat\Lambda_i^{'b}=\hat
U_b(t_0,t_1)\hat\Lambda_i^{b}\,\hat U_b^\dag(t_0,t_1)$, where $\hat U_b(t_0,t_1)$ is the unitary time
displacement operator of subsystem $b$ between times $t_0$ and $t_1$. The probability of the joint event
$(i,j,k)$ occurring is then, from Eqn~(\ref{2.1})
\begin{equation}\label{2.34}
  \pr(i,j,k)=\frac{\textrm{Tr}_{ab}[\hat\Lambda_j^a\otimes\hat\Lambda'{}_i^b\hat\Gamma^{ab}_k]}
  {\textrm{Tr}_{ab}[\hat\Lambda^a\otimes\hat\Lambda'{}^b\hat\Gamma^{ab}]}
\end{equation}
where $\hat\Lambda^{a}=\sum_j\hat\Lambda^{a}_j$, $\hat\Lambda^{'b}=\sum_i\hat\Lambda^{'b}_i$ and
$\hat\Gamma^{ab}=\sum_k\hat\Gamma^{ab}_k$ and the trace is over the state space of both subsystems. Following similar
steps to those presented in the preceding subsection, we arrive at an expression for the retrodictive probability
$\pr (i|j,k)$ that the earliest preparation event was $i$, given the combined outcome $(j,k)$ of the later preparation
event and the final measurement event was observed, to be
\begin{equation}\label{2.35}
  \pr (i|j,k)=\frac{\textrm{Tr}_{b}[\hat\Lambda_i^b\hat U^\dag_b(t_0,t_1)\hat\Phi_{jk}^b\hat U_b(t_0,t_1)]}
  {\textrm{Tr}_{b}[\hat\Lambda^b\hat U^\dag_b(t_0,t_1)\hat\Phi_{jk}^b\hat U_b(t_0,t_1)]}.
\end{equation}
where
\begin{equation}\label{2.36}
\hat\Phi_{jk}^b=\frac{\textrm{Tr}_{a}[\hat\Lambda_j^a\hat\Gamma^{ab}_k]}{\textrm{Tr}_{ab}[\hat\Lambda^a\hat\Gamma^{ab}]}
\end{equation}
is an operator acting solely in the state space of subsystem $b$. From (\ref{2.36}) it is apparent that
$\hat\Phi^b_{jk}$ is a non-negative operator and, as such, this allows an alternative interpretation to be
offered to the configuration. The probability of Eqn~(\ref{2.35}) is identical to that which would be
obtained if the dynamics were decomposed into a single preparation event at time $t_0$ associated with the
PDO $\hat\Lambda_i^b$ and a measurement event associated with the MDO $\hat\Phi_{jk}^b$ at time $t_1$. The
measurement event is conditioned on the combined outcome $(j,k)$. As such, we interpret the second
preparation event occurring at $t_1$ to be part of the measurement event. If we wish to associate a
retrodictive density operator to the measurement event instead of the MDO, we can define
\begin{equation}\label{2.37}
  \hat\rho_{jk}^{\textrm{ret}}=\frac{\hat\Phi_{jk}^b}{\textrm{Tr}_{b}[\hat\Phi_{jk}^b]}
\end{equation}
where the mode label $b$ is dropped for notational convenience. Such a definition is consistent with all the
probability formulae and thus provides a valid alternative interpretation to the arrangement proposed above.
The interesting feature of this retrodictive state is that it is generated conditioned on the
\emph{preparation} outcome $j$ which we \emph{do} have control over. This gives us \emph{some} control over
the retrodictive states that can be prepared but not total control as causality must still be preserved. In
this situation causality is upheld because the generation of the retrodictive state is also conditioned on
the measurement outcome $k$ which we do not have control over. Indeed, we see by summing the MDO
$\hat\Phi^b_{jk}$ in Eqn~(\ref{2.36}) over $j,k$ that the total MDO $\hat\Phi^b$ is proportional to the
identity operator acting on subsystem $b$, showing that this interpretation is consistent with causality. The
control that comes about from the preparation event $j$ can generally be used to select the \emph{type} of
retrodictive states that are being created, rather that the specific state itself. It is this form of control
that is exploited throughout this thesis to engineer retrodictive states which cannot be generated by simple
measurement devices alone.

%% file: Chap3.tex
\chapter{Retrodictive state engineering}\label{chap3}

It has long been known that any physical observable can be represented mathematically by a set of POM
elements. However, not until the projection synthesis technique introduced by Barnett and Pegg
\cite{Barnett96,Pegg97}, and later generalised by Phillips, Barnett and Pegg \cite{Phillips98}, was it
illustrated how any such POM element could be synthesised experimentally. In terms of the retrodictive
formalism, the technique can be regarded as starting with the selection of a standard POM element
conditioned, typically, on a photodetection reading. By way of a beam-splitter and, generally, a
non-classical reference state, the standard POM element is transformed non-unitarily backwards in time to a
POM element representing the observable of interest. Viewed in this way projection synthesis is seen
essentially as a form of retrodictive state engineering. Despite the theoretical implications of this
technique, it does suffer from the practical drawback that it requires a non-classical reference state to
generate the non-unitary transformation.

In this chapter, we extend the projection synthesis technique of Barnett and Pegg to include general linear
optics of multiports, and find that a \emph{coherent} states is all that is needed to synthesise \emph{any}
general optical POM element from a specific photodetection reading. This work provides a general technique
for engineering any retrodictive optical state with a finite number of photon number-state components using
currently available technology. It expands upon the work published by D. T. Pegg and myself
\cite{Pregnell04}.

I begin this analysis with a brief mathematical summary of a useful representation for a state expressible in
a finite dimensional Hilbert space.

\section{Factorizing states in finite dimensional Hilbert spaces}\label{sec3.1}

In general, any pure state can be expressed as a linear combination of basis states
$|\psi\rangle=\sum_{n=0}^\infty\psi_n|n\rangle$, where here $|n\rangle$ is an eigenstate of the photon number operator
and satisfies the orthonormality condition $\langle n|m\rangle=\delta_{n,m}$. By expressing each photon number state in
terms of the vacuum state and the creation operator $\hat a^\dag$ as $|n\rangle=(n!)^{-1/2}\left(\hat
a^\dag\right)^n|0\rangle$, it is possible to represent any state as an infinite summation of integer powers of the
creation operator acting on the vacuum state. However, to any desired degree of accuracy, we can truncate this
summation at some finite integer, N, to give
\begin{equation}\label{3.1}
  |\psi\rangle=\sum_{n=0}^N\frac{\psi_n}{\sqrt{n!}}\left({\hat a^\dag}\right)^n|0\rangle.
\end{equation}

Alternatively, any state which exists in a finite dimensional Hilbert space can be created from the vacuum by repeated
application of the operator $(\hat a^\dag-\beta_i^*)$ to give
\begin{equation}\label{3.2}
  |\psi\rangle=\kappa\left[\prod_{i=1}^N\left({\hat a^\dag-\beta_i^*}\right)\right]|0\rangle
\end{equation}
where $\kappa$ is a normalisation constant and $\beta_i$ is a c-number. To relate the two expressions of $|\psi\rangle$
given above, we act on both sides of (\ref{3.1}) and (\ref{3.2}) with a coherent state $\langle\gamma|$ to give
\begin{equation}\label{3.3}
  \kappa\prod_{i=1}^N\left({\gamma^*-\beta_i^*}\right)=
  \sum_{n=0}^N\frac{{\psi}_n}{\sqrt{n!}}\left({\gamma^*}\right)^n.
\end{equation}
This expression effectively factorizes the polynomial of degree $N$ on the right hand side and so, by setting
$\gamma=\beta_i$, the $N$ values of $\beta_i$ are the $N$ complex roots of the equation in $\gamma^*$ \cite{Clausen00}
\begin{equation}\label{3.4}
  \sum_{n=0}^N\frac{{\psi}_n}{\sqrt{n!}}\left({\gamma^*}\right)^n=0
\end{equation}
The normalisation constant $\kappa$ can be found after substituting the $N$ values of $\beta_i$ back into
(\ref{3.2}). For the remainder of this thesis I will refer to this equation as the characteristic polynomial.
It is an important expression which relates the coefficients $\psi_n$ in one representation to the
coefficient $\beta_i$ in another. As a matter of interest, it is worth mentioning that the complex
coefficients $\beta_i$ correspond to the $N$ zeros in the Q-function representation of an $N+1$ dimensional
state and are a necessary and sufficient set of parameters to describe any general quantum state in an $N+1$
dimensional Hilbert space.

\section[Discrete unitary transformations]{Experimental realization of any discrete unitary transformation}\label{sec3.2}

A beam-splitter unitarily transforms two input modes of a travelling optical field to two output modes. This
transformation is an example of the well known $U(2)$ group \cite{Yurke86}. The natural extension to this is
a multiport device consisting of $N+1$ input modes and $N+1$ output modes where a $(N+1)$-dimensional unitary
matrix $\mathbf{U}(N+1)$ describes the linear transformation of the mode operators. Recently, Reck \emph{et
al.} \cite{Reck94} have shown how any such transformation is equivalent to specific arrays of beam-splitters
and phase-shifters. Such a proof opens the doors to more general experiments, including the recently proposed
linear optics quantum computation scheme of Knill, Laflamme and Milburn \cite{Knill01}. Here we outline the
proof introduced in \cite{Reck94} with the intent to use such a device to generalise the projection synthesis
technique of Pegg and Barnett.

Consider a unitary `rotation' matrix $\mathbf{R}^\dag(N+1)$ that transforms a (normalised) arbitrary row vector in a
$(N+1)$-dimensional vector space into a unit vector in the same space,
\begin{equation}\label{3.5}
  \begin{pmatrix}
    0 \\
    0 \\
    0 \\
    \vdots \\
    0 \\
    e^{i\delta_N} \
  \end{pmatrix}^T
  =
  e^{i\delta_N}\begin{pmatrix}
    ie^{i\phi_0}\sin\theta_0 \\
    ie^{i\phi_1}\sin\theta_1\cos\theta_0 \\
    ie^{i\phi_2}\sin\theta_2\cos\theta_1\cos\theta_0 \\
    \vdots \\
    ie^{i\phi_N}\sin\theta_N\cos\theta_{N-1}\hdots\cos\theta_0 \\
    \cos\theta_{N}\cos\theta_{N-1}\hdots\cos\theta_0
  \end{pmatrix}^T
  \cdot\mathbf{R}^\dag(N+1)
\end{equation}
where the arbitrary vector is represented in generalised spherical coordinates and the phase factor $e^{i\delta_N}$ is
included for generality. By writing the $(N+1)^\textrm{th}$ row of a unitary matrix $\mathbf{U}(N+1)$ in terms of the
generalised spherical coordinates, the action of the rotation matrix $\mathbf{R}^\dag(N+1)$ can be seen to partially
diagonalise the unitary $\mathbf{U}(N+1)$ into a reducible matrix
\begin{equation}\label{3.6}
  \mathbf{U}(N+1)\cdot\mathbf{R}^\dag(N+1)=
  \begin{pmatrix}
  {} & {} & \vline &  \\
  {} &\mathbf{U}(N)& \vline & 0\\
  {} & {} & \vline & {} \\\hline
  {}  & 0  & \vline & e^{i\delta_N}
  \end{pmatrix}.
\end{equation}
Repeated application of rotation matrices of successively lower dimensions can then be used to completely diagonalises
the unitary $\mathbf{U}(N+1)$
\begin{equation}\label{3.7}
  \mathbf{U}(N+1)\cdot\mathbf{R}^\dag(N+1)\cdot\mathbf{R}^\dag(N)\hdots\mathbf{R}^\dag(2)=\mathbf{D}
\end{equation}
where $\mathbf{D}$ is a diagonal matrix with elements $e^{i\delta_n}$ along the leading diagonal. Inverting
Eqn~(\ref{3.7}) gives a factorised expression for $\mathbf{U}(N+1)$ in terms of rotational matrices and a diagonal
matrix as
\begin{equation}\label{3.8}
  \mathbf{U}(N+1)=\mathbf{D}\cdot\mathbf{R}(2)\hdots\mathbf{R}(N)\cdot\mathbf{R}(N+1).
\end{equation}

The achievement of Reck \emph{et al.} was to realise that the necessary rotational matrices $\mathbf{R}(m)$,
$m=2,3,\dots,N+1$, and the diagonal matrix $\mathbf{D}$ can be constructed from selected arrays of
beam-splitters and phase-shifters. The unitary transformation can be physically implemented then by
successively applying each array of beam-splitters and phase-shifters according to Eqn~(\ref{3.8}). To
achieve this physical realisation we first need to consider the effect a lossless beam-splitter has on the
input and output modes of a travelling optical field.

\begin{figure}
\begin{center}
\includegraphics{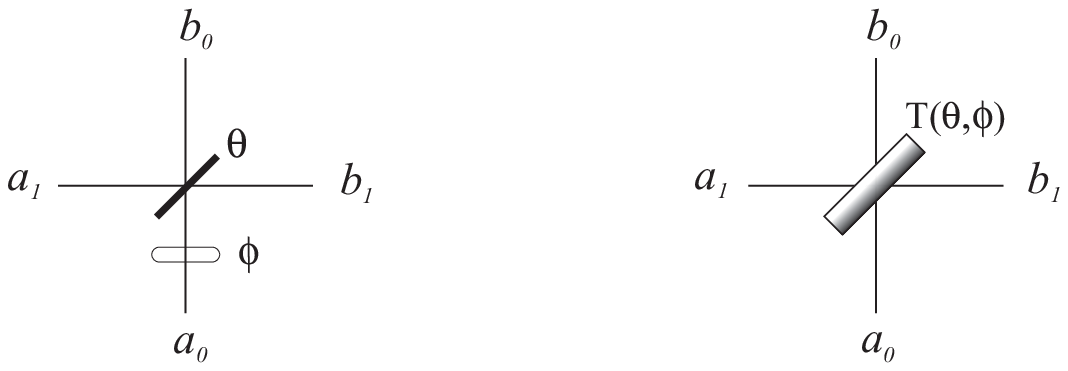}
\caption[Basic optical element necessary to generate arbitrary linear transformations]{Basic element of any
general unitary transformation of mode operators. On the left is a beam-splitter with transmittancy
$\cos\theta$ and phase-shifter generating a phase shift of $\phi$ in input mode 0. On the right is a
shorthand notation for the elements on the left. Note that the blackened side contains the phase-shifter at
the input.} \label{fig3.1}
\includegraphics{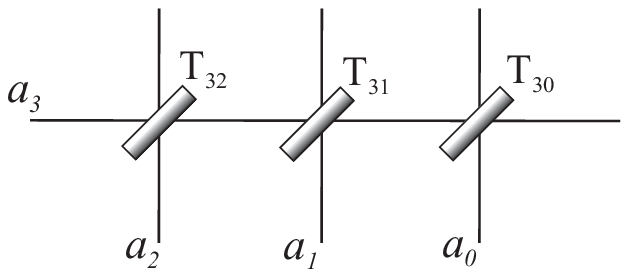}
\caption[A linear array of optical elements necessary to generate a rotational transformation in four
dimensions]{A linear array of optical elements. Such an array is found to produce a general rotation
$\mathbf{R}(4)$ necessary to partially diagonalise a unitary matrix $\mathbf{U}(4)$.} \label{fig3.2}
\includegraphics{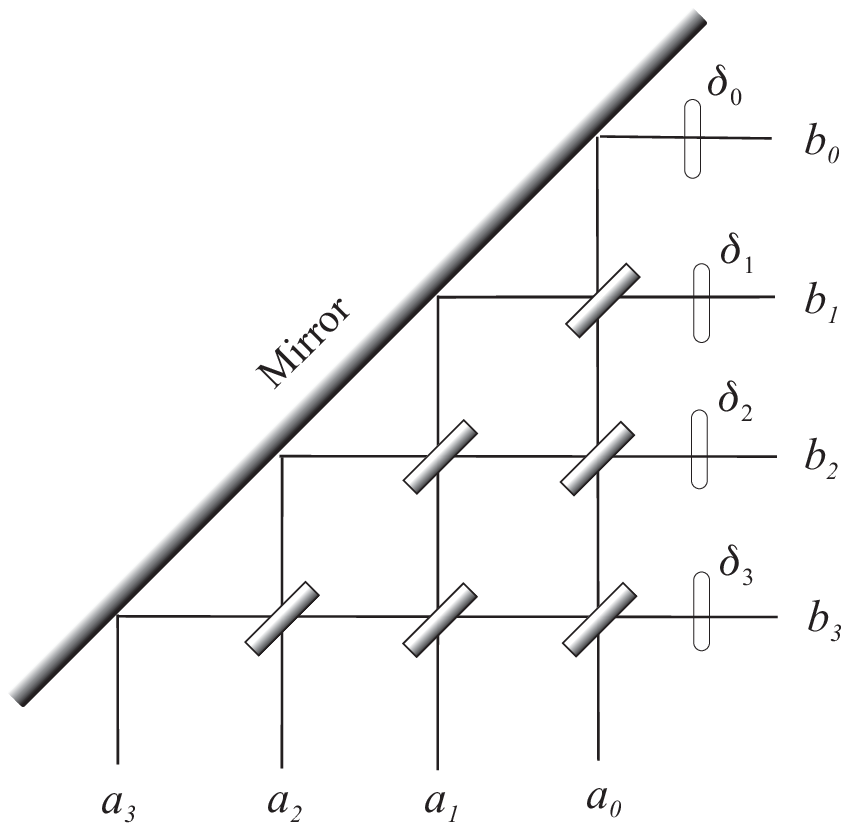}
\caption[A general linear optical transformation of four modes]{A general linear optical transformation of
four modes. The first row of elements from the bottom performs the rotation $\mathbf{R}(4)$, while the second
row performs the rotation $\mathbf{R}(3)$ and so on. The four phase-shifters, as well as off-setting any
phase shift introduced by the mirror, provide the diagonal transformation $\mathbf{D}$.} \label{fig3.3}
\end{center}
\end{figure}

A lossless beam-splitter can be modelled by a linear, unitary transformation of the mode operators $(\hat a_0^\dag,\hat
a_1^\dag)^T$ and $(\hat b_0^\dag,\hat b_1^\dag)^T$ for the appropriate input and output modes respectively. For the
remainder of this thesis I take this transformation to be
\begin{equation}\label{3.9}
  \begin{pmatrix}
  \hat b_0^\dag\\
  \hat b_1^\dag
  \end{pmatrix}
  =
  \begin{pmatrix}
  \cos\theta  & i\sin\theta\\
  i\sin\theta & \cos\theta
  \end{pmatrix}
  \begin{pmatrix}
  \hat a_0^\dag\\
  \hat a_1^\dag
  \end{pmatrix},
\end{equation}
where $t=\cos\theta$ is the transmittance and $r=i\sin\theta$ is the reflectance. It should be mentioned that
this transformation is not unique, indeed Reck \emph{et al.} select a different one in \cite{Reck94}, however
so long as the transformation matrix is unitary the net results will be equivalent. The above transformation
is chosen for two reasons: firstly, in the limit where the beam-splitter becomes totally transmissive the
transformation should approach the identity operator, and secondly, any phase change occurring upon
reflection should be identical on both sides of the beam-splitter. To ensure unitary evolution, a $\pi/2$
phase change is then necessary upon reflection from either side of the beam-splitter. By inserting a
phase-shifter at the input to mode $0$, we can construct a general unitary transformation in $U(2)$ that will
become the basic building block for this scheme. We represent such a element in Figure~\ref{fig3.1}, and
denote the transformation matrix for such an element as
\begin{equation}\label{3.10}
  \mathbf{T}=
  \begin{pmatrix}
  e^{i\phi}\cos\theta  & i\sin\theta\\
  ie^{i\phi}\sin\theta & \cos\theta
  \end{pmatrix}.
\end{equation}

Consider the case of a multiport device where there are $N+1$ input modes and $N+1$ output modes. A general
beam-splitter described by the transformation in Eqn~(\ref{3.10}), combining modes $p$ and $q$, can be
represented by a matrix $\mathbf{T}_{pq}$ which is an $(N+1)$-dimensional identity matrix with the elements
$I_{pp}$, $I_{pq}$, $I_{qp}$ and $I_{qq}$ replaced by the corresponding elements of $\mathbf{T}$. This matrix
will act on the appropriate 2-dimensional subspace, leaving a $(N-1)$-dimensional subspace unchanged. Then by
induction, it is straightforward to show that a rotation matrix introduced in Eqn~(\ref{3.5}) can be
constructed from successive transformation matrices $\mathbf{T}_{p,q}$ as
\begin{equation}\label{3.11}
  \mathbf{R}^\dag(N+1)=\prod_{n=1}^N \mathbf{T}^\dag_{N,N-n}.
\end{equation}
To illustrate, take for example the rotational matrix
$\mathbf{R}(4)=\mathbf{T}_{3,0}\cdot\mathbf{T}_{3,1}\cdot\mathbf{T}_{3,2}$. Physically this is constructed from a
linear array of three beam-splitters, each with one mode in common which we label as mode $3$. Such an arrangement is
illustrated in Figure~\ref{fig3.2}. Then, by progressively applying linear arrays of beam-splitters, we can construct
the the unitary matrix of interest. To implement the diagonal matrix $\mathbf{D}$ a phase shift at the exit of each the
the linear arrays is necessary. The experimental setup of a general $4\times 4$ unitary transformation is illustrated
in Figure~\ref{fig3.3}.
Extending this arrangement to $(N+1)$-dimensions is straightforward. It should be noted that only
$\binom{N+1}{2}=\frac{N(N+1)}{2}$ beam-splitter devices are required to construct a general
$(N+1)$-dimensional unitary transformation. Finally, a general algorithm introduced by Reck \emph{et al.}
\cite{Reck94} has been presented in this section. Such an algorithm allows any finite dimensional unitary
transformation to be constructed from a finite number of beam-splitters and phase-shifters.

\subsection{Generalised 50/50 beam-splitter}\label{sec3.2.1}
A transformation that will recur throughout this thesis is the generalised 50/50 beam-splitter. Such a device
takes a single photon at the input and distributes it evenly across all modes such that a measurement will find, with
equal probability, the photon in any one of the $N+1$ output modes. Conversely, if a single photon is detected in any
one of the outputs, then it is equally likely to have come from any one of the input modes. Mathematically, I describe
such a transformation by the unitary matrix $\mathbf{\Omega}(N+1)$, with elements
\begin{equation}\label{3.12}
  \Omega_{n,m}=\frac{\omega^{nm}}{\sqrt{N+1}},
\end{equation}
where $\omega=\exp[i2\pi/(N+1)]$ is the $(N+1)^{\textrm{th}}$ root of unity. The input and output mode operators
related by such a transformation then form a discrete Fourier transform (DFT) pair. As such, for the remainder of this
thesis I will refer to the transformation described by Eqn~(\ref{3.12}) as a discrete Fourier transform (DFT) in
$(N+1)$-dimensions. As a working example to illustrate the iterative procedure of Reck \emph{et al.} described in the
preceding section, I will derive the DFT in 4-dimensions,
\begin{equation}\label{3.13}
  \mathbf{\Omega}(4)=\frac{1}{2}\left(
  \begin{array}{rrrr}
    1 & 1 & 1 & 1 \\
    1 & i & -1 & -i \\
    1 & -1 & 1 & -1 \\
    1 & -i & -1 & i
  \end{array}\right),
\end{equation}
which will be used later in this thesis.

Following the procedure outlined above, the elements in the last row of $\mathbf{\Omega}(4)$ are equated to the
corresponding elements of the generalised unit vector in Eqn~(\ref{3.5}), giving the set of four equations
\begin{eqnarray}\label{3.14}
  \frac{1}{2} &=&\exp[i(\phi_0+\pi/2+\delta_3)]\,\sin\theta_0\nonumber\\
  -\frac{i}{2}&=&\exp[i(\phi_1+\pi/2+\delta_3)]\,\sin\theta_1\cos\theta_0\nonumber\\
  -\frac{1}{2}&=&\exp[i(\phi_2+\pi/2+\delta_3)]\,\sin\theta_2\cos\theta_1\cos\theta_0\nonumber\\
  \frac{i}{2} &=&\exp[i\delta_3]\,\cos\theta_2\cos\theta_1\cos\theta_0
\end{eqnarray}
that can be solved simply to give
\begin{eqnarray}\label{3.15}
  &\phi_0=\pi & \sin\theta_0=1/\sqrt{4}\nonumber\\
  &\phi_1=\pi/2 & \sin\theta_1=1/{\sqrt{3}}\nonumber\\
  &\phi_2=0 & \sin\theta_2=1/{\sqrt{2}}
\end{eqnarray}
with $\delta_3=\pi/2$. The rotation matrix $\mathbf{R}(4)=\mathbf{T}_{3,0}\cdot\mathbf{T}_{3,1}\cdot\mathbf{T}_{3,2}$
can then be constructed from Eqn~(\ref{3.11}) using the values in Eqn~(\ref{3.15}), where the reflection coefficient
$\sin\theta_i$ and phases $\phi_i$ are terms belonging to the matrix $\mathbf{T}_{3,i}$. Indeed, the rotation matrix
serves to partially diagonalise $\mathbf{\Omega}(4)$
\begin{equation}\label{3.16}
  \mathbf{\Omega}(4)\cdot\mathbf{R}^\dag(4)=\left(
  \begin{array}{rrrr}
    -1/\sqrt 3 & e^{-i\eta}\sqrt{5/12} & e^{-i\pi/4}/2 & 0 \\
    -1/\sqrt 3 & 1/\sqrt{6} & -1/\sqrt 2 & 0 \\
    -1/\sqrt 3 & e^{i\eta}\sqrt{5/12} & e^{i\pi/4}/2 & 0 \\
    0 & 0 & 0 & i
  \end{array}\right)
\end{equation}
leaving a matrix in a 3-dimensional sub-space, where $e^{i\eta}=(3i-1)/\sqrt{10}$. Repeating this procedure a
further two times is sufficient to find $\mathbf{R}(3)$ and $\mathbf{R}(2)$ thereby completely diagonalising
$\mathbf{\Omega}(4)$. Once this is completed, the DFT may be expressed as a product of transformation
matrices according to Eqns (\ref{3.8}) and (\ref{3.11}) as
\begin{equation}\label{3.17}
  \mathbf{\Omega}(4)=\mathbf{D}\cdot\mathbf{T_{1,0}}\cdot\mathbf{T_{2,0}}\cdot\mathbf{T_{2,1}}\cdot\mathbf{T_{3,0}}\cdot
  \mathbf{T_{3,1}}\cdot\mathbf{T_{3,2}}
\end{equation}
where the reflection coefficients $r_{ij}=\sin\theta_{ij}$ and phases $\phi_{ij}$ and $\delta_j$ associated with with
the transformation matrices $\mathbf{T}_{ij}$ and $\mathbf{D}$ are listed in Table \ref{tab3.1}.
\begin{table}\begin{center}
\begin{tabular}{|c|c||r@{\,}l r@{\,}l r@{\,}l|c|}\hline\hline
         \multicolumn{2}{|c||}{}              &                             \multicolumn{6}{c|}{$j$}                          &                \\ \cline{3-8}
 \multicolumn{2}{|c||}{ $t_{ij},\phi_{i,j}$}  & \multicolumn{2}{c}{2}&\multicolumn{2}{c}{1}      &\multicolumn{2}{c|}{0}      &   $\delta_i$   \\\hline\hline
               &      3                       & $1/\sqrt 2$,&$0$ &  $1/\sqrt 3$, & $\pi/2$       & $1/\sqrt 4$,& $\pi$        &  $\pi/2$       \\
     $i$       &      2                       &             &    &  $\sqrt{5/8}$,& $\eta-3\pi/4$ & $1/\sqrt 3$,& $\pi/4$      &  $\pi/4$       \\
               &      1                       &             &    &               &               & $1/\sqrt 2$,& $\eta-3\pi/4$&  $\pi-\eta$    \\
               &      0                       &             &    &               &               &             &              &  $3\pi/4-\eta$ \\\hline\hline
\end{tabular}
\caption[Reflection coefficients and phases for the optical elements necessary to produce a DFT in
4-dimensions]{Reflection coefficients $t_{ij}=\sin\theta_{i,j}$ and phases $\phi_{ij}$ and $\delta_j$ associated with
with the transformation matrices $\mathbf{T}_{ij}$ and $\mathbf{D}$. The collection of optical elements described by
this table is sufficient to realise the transformation $\mathbf{\Omega}(4)$, where the parameter $\eta$ is defined as
$\exp(i\eta)=(3i-1)/\sqrt {10}$} \label{tab3.1}
\end{center}\end{table}

It should be noted that the factorisation of Reck \emph{et al.} is not unique. In fact, we showed in
\cite{Pregnell03a} how the 8-port interferometer used by Noh, Foug\'{e}res and Mandel
\cite{Noh91,Noh92b,Noh93a,Fougeres94} to measure operational phase does, with some additional phase-shifters,
reconstruct the DFT in 4-dimensions. Not only are the values of the transmission coefficients different for
each beam-splitter, but the 8-port interferometer produces the desired transformation using only four
beam-splitters, a reduction of two over the general technique.

For a general DFT in $(N+1)$-dimension, T\"{o}rm\"{a} and Jex \cite{Torma95} have shown how plate
beam-splitters\footnote{A plate beam-splitter is an optical element similar to a beam-splitter, however, the
transmitivity of the element varies along the length of the device. For a review see, for example, \cite{Torma95}.} can
be used to decrease the number of optical elements necessary in the construction. In reference \cite{Torma95}, it was
shown how a general DFT can be constructed using plate beam-splitters, where the number of optical elements scales as
$N+1$, where as the general technique requires $(N+1)^2$, a square root reduction.

\section{Retrodictive state engineering}\label{sec3.3}

\begin{figure}
\begin{center}
\includegraphics{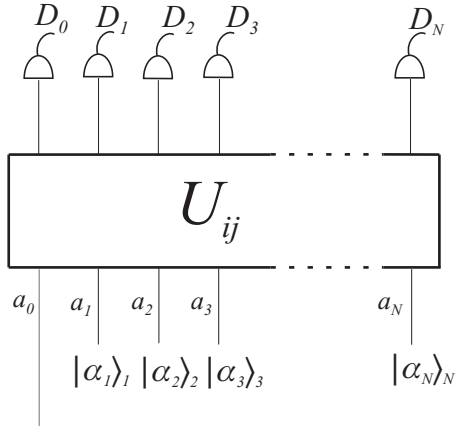}
\caption[Multiport device capable of producing any general retrodictive state in $N+1$ dimensions]{A
multiport device capable of producing \emph{any} retrodictive state that can be written as a superposition of
$N+1$ photon number states. The desired state is produced at the input mode 0 of the multiport when each
photo-detector at the output registers a single photocount, except the one in mode 0, which detects the
vacuum. The reference field $|\alpha_i\rangle$ in input mode $i$ is in a coherent state.} \label{fig3.4}
\end{center}
\end{figure}

Consider the apparatus illustrated in Figure \ref{fig3.4};
it is a multiport device consisting of $N+1$ input ports and $N+1$ output ports. In all output ports a
photodetector capable of distinguishing between zero, one, and more than one photon is positioned. In $N$ of
the input ports, a coherent state $|\alpha_m\rangle_m$ with $m=1,2,\dots,N$ is present, where the label $m$
numbers the corresponding input port. For any photodetector reading at the detectors, there will be some
retrodictive state that propagates backwards in time out of the free input port, which we label as port 0.
For a given photocount pattern, the specific form of the retrodictive state will depend on the construction
of the device and the amplitudes of the $N$ coherent input states. To controllably engineer different
retrodictive states, both the construction of the device could be altered in addition to changing the
amplitudes of the $N$ coherent reference states. However, to maintain an emphasis on a physically
implementable measuring device, we will not consider altering the multiport device to engineer different
retrodictive states, rather we shall consider the more practical alternative of altering the amplitudes of
the $N$ coherent reference states. This leaves us with $N$ controllable parameters, so at most we could hope
to engineer an arbitrary retrodictive state of $N+1$ dimensions as in Eqn~(\ref{3.2}), where the additional
free parameter is constrained by the normalisation condition.

By energy conservation, to obtain a retrodictive state in port 0 containing at most $N$ photons we need to
register a photocount sequence tallying $N$ photocounts. As we wish to use photodetectors that can only
discriminate between zero, one and many photocounts, resembling that of practical photodetectors, the only
possible detection sequence available is when all $N$ photodetector registers a single photocount and one
detector, which we take without loss of generality to be $D_0$, detects no photocounts. For now we consider
the photodetectors to have unit quantum efficiency and consider in Appendix~\ref{appendixA} the implications
of nonunit efficiency. For this case, the POM element for this detection event is simply the tensor product
of the single mode POM elements,
\begin{equation}\label{3.18}
  \hat\Pi(0)=|0\rangle_0{}_0\langle 0|\prod_{m=1}^{N}|1\rangle_m{}_m\langle 1|,
\end{equation}
where the index on the POM element labels the mode where no photocounts where detected. We can write this as
a projector $\hat\Pi(0)=|\Psi\rangle\langle\Psi|$, where
\begin{equation}\label{3.19}
  |\Psi\rangle=|0\rangle_0\prod_{m=1}^{N}|1\rangle_m
\end{equation}
is a multimode state. Following the formalism introduced in Chapter~\ref{chap2}, we define a multimode
retrodictive state at the output ports of the device as the MDO $|\Psi\rangle\langle\Psi|$. We note that this
state requires no renormalisation as its trace is already unity. The evolution of this state backwards
through the multiport device can be represented by a unitary operator, $\hat S^\dag$, acting in the total
Hilbert space of the $N+1$ optical modes. We restrict the action of the device to linear transformations of
the mode operators
\begin{equation}\label{3.20}
  \hat S^\dag\hat a^\dag_m\hat S=\sum_{j=0}^N U_{mn}^*\hat a^\dag_n
\end{equation}
where $U_{nm}$ is an element of a unitary matrix. Such transformations were discussed in the preceding section, where
it was shown how any finite dimensional unitary transformation of mode operators could be constructed from
beam-splitters and phase-shifters.

At the entry of the device the retrodictive state $|\Psi\rangle\langle\Psi|$ has evolved backwards in time
\begin{equation}\label{3.21}
  \hat\rho^\textrm{ret}=\hat S^\dag|\Psi\rangle\langle\Psi|\hat S.
\end{equation}
Also, the PDO describing the preparation event $\hat\Lambda=\hat\Lambda_0\otimes\hat\Lambda_{1,2,\dots,N}$
can be introduced, where
\begin{equation}\label{3.22}
  \hat\Lambda_{1,2,\dots,N}=\prod_{m=0}^{N}|\alpha_m\rangle_m{}_m\langle\alpha_m|
\end{equation}
represents the $N$ coherent reference states and $\hat\Lambda_0$ is the PDO for mode 0. As discussed in
Section~\ref{sec2.6.2}, by redefining the boundaries between the preparation and measurement apparatuses we
can interpret all of Figure~\ref{fig3.4} except input mode 0 as a measuring device. Generalising the
derivation of Section~\ref{sec2.6.2} to $N+1$ systems, we project the $N+1$ mode retrodictive state
$\hat\rho^\textrm{ret}$ onto the $N$ mode PDO $\hat\Lambda_{1,2,\dots,N}$ to produce a single mode MDO for
this enlarged measuring apparatus as,
\begin{equation}\label{3.23}
  \hat\Gamma_0(0)=\Tr{\hat\rho^\textrm{ret}\hat\Lambda_{1,2,\dots,N}}=|\tilde{\psi}\rangle_0{}_0\langle\tilde{\psi}|.
\end{equation}
where the trace is taken over all modes except mode 0. It follows then that the normalised retrodictive state
$\hat\rho^\textrm{ret}_0$, at the entry of the device in input port 0, generated by the measurement event
associated with this MDO is $\hat\Gamma_0(0)/\tr{\hat\Gamma_0(0)}$. This propagates backwards in time from
the measurement event. Because both $\hat\rho^\textrm{ret}$ and $\hat\Lambda_{1,2,\dots,N}$ are not mixed,
the single mode retrodictive state is a projection operator formed from the unnormalised state
\begin{equation}\label{3.24}
  |\tilde{\psi}\rangle_0=\left(\prod_{m=1}^{N}{}_m\langle\alpha_m|\right)\hat S^\dag|\Psi\rangle,
\end{equation}
which is the (unnormalised) retrodictive state at the input port 0 of the device, propagating backwards in
time away from the measurement event. Equation~(\ref{3.23}) can be viewed as the time reversal of the
predictive case where some subsystem of the total system is measured, resulting in an (unnormalised)
predictive state propagating forward in time representing the state of a smaller system.

To evaluate Eqn~(\ref{3.24}), we write the multimode retrodictive state in terms of the creation operators acting on
the vacuum state
\begin{equation}\label{3.25}
  |\Psi\rangle=\left(\prod_{m=1}^{N}\hat a^\dag_m\right)|0\rangle
\end{equation}
where $|0\rangle=|0\rangle_0\otimes|0\rangle_1\otimes\dots|0\rangle_N$ is the multimode vacuum. The evolution of this
state backwards in time through the optical elements results in a general multimode entangled state at the input of the
device, given by
\begin{equation}\label{3.26}
  \hat S^\dag|\Psi\rangle=\left[\prod_{i=1}^{N}\left({\hat S^\dag\hat a^\dag_i\hat S}\right)\right]\hat S^\dag|0\rangle
  =\left[\prod_{i=1}^{N}\left({\sum_{j=0}^N U_{ij}^*\hat a^\dag_j}\right)\right]|0\rangle
\end{equation}
where, by energy conservation, the vacuum state transforms to the vacuum state and Eqn~(\ref{3.20}) has been
used to transform the mode operators. Projecting onto the coherent reference states gives the single mode
retrodictive state as
\begin{equation}\label{3.27}
  |\tilde{\psi}\rangle_0
  =\left({\prod_{j=1}^N{}_{j}\langle\alpha_j|0\rangle_{j}}\right)\left[\prod_{i=1}^{N}
  \left({U_{i0}^*\hat a_0^\dag+\sum_{j=1}^N U_{ij}^*\alpha^*_j}\right)\right]|0\rangle_{0}.
\end{equation}
For simplicity, we define the parameter $\beta_i$, for $i=0,1,\dots,N$ as
\begin{equation}\label{3.28}
  U_{i0}\beta_i=-\sum_{j=1}^NU_{ij}\alpha_j
\end{equation}
and
\begin{equation}\label{3.29}
  \bar\kappa=\exp{\left(-\half\sum_{j=1}^N|\alpha_j|^2\right)}.\prod_{i=1}^N U_{i0}^*
\end{equation}
which simplifies Eqn~(\ref{3.27}) to a form resembling that introduced in Section~{\ref{sec3.1}}
\begin{equation}\label{3.30}
  |\tilde{\psi}\rangle_0=\bar{\kappa}\left[\prod_{i=1}^N\left({\hat a^\dag_0-\beta_i^*}\right)\right]|0\rangle_{0},
\end{equation}
which can represent any state in a $N+1$ dimensional Hilbert space provided arbitrary control over the
parameters $\beta_i$ is available. It should be noted that $\bar\kappa$, as opposed to $\kappa$, is not a
normalisation constant and is constrained by the elements of the unitary transformation matrix and the
amplitudes of the coherent reference states as defined by Eqn~(\ref{3.29}).

To engineer a specific retrodictive state, we need to determine the amplitudes of the $N$ coherent reference states
$\alpha_m$. To do this we need to invert Eqn~(\ref{3.28}). A simple way of doing this is to make use of the unitary
nature of the matrix elements $U_{ij}$ by incorporating two additional variables, $\alpha_0$ and $\beta_0$. To make
sure this does not alter the set of $N+1$ equations we set the value of $\alpha_0$ to be zero. The set of equations can
now be written as
\begin{equation}\label{3.31}
  U_{i0}\beta_i=-\sum_{j=0}^NU_{ij}\alpha_j\quad\quad \mbox{for }i=0,1,\dots,N
\end{equation}
which can be inverted simply to give
\begin{equation}\label{3.32}
  \alpha_j=-\sum_{i=0}^NU_{ij}^*U_{i0}\beta_i\quad\quad \mbox{for }j=0,1,\dots,N.
\end{equation}
The constraint on $\alpha_0$ imposes a restriction on the set of values taken by $\beta_i$,
\begin{equation}\label{3.33}
  \sum_{i=0}^N|U_{i0}|^2\beta_i=0.
\end{equation}
As the $N$ values of $\beta_i$, $i=1,2,\dots,N$, are predetermined by the retrodictive state that we wish to
engineer, this equation serves to define the value of the additional variable $\beta_0$. We note, from
Eqn~(\ref{3.32}), that there is still enough flexibility in the $N$ coherent amplitudes $\alpha_j$ to
generate any set of $N$ values of $\beta_i$.

So we find with control of $N$ coherent reference states we can, upon detection of a specific photocount
pattern, engineer \emph{any} retrodictive state that can be sufficiently well approximated in an $N+1$
dimensional Hilbert space. The unitary evolution of the device required is represented by a linear
transformation of the mode operators and can be implemented physically by beam-splitters and phase-shifters
alone. Surprisingly, almost any lossless multimode device can be used since the values of the coherent
amplitudes can be adjusted accordingly.

\subsection{State optimisation}\label{sec3.3.1}

While the above multiport device is capable of producing a large set of retrodictive states, sometimes only a
particular state or states derived from it by, for example, a phase shift, is all that may be required. In
such cases it is worth optimising the probability with which the apparatus can generate the desired state.
Since the apparatus can be arranged so to produce any retrodictive state whenever the photocount sequence
$(0,1,\dots,1)$ is observed at the detectors $D_0$, $D_1$, etc, this problem is equivalent to maximizing the
probability with which such a sequence is observed at the readouts of the detectors. This is just the
probability of a measurement event for which the corresponding MDO is given by Eqn~(\ref{3.23}). This
probability for any input state will be proportional to $|\bar\kappa|^2$ as can be seen from (\ref{3.30}). So
the problem of maximising the probability in which the apparatus can produce a single retrodictive state
associated with the detection sequence $(0,1,\dots,1)$, reduces to optimising the variable $|\bar\kappa|^2$,
where $\bar\kappa$ is given by (\ref{3.29}).

It is rather a surprising result that almost any lossless multimode device can be used to construct a general
retrodictive state. In fact, the only necessary condition that must be satisfied by the unitary evolution of
the device is a photon entering the device from input port 0 must have a nonzero probability of exiting the
device at each of the $N+1$ output ports. This condition can be stated as $|U_{i0}|\neq0$ for all $i$, and
can be derived from Eqn~(\ref{3.29}) where if $|U_{i0}|$ was zero, for any $i=1,2\dots,N$, the constant
$\bar\kappa$ would be zero. The probability for detecting the state $|\psi\rangle$ would then be zero,
meaning that this state can never be detected with this apparatus. Also if $|U_{00}|$ were to be zero we
would, from Eqn~(\ref{3.33}), lose the freedom in choosing one of the variables $\beta_i$ for $i=1,2,\dots,N$
which would imply restrictions on the allowed retrodictive states that can be engineered.

In determining the value of $\bar\kappa$, it is necessary to evaluate the sum of the squares of the field strengths
$\alpha_j$. From Eqn~(\ref{3.32}) this can be simply expressed as a linear combination of the square of the
characteristic roots,
\begin{equation}\label{3.34}
  \sum_{j=0}^N|\alpha_j|^2=\sum_{i=0}^N|U_{i0}|^2|\beta_i|^2,
\end{equation}
so we can explicitly write $|\bar\kappa|^2$ as
\begin{equation}\label{3.35}
  |\bar\kappa|^2=\exp{\left(-\sum_{i=0}^N|U_{i0}|^2|\beta_i|^2\right)}\cdot\prod_{i=1}^N|U_{i0}|^2.
\end{equation}
As the values of $\beta_i$ are set by the specific state we wish to engineer, it is interesting to note that
$|\bar\kappa|^2$ depends only on the elements in the first column of the transformation matrix which we do have freedom
in choosing, provided
\begin{equation}\label{3.36}
  \sum_{i=0}^N|U_{i0}|^2=1 \quad\mbox{and}\quad |U_{i0}|\neq0
\end{equation}
where the last condition summarises the argument given earlier. Optimising over these variables then implies
\begin{equation}\label{3.37}
  \frac{\partial {|\bar\kappa|^2}}{\partial{x_i}}=0, \quad\quad i=0,1,\dots,N
\end{equation}
where $x_i=|U_{i0}|^2$. Taking Eqn (\ref{3.36}) as a constraint, this problem is best handled using Lagrange
multipliers. The end result will be a set of $N+1$ values for the transmission coefficients which will optimise the
probability in which the apparatus can detect the state $|\psi\rangle$.

\section{Retrodictive phase state with linear optics}\label{sec3.4}

We found that the probability of producing the retrodictive state associated with the detection sequence
$(0,1,\dots,1)$ is proportional to the constant $|\bar\kappa|^2$. We see from Eqn~(\ref{2.5}) that it also
depends upon the input state to the apparatus. It is useful then to define a measure independent of the
initial state that represents the efficiency of production of a retrodictive state. To construct such a
measure, we first consider a von~Neumann type measurement \cite{vonNeumann55} corresponding to the projection
$|\psi\rangle\langle\psi |$, where $|\psi\rangle$ is the normalised version of $|\tilde\psi\rangle$ given by
(\ref{3.2}). We take as our standard a von~Neumann measuring device whose POM incudes this projector. We
define our measure $P_\psi$ then as the ratio of the probability of observing the photodetection sequence
$(0,1,\dots,1)$ to the probability of observing the outcome associated with the state $|\psi\rangle$ with the
von~Neumann measuring device $\pr(\psi)$, with the same input state. From (\ref{2.5}), (\ref{3.2}) and
(\ref{3.30}), we see that this relative probability is equal to $|\bar\kappa/\kappa|^2$, where $\kappa$ is a
normalisation constant defined in (\ref{3.2}).

As mentioned earlier one of the applications of retrodictive state engineering is for projection synthesis.
In this case the retrodictive state production efficiency $P_\psi$ has the following interpretation. When the
generated retrodictive state is used as an MDO for a measurement of a particular observable, $P_\psi$ is the
ratio of the probability of a successful detection event to the corresponding probability for the case where
the POM element is an ideal von~Neumann projector. Thus the relative probability $P_\psi$ is also a measure
of the efficiency of the projection synthesis.

With this defined, we illustrate the effect the chosen transformation can have on the efficiency of the
multiport device when used for projection synthesis. For this illustration, we calculate the efficiency
$P_\psi$ of generating the retrodictive state $3^{-1/2}(|0\rangle+|1\rangle+|2\rangle)$ for three different
multiport configurations. The first multiport considered is the simplest experimental configuration,
consisting of three input ports and three output port formed by a linear array of two beam-splitters with a
single mode in common. The t:r ratio of each beam-splitter was some what arbitrarily chosen as 1:1. The
tradeoff between ease of construction and efficiency is evident in this design, giving the relative
probability $P_\psi$ as a small $0.8\%$. This can be compared to the optimal configuration which is derived
using Lagrange multipliers and can generate the retrodictive state with an efficiency $P_\psi$ of $14.9\%$.
Such a configuration is compared, out of interest, to the DFT in 3-dimensions which is found to have an
efficiency of $13.3\%$, almost as good as for the optimal configuration \cite{Pregnell04}.

The procedure to generate the required retrodictive state is to detect one photon at $D_1$ and $D_2$, and no
photons at $D_0$. Solving the characteristic polynomial for $\tilde \psi _0=\tilde\psi_1=\tilde\psi_2=1$
gives
\begin{equation}\label{3.38}
  \beta_1=\frac{-1}{\sqrt{2}}+i\sqrt{\sqrt{2}-1/2},\quad\quad \beta_2=\beta_1^*
\end{equation}
The coherent stare amplitudes $\alpha_1$ and $\alpha_2$ are then calculated from Eqn~(\ref{3.32}) once the unitary
transformation $\textbf{R}$ is specified. The following three subsections outline the procedure of such a calculation
for the three transformations mentioned above.

\subsection{Simple configuration}\label{sec3.4.1}
\begin{figure}
\begin{center}
\includegraphics{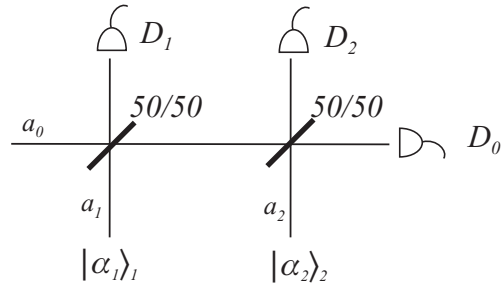}
\caption[Generating a truncated phase state of three dimensions with two beam-splitters]{The simplest optical
arrangement to produce the retrodictive state proportional to $|0\rangle+|1\rangle+|2\rangle$ at the input to
mode 0. The reference fields $|\alpha_1\rangle_1$ and $|\alpha_2\rangle_2$ are in coherent states while
photo-detectors $D_0$, $D_1$ and $D_2$ are in each output mode of the two 50/50 beam-splitters. When the
photo count sequence (0,1,1) is detected at detectors $D_0$, $D_1$ and $D_2$ respectively, the desired
retrodictive is produced.} \label{fig3.5}
\end{center}
\end{figure}

Consider the two 50/50 beam-splitters, $BS1$ and $BS2$, illustrated in Figure~\ref{fig3.5}.
The beam-splitters are arranged such that one of the output modes of $BS1$, labelled $a_0$, is the input mode
to $BS2$. Accordingly, the second input mode of $BS1$ is labelled $a_1$ while the second input mode of $BS2$
is $a_2$. To each of the three outputs a photocounter is positioned. Each photocounter is labelled $D_0, D_1,
D_2$ corresponding to the appropriate mode which it is positioned.

As introduced in Section~\ref{sec3.2} above, the transformation matrix for $BS1$, a symmetric 50/50 beam-splitter
coupling modes $a_0$ and $a_1$, is written as
\begin{equation}\label{3.39}
  \textbf{T}_1=\frac{1}{\sqrt{2}}\left(
  \begin{array}{rrr}
    1  & i & 0 \\
    i & 1 & 0 \\
    0  & 0 & \sqrt {2}
  \end{array}
  \right),
\end{equation}
with $\cos\theta=\sin\theta=1/\sqrt 2$ in Eqn~(\ref{3.9}). Similarly a transformation matrix $\textbf{T}_2$ for $BS2$
can be assigned. Since the beam-splitters act in succession, the joint transformation the two beam-splitters have upon
the three mode incident field is found by the matrix product of $\mathbf{T}_2$ and $\mathbf{T}_1$ which is simply
calculated to be
\begin{equation}\label{3.40}
  \textbf{U}=\textbf{T}_2\textbf{T}_1=\frac{1}{2}\left(
  \begin{array}{rrr}
    1  & i & i\sqrt {2} \\
    i\sqrt {2} & \sqrt {2} & 0 \\
    i  & -1 & \sqrt {2}
  \end{array}
  \right)
\end{equation}

With these values of $|U_{i0}|^2$, the value of $|\bar\kappa|^2$ is calculated for this optical configuration
from Eqn~(\ref{3.35}) to be $1.36\times 10^{-3}$, where the value of $\beta_1$ and $\beta_2$ is given above.
This gives the relative probability $P_\psi$ of producing the retrodictive phase state as $0.8\%$. The
amplitudes of the coherent fields required at the inputs of this apparatus can be calculated from
Eqn~(\ref{3.32}) to be $\alpha_1=1/\sqrt{2}$ and $\alpha_2=-i2\sqrt{\sqrt{2}-1}$. This agrees with the
results of Clausen \emph{et al.} \cite{Clausen00} who studied this particular example from a different
viewpoint.

\subsection{Optimal configuration}

In general there will exist some 6-port configuration of optical elements that will optimise the probability
in which the retrodictive truncated phase state will be produced. It will consist of at most six
beam-splitters and an equal number of phase-shifters. To determine the best transformation describing such a
configuration it is necessary to optimise the parameter $|\bar\kappa|^2$ in Eqn~(\ref{3.35}) over the
transformation matrix elements $|U_{i0}|^2$, $i=0,1,2$. For notational convenience we denoted these variables
as $x_i$ as introduced in Section~\ref{sec3.3.1}. Writing $\beta_1=|\beta|e^{i\theta}$, Eqn~(\ref{3.35})
becomes
\begin{eqnarray}
  |\bar\kappa|^2&=&x_1x_2\exp\left[{-|\beta|^2(x_1+x_2)-\frac{|\beta|^2}{x_0}(x_1^2+x_2^2+2\cos{2\theta}x_1x_2)}\right]\nonumber\\
         &=&x_1x_2\exp(\Theta)\label{3.41}
\end{eqnarray}
where
\begin{equation}\label{3.42}
  \Theta=-\sqrt{2}\left({x_1+x_2+\frac{x_1^2+x_2^2+2\cos{2\theta}x_1x_2}{x_0}}\right)
\end{equation}
with $|\beta|^2=\sqrt 2$ and $\cos{2\theta}=1/\sqrt{2}-1$. The function to optimise, $f(x_0,x_1,x_2)=|\bar\kappa|^2$,
is a function of three variables which are subject to the normalisation constraint $\phi(x_0,x_1,x_2)=x_0+x_1+x_2=1$. A
solution to such a problem is easily obtained by method of Lagrange multiplier \cite{Boas83}. The method
involves finding the minimum of the function
\begin{equation}\label{3.43}
  F(x_0,x_1,x_2)=f(x_0,x_1,x_2)+\lambda\phi(x_0,x_1,x_2),
\end{equation}
over the unknown variables $x_i$, with $\lambda$ referred to as a Lagrange multiplier. The minimum is obtained when
$\partial F/\partial x_i=0$ giving the following set of three equations,
\begin{eqnarray}
    \frac{\partial F}{\partial x_0}&=&f\frac{\partial \Theta}{\partial x_0}+\lambda=0\label{3.44} \\
    \frac{\partial F}{\partial x_1}&=&\frac{f}{x_1}+f\frac{\partial \Theta}{\partial x_1}+\lambda=0 \label{3.45}\\
    \frac{\partial F}{\partial x_2}&=&\frac{f}{x_2}+f\frac{\partial \Theta}{\partial x_2}+\lambda=0\label{3.46}
\end{eqnarray}
where
\begin{eqnarray}
    \frac{\partial \Theta}{\partial x_0}&=&\frac{\sqrt{2}}{x_0^2}(x_1^2+x_2^2+2\cos{2\theta}x_1x_2)\label{3.47}\\
    \frac{\partial \Theta}{\partial x_1}&=&\frac{-\sqrt{2}}{x_0}({x_0+2x_1+2\cos{2\theta}x_2})\label{3.48}\\
    \frac{\partial \Theta}{\partial x_2}&=&\frac{-\sqrt{2}}{x_0}({x_0+2x_2+2\cos{2\theta}x_1})\label{3.49}.
\end{eqnarray}
Subtracting Eqn~(\ref{3.45}) from Eqn~(\ref{3.46}) after substituting Eqns~(\ref{3.48}) and (\ref{3.49}) gives the
expression
\begin{equation}\label{}
  (x_1-x_2)\cdot\left[{\frac{1}{x_1x_2}+\frac{2(2\sqrt 2-1)}{x_0}}\right]\cdot f=0.
\end{equation}
It is not desirable for $f=0$, as this would imply $|\bar\kappa|^2=0$, giving the probability for a successful
detection as zero. The variables $x_i$ are defined to be non-negative, making it impossible for the term in the square
brackets to be zero. The minimum is then achieved only when $x_1=x_2$. By again subtracting, this time Eqn~(\ref{3.44})
from Eqn~(\ref{3.45}), the following third order polynomial in $x_1$ can be be derived
\begin{equation}\label{}
  (4\sqrt{2}-2)x_1^3-(4\sqrt{2}+2)x_1^2+(4+\sqrt 2)x_1-1=0,
\end{equation}
where the normalisation constraint has been used to remove the variable $x_0$. Numerical methods provided a single real
root for this expression in the domain $0\le x_1\le 0.5$ as $x_1=0.2820_5$. The first column of the transformation
matrix which optimises the probability in which a truncated phase state can be detected is then
\begin{eqnarray}\label{3.52}
|U_{00}|^2&=&0.4359_1\nonumber\\
|U_{10}|^2&=&0.2820_5\nonumber\\
|U_{20}|^2&=&0.2820_5.
\end{eqnarray}
which gives the optimal efficiency $P_\psi$ of producing the retrodictive truncated phase state using any
multiport configuration as $14.92\%$ . It is interesting to note that the other six matrix elements along
with the three unspecified phases of $U_{i0}$ are not necessary in optimising $P_\psi$. Since these elements
are unspecified the physical realisation of the optimal optical multiport is not unique. As such these
elements will generally be set by the experimenter when realising the particular optical configuration most
suited to them provided Eqn~(\ref{3.52}) holds. Once realised, then the amplitudes of the necessary coherent
reference states can be derived from Eqn~(\ref{3.32}).

\subsection{Discrete Fourier Transform}

It is interesting to note that there is an intimate relation between the truncated phase state that we are trying to
generate (in retrodiction) and the photon number eigenstates that we are generating it from. In a 3-dimensional Hilbert
space these sets of states form what is know as a discrete Fourier transform pair. That is they satisfy the condition
\begin{equation}\label{3.53}
  \langle n|\theta_m\rangle=\frac{\exp(i2\pi n m/3)}{\sqrt3},\quad\mbox{with}\quad n,m=0,1,2,
\end{equation}
where $|n\rangle$ is a photon number eigenstate and $|\theta_m\rangle$ is a phase state in three dimensions.
Both sets of states form an orthonormal basis in the three dimensional Hilbert space. With this in mind, it
then is natural to consider how effective a discrete Fourier transform multiport device is at transforming
the three mode photon number state $|0\rangle_0|1\rangle_1|1\rangle_2$ into a single mode truncated phase
state.

The optical multiport of interest was introduced in Section~\ref{sec3.2.1} as a generalisation of a 50/50
beam-splitter. The matrix elements of the transformation matrix are given by Eqn~(\ref{3.12}) which are
proportional to powers of $\omega=\exp(i2\pi/3)$. With the matrix elements defined, the probability in which
a truncated phase state can be detected using this optical configuration is calculated to be $13.33\%$
provided the amplitude of the two coherent reference states required at the inputs of the multiport are
$\alpha_1=-1.2591$ and $\alpha_2=-0.1551$.

From the symmetry involved in such a transformation and the intimate relation existing between photon number
states and truncated phase states it is not surprising that this optical configuration is near optimal.
Perhaps what is even more surprising is that the DFT is \emph{not} the optimal configuration as one might be
tempted to think. A closer look at the physical process involved suggests why this is not the case.

The optical multiport, in conjunction with the non-unitary projection onto the coherent reference states,
takes a 3-mode photon number state and transforms it into a single mode retrodictive phase state. The process
is conditioned on observing the particular photocount pattern $(0,1,1)$ at the three photo-detectors. Such a
phot-count pattern inadvertently introduces an asymmetry into the variable $|\bar\kappa|^2$ which we aim to
optimise, with respect to the transformation matrix elements $|U_{i0}|^2$. Accordingly, the optimal optical
arrangement needs to contain a bias in the transformation matrix elements to account for the inherent
asymmetry introduced by the detection event. It is then clear why the unbiased DFT is not the optimal optical
configuration. However, in the example considered in this section the variable $|\bar\kappa|^2$ is symmetric
with respect to the transformation matrix elements $|U_{10}|^2$ and $|U_{20}|^2$ leading to an optimal
configuration where $|U_{10}|^2=|U_{20}|^2$ as derived in the preceding subsection. As the DFT satisfies this
condition it is now clear why such a transformation is \emph{near} optimal, but not exactly optimal.

\section{Generalised measuring apparatus}
In what has preceded, only the retrodictive states corresponding to particular photocount patterns were
studied. As all detection events gives rise to some retrodictive state there is a large number of
retrodictive states not considered. In what follows, we generalise the scheme presented above to include all
detection events and identify the class of retrodictive states that can be generated by a given multiport
configuration.

Consider the optical multiport introduced in Section~\ref{sec3.3}. Injected to all but input mode zero was a
coherent reference state with a controllable amplitude and positioned at each of the $N+1$ output modes is a
photocounting device. To study the potential generation of novel retrodictive states we now consider the
idealised situation where the photocounting device can discriminate between $0,1,2,\dots,M$ photons, where
$M$ is some arbitrary number. Take the situation where by the $i$-th photodetector, $D_i$, registers $n_i$
photons, with $n_i\le M$. Based on the outcome of this measurement event we assign the multimode retrodictive
state
\begin{equation}\label{3.54}
  |\Psi\rangle=\prod_{i=0}^N|n_i\rangle_i=\prod_{i=0}^{N}\frac{(\hat a^\dag_i)^{n_i}}{\sqrt{n_i!}}|0\rangle,
\end{equation}
at the output of the multiport device. The reasoning of such an assignment follows that of the particular
case introduced in Section~\ref{sec3.3}. With the same unitary operator $\hat S^\dag$ responsible for the
evolution of this state backwards in time it becomes a straightforward calculation, akin to that presented
earlier, to derive the single mode retrodictive state $|\tilde\psi\rangle_0$ as
\begin{equation}\label{3.55}
  |\tilde\psi\rangle_0=\bar\kappa\left[\prod_{i=0}^N\left({\hat a^\dag_0-\beta_i^*}\right)^{n_i}\right]|0\rangle
\end{equation}
resulting in a more general expression for $|\bar\kappa|^2$ as
\begin{equation}\label{3.56}
  |\bar\kappa|^2=\exp{\left(-\sum_{i=0}^N|U_{i0}|^2|\beta_i|^2\right)}\cdot\prod_{i=0}^N\frac{|U_{i0}|}{n_i!}^{2n_i}.
\end{equation}
The variable $\beta_i$ remains unchanged to that previously defined in Eqn~(\ref{3.31}) and is also subject to the same
constraint given by Eqn~(\ref{3.56}) which is restated here for convenience as
\begin{equation}\label{3.57}
  \sum_{i=0}^N|U_{i0}|^2\beta_i=0.
\end{equation}

By allowing for more general detection events we generalise the retrodictive states defined by
Eqn~(\ref{3.55}) to include some states of higher dimensions. This happens in two ways. Firstly, by
considering detection events other than zero photons at at detector $D_0$ we introduce terms that are powers
of $({\hat a^\dag_0-\beta_0^*})$ into the retrodictive state. Since the parameter $\beta_0$ is constrained by
Eqn~(\ref{3.57}) above this does not provide an additional degree of freedom necessary to produce arbitrary
states in $N+2$ dimensions. However, by selecting particular transformation coefficients there is some
freedom in choosing particular values of $\beta_0$ which will allow for the generation of some states in
higher dimensions. Secondly, by allowing for multi-photon detection events at all other detectors we
introduce powers in terms like $({\hat a^\dag_0-\beta_i^*})$ which raise the dimensions of the retrodictive state space but do not introduce additional free parameters necessary to provide general states of higher dimensions. It
is quite remarkable that the set of all possible retrodictive states that can be constructed by this
multiport device are characterised by the set of parameters $\beta_i$ which are chosen subject to the
constraint of Eqn~(\ref{3.57}). Even more remarkable is the ease at which these parameters are physically
realised through the amplitudes of the coherent reference states and the chosen multiport device.

To illustrate how the apparatus can generate retrodictive states with higher number state components it will
be shown how the state proportional to $|0\rangle-|N+1\rangle$ can be generated from an $N+1$ multiport
device. Such a state is a specific example of a general quantum state in $N+2$ dimensions. This state is
interesting not just for its extreme non-classical nature but also because it has applications in parameter
estimation techniques. It was shown in \cite{Wiseman04} that such a state, could it be produced, is
necessary to estimate with minimum uncertainty the quantity $\theta=\omega t$ of a harmonic oscillator, where
$\omega$ is the angular frequency of the oscillator and $t$ is the time over which the system oscillates.

Solving the characteristic polynomial for $\tilde\psi_0=-\tilde\psi_{N+1}=1$ gives the $N+1$ solutions
\begin{equation}\label{3.58}
  \beta_n=\sqrt[2(N\!+\!1)]{(N+1)!}\,\,e^{i n\delta\!\theta},
\end{equation}
with $\delta\!\theta=2\pi/(N+1)$ and $n=0,1,\dots,N$. The solution allows the state to be written in factored form as
\begin{equation}\label{3.59}
  |0\rangle-|N+1\rangle=\bar\kappa\prod_{n=0}^{N}\left(\hat a^\dag-|\beta|e^{in\delta\!\theta}\right)|0\rangle
\end{equation}
with $|\beta|=\sqrt[2(N\!+\!1)]{(N+1)!}$. By comparison to the general expression for a retrodictive state,
Eqn~(\ref{3.55}), the desired retrodictive state is generated from this apparatus when each of the $N+1$
photo-detectors detect a single photon.

With $\beta_i$ defined, Eqn~(\ref{3.57}) acts as a constraint on the matrix elements of the optical multiport
device. To satisfy such a constraint it is necessary for the transformation to have matrix elements of equal
magnitude, with the DFT being one such example. Taking the DFT to be the optical transformation, the coherent
reference states required at the input of the multiport are $\alpha_n=-\delta_{n,1}|\beta|$. This equates
physically to a simple configuration that has a vacuum state at the input of modes $2,3,\dots,N$ and a
coherent state with an amplitude of $-|\beta|$ at input mode one. The efficiency with which this apparatus
can generate such a retrodictive state is then
\begin{equation}\label{3.60}
  P_\psi=2\exp(-|\beta|^2)\frac{(N+1)!}{(N+1)^{N+1}},
\end{equation}
which unfortunately decreases exponentially with the number of dimensions $N+1$.

In summary, what has been presented in this chapter is a general apparatus that can generate \emph{any}
retrodictive state which can be represented in a finite dimensional Hilbert space. Such an apparatus can be
used to generalise the projection synthesis technique of Barnett and Pegg to include optical multiport
devices. The most notable advantage of such a generalisation is to replace the non-classical reference state
required in the original projection synthesis technique with simply prepared coherent reference states. Such
a substitution allows for the generation of extremely non-classical retrodictive states from very classical
coherent reference states. In addition, this technique only requires photo-detectors that can discriminate
between none, one and many photons in a single mode which is more like realistic detectors, where as the
original technique requires some detection mechanism that can detect $N$ photons in a single mode. Overall,
the technique presented here is far more practical than the original projection synthesis technique whilst
still being able to generate all states in a Hilbert space with a dimension equal to the number of inputs,
and even some states in a higher dimensional space than this.

Finally we note that the required $N$ coherent input states can be generated from a single coherent state
field by means of a suitable linear array of beam-splitters with phase-shifters in their outputs and vacuum
states in all remaining input ports. This array can be incorporated into the general multiport device giving
a larger multiport that can generate the required retrodictive states with just \emph{one} non-vacuum
coherent state input. This provides a method of projection synthesis using just one coherent reference state
replacing the exotic reference state of the original projection synthesis device \cite{Barnett96}.

%% file: Chap4.tex
\chapter[Simplifying experiments with retrodictive states]{Simplifying experiments with exotic retrodictive probe states}\label{chap4}

Different measurement schemes aim to measure different properties
of a quantum state of light. While some measurement schemes, such
as homodyne \cite{Shapiro79,Yuen83a} and heterodyne
\cite{Walker84,Wiseman93} detection, are physically realisable
there are many which are not. Such proposals
\cite{Pegg99,Steuernagel95}, while still theoretically
interesting, generally require some form of state preparation
which is not achievable with present technology. However, while
the list of predictive states which can be prepared in the
laboratory is relatively small, the list of retrodictive states is
significantly larger as we have seen in the previous chapter. The
asymmetry in generating more exotic retrodictive states than the
predictive counterpart is inherent in the ease at which a photon
number state can be measured as opposed to prepared. With this in
mind we ask the question: is it possible to modify these proposed
experiments such that the necessary state preparation requires an
exotic \emph{retrodictive} quantum state rather than a predictive
state, thereby making the scheme more realisable with current
technology?

In this chapter we take two such proposals which are proving
difficult to implement physically and redesign them to utilise the
non-classical properties of retrodictive quantum states. The first
such proposal, introduced by Pegg, Phillips and Barnett
\cite{Pegg99}, was designed to measure the the phase variance of
light. The scheme was novel but would be difficult to implement as
it requires a two component probe field of the form
$c_0|0\rangle+c_1|1\rangle$. Even using the quantum scissors
device of \cite{Pegg98} to generate such a state by truncating a
coherent state is by no means trivial \cite{Babichev03}. The other
proposal is that of Steuernagel and Vaccaro \cite{Steuernagel95}
to measure directly the density matrix element
$\rho_{N,N\!+\!\lambda}$ of light. Again, the proposal requires a
two-component probe field of the form $|0\rangle+|\lambda\rangle$
to perform the experiment. In general, no such predictive state is
experimentally available. 

In this chapter we propose a single experiment which is capable of measuring both the density matrix elements
\cite{Pregnell02b} $\rho_{N,N\!+\!\lambda}$ in addition to the phase moments $\langle \cos
(\lambda\theta)\rangle$ and $\langle \sin (\lambda\theta)\rangle$ of light \cite{Pregnell01, Pregnell02a}.
The experiment is simple, consisting of only two beam-splitters and one phase-shifter. Remarkably, the only
predictive states required for this proposal are a vacuum state and a reference state which is easily derived
from a coherent state. The experiment has been designed to utilise the non-classical features of retrodictive
states thus removing the emphasis of the two component predictive probe field necessary in the previous
proposals. Because of the inability of detectors to reliably discriminate between large photon numbers in
short time intervals, this the scheme is practical only for relatively weak fields.

\section{Phase variance}

While there are some differences in various theoretical quantum descriptions of the phase of light, a common
feature is that there should be some uncertainty relation between photon number and phase.  Thus the quantum
nature of phase should be manifest as an uncertainty, that is as a non-zero variance in the phase probability
distribution. This uncertainty should be most pronounced for states of light with very small photon number
variances as must pertain, for example, to states that do not differ very much from the vacuum.  By contrast,
strong coherent states of light, which approximate classical states, should have sharply defined values of
phase. For this reason experimental investigations into the quantum nature of the phase of light
\cite{Gerhardt73,Gerhardt74,Noh91,Noh92a,Noh92b,Noh93a} have paid particular attention to finding the width
of the phase distribution of states of light with low mean photon number. As the variance of the phase angle
$\theta$ itself depends critically on the $2\pi$ window assigned to its range of values, such experiments are
usually directed at measuring the phase cosine and sine variances $(\Delta\cos\theta)^2$ and
$(\Delta\sin\theta)^2$. For small phase variances one might expect from expanding the classical series that
$(\Delta\cos\theta)^2+(\Delta\sin\theta)^2\approx(\Delta\theta)^2$. Simple balanced homodyne techniques,
sometimes referred to as phase measurements, can be used to obtain a distribution of a suitably defined
operational, or measured, sine and cosine of phase \cite{Barnett86}. For states with a small enough phase
variance, this distribution can give a very good approximation to the canonical phase distribution
\cite{Vaccaro94} where the canonical phase is defined as the complement of the photon number
operator\footnote{Equivalently, canonical phase can be defined as the quantity possessing a probability
distribution that is invariant under a photon number shift. For further definitions in regards to defining
canonical phase see, for example, \cite{Leonhardt95}. We return to discussing various concepts of phase in
section \ref{sec5.1.1}.} and can be described mathematically by the formalism in \cite{Pegg88,
Barnett89,Pegg89}. For weak fields in the quantum regime, however, which by necessity have broader phase
distributions, significant divergences occur between the canonical and operational phase distributions. This
is also true for the operational phase defined and measured by Noh \emph{et al.}
\cite{Noh91,Noh92a,Noh92b,Noh93a} whose distribution width for coherent states has a maximum divergence from
that of the canonical distribution for mean photon numbers around unity. More recently other techniques have
also been suggested that focus on measuring directly the phase properties of weak fields. These include
projection synthesis \cite{Barnett96, Pegg97} for measuring the canonical phase distribution and a
two-component probe technique \cite{Pegg99} for measuring the canonical phase cosine or sine variance. As
previously mentioned these techniques rely on engineering specifically tailored (predicitve) probe states
that, although possible in principle, will be very difficult in practice and have so far not been produced.
Thus on one hand there are techniques that use easily prepared states but which do not measure the canonical
phase variances and on the other there are techniques that measure canonical phase variances but rely on
exotic quantum states.

In this section we examine the possibility of measuring the canonical phase cosine and sine variances of
optical fields, with a particular interest in weak fields,  by using input states which are easily produced
in the laboratory, that is, coherent states. We find that, even though the two-component predictive probe
states needed for the technique of \cite{Pegg99} are effectively not available at present, it is not
difficult to use a \emph{retrodictive} two-component probe state for our purposes.

\subsection{Mean of the phase sine}

Different quantum descriptions of phase will yield different values of $\langle \cos (\lambda\theta)\rangle$ and
$\langle\sin (\lambda\theta)\rangle$. For canonical phase,
\begin{equation}\label{4.1}
  \langle \cos (\lambda\theta)\rangle={\mbox{$\frac{1}{2}$}}\sum_{p=0}^\infty\left(\rho_{p,p+\lambda}+\rho_{p+\lambda,p}
  \right)
\end{equation}
and
\begin{equation}\label{4.2}
  \langle \sin (\lambda\theta)\rangle={\mbox{$\frac{i}{2}$}}\sum_{p=0}^\infty\left(\rho_{p,p+\lambda}-\rho_{p+\lambda,p}
  \right)
\end{equation}
where $\rho_{n,m}$ are elements of the optical density matrix. One method of obtaining Eqns~(\ref{4.1}) and (\ref{4.2})
is the limiting procedure of \cite{Pegg88, Barnett89, Pegg89}. Here
\begin{equation}\label{4.3}
  \langle\cos(\lambda\theta)\rangle=\lim_{s\rightarrow\infty}\langle\cos(\lambda\hat\phi_\theta)\rangle
\end{equation}
where $\hat\phi_\theta$ is the Hermitian phase operator acting on a $(s+1)$-dimensional Hilbert space. For a mixed
state $\hat\rho$,
\begin{equation}\label{4.4}
  \langle\cos(\lambda\hat\phi_\theta)\rangle=\Tr{\hat\rho_s\cos(\lambda\hat\phi_\theta)}
\end{equation}
where $\hat\rho_s$ is the truncation of $\hat\rho$ onto the $(s+1)$-dimensional subspace. Within this formalism, where
the operators $\cos\hat\phi_\theta$ and $\sin\hat\phi_\theta$ commute, it is possible to derive the relations
\begin{equation}\label{4.5}
  \langle\cos^2\theta\rangle= \mbox{$\frac{1}{2}$}\left[1+\langle\cos(2\theta)\rangle\right]
\end{equation}
which will be used later in calculating the variance in the phase sine and cosine. Equations~(\ref{4.1}) and
(\ref{4.2}) are the same results that would have been obtained by using
\begin{eqnarray}
  \langle\cos(\lambda\theta)\rangle &=& \Tr{\hat\rho\,\hat C_\lambda}\label{4.6}\\
  \langle\sin(\lambda\theta)\rangle &=& \Tr{\hat\rho\,\hat S_\lambda}\label{4.7}
\end{eqnarray}
where the operators
\begin{eqnarray}
  \hat C_\lambda ={\mbox{$\frac{1}{2}$}}\sum_{p=0}^\infty|p+\lambda\rangle\langle p\,|+|p\rangle\langle p+\lambda|
  \label{4.8}\\
  \hat S_\lambda ={\mbox{$\frac{i}{2}$}}\sum_{p=0}^\infty|p+\lambda\rangle\langle p\,|-|p\rangle\langle p+\lambda|
  \label{4.9}
\end{eqnarray}
act on the usual infinite-dimensional Hilbert space.

\begin{figure}
\begin{center}
\includegraphics{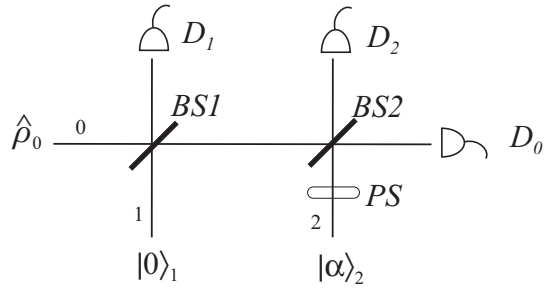}
\caption[Experimental apparatus for measuring the sine and cosine variances]{Experimental apparatus for
measuring the sine and cosine variances. The controllable reference state in input mode 2 of \emph{BS2} is a
coherent state while the state to be measured $\hat\rho_0$ in input mode 0 of \emph{BS1}. In the other input
mode of \emph{BS1}, mode 1, there is a vacuum state. There is also a photo-detector in each of the three
output modes.} \label{fig4.1}
\end{center}
\end{figure}

Our proposed measurement technique uses the beam-splitter arrangement shown in Figure~\ref{fig4.1}.
A controllable reference field in a coherent state
$|\alpha\rangle_2=\sum_n a_n|n\rangle_2$ is in the input mode $2$
of a 50/50 symmetric beam-splitter $BS2$. The state of the system
$\hat\rho_0$ to be measured is in the input mode $0$ of
beam-splitter $BS1$ and a vacuum state is in input mode $1$ of
$BS1$. The transmission and reflection coefficients of $BS1$
remain, for now, unspecified. Photon detectors $D_2$, $D_0$ and
$D_1$ are in the output mode $2$ and output mode $0$ of $BS2$ and
in the output mode $1$ of $BS1$. We shall assume for now that
these detectors can count photons with perfect efficiency, no dark
counts and negligible dead time. We address such an assumption in
Appendix~\ref{appendixA} where it is shown how to correct for some
of these imperfections. It should be noted that this technique is
similar in principle to the projection synthesis of
\cite{Barnett96,Pegg97,Phillips98} in which the unknown quantity
is measured by observing specific relative frequencies of
particular event in an experiment. As such, multiple copies of the
state $\hat\rho_0$ will be necessary to obtain accurate sampling.

In the generalised measurement theory introduced in Chapter~\ref{chap2}, a set of MDOs is assigned to
describe the total measurement procedure which is a two-step process incorporating both measurement and
recording of the desired outcome. With appropriate normalisation of the MDO corresponding to a particular
measurement result, a retrodictive state can be assigned to the field immediately prior to the measurement
event. This state evolves backwards in time until the field interacts with the preparation apparatus. The
preparation apparatus is described in general by a set of PDOs with elements corresponding to possible
outcomes of the preparation event. The probability of the joint preparation and measurement event occurring
is given by the projection of the evolved PDO onto the associated MDO, that is, by the trace of the product
of the MDO and the PDO. To avoid unnecessary complications, we assume the PDO we assign to the input fields
of the device in Figure~\ref{fig4.1} describe the fields at their entry to the beam-splitters.  The free
evolution in the intermediate mode $0$ between the beam-splitters only changes the phase of the field in this
mode so, by choosing the distance between beam-splitters to be an integer number of wavelengths, we can
ignore this evolution. In practice, even if this is not the case such a phase shift can be compensated by
adjusting the phase of the reference field $|\alpha\rangle_2$. Finally we can ignore the free evolution in
all the output modes as these do not affect the photocount probabilities. We denote the total (forward time)
unitary operator for the actions of beam-splitters $BS1$ and $BS2$ as $\hat S=\hat S_2\otimes\hat S_1$ where
$\hat S_2$ acts on states in modes $2$ and $0$, and $\hat S_1$ acts on states in modes $1$ and $0$.

Since the preparation procedure is always the same there is only one PDO, so we denote this by $\hat\Lambda$.
The initial combined PDO for the three input field is the tensor product of the individual density matrices
is then
\begin{equation}\label{4.10}
  \hat\Lambda=\hat\rho_0\otimes|0\rangle{}_1{}_1\langle0|\otimes|\alpha\rangle{}_2{}_2\langle\alpha|.
\end{equation}
The MDO $\hat\Gamma(n_0,n_1,n_2)$ for the detection of $n_0$, $n_1$ and $n_2$ photons in output modes $0$,
$1$ and $2$ respectively is, up to an arbitrary constant which is set to unity, equivalent to the POM element
\begin{equation}\label{4.11}
  \hat\Pi(n_0,n_1,n_2)=|n_0\rangle{}_0{}_0\langle n_0|\otimes|n_1\rangle{}_1{}_1\langle n_1|\otimes|n_2
  \rangle{}_2{}_2\langle n_2|.
\end{equation}
as the detection event is unbiased. The probability for the detection of $n_0$, $n_1$ and $n_2$ photons at detectors
$D_0$, $D_1$ and $D_2$ respectively is, from Eqn~(\ref{2.6}) with $\hat\Gamma=\hat 1$,
\begin{equation}\label{4.12}
  \pr(n_0,n_1,n_2)=\Tr{\hat\Lambda\hat S^\dag\hat\Pi(n_0,n_1,n_2)\hat S},
\end{equation}
where the trace is taken over all modes. Substituting from Eqns~(\ref{4.10}) and (\ref{4.11}) this can be rewritten as
\begin{equation}\label{4.13}
  \pr(n_0,n_1,n_2)=\Trs{0}{\hat\rho_0\hat\Gamma_0(n_0,n_1,n_2)}.
\end{equation}
Here $\hat\Gamma_0(n_0,n_1,n_2)$ is the MDO for the measurement event where the measuring device which is
defined as everything in Figure~\ref{fig4.1} except the state to be measured, $\hat\rho_0$. This can be
expressed as
\begin{eqnarray}\label{4.14}
\hat\Gamma_0(n_0,n_1,n_2)&=&\Trs{1,2}{|0\rangle{}_1{}_1\langle0|\otimes|\alpha\rangle{}_2{}_2\langle\alpha|\hat S^\dag
  \hat\Pi(n_0,n_1,n_2)\hat S}\\
  &=&|\tilde{\psi}\rangle{}_0{}_0\langle\tilde{\psi}|\nonumber,
\end{eqnarray}
where, from Eqn~(\ref{4.11}),
\begin{equation}\label{4.15}
  |\tilde{\psi}\rangle{}_0={}_1\langle 0|\hat S_1^\dag|n_1\rangle{}_1\,{}_2\langle\alpha|\hat
  S_2^\dag|n_0\rangle{}_0|n_2\rangle{}_2.
\end{equation}
The state $|\tilde{\psi}\rangle{}_0$ is interpreted as the unnormalised retrodictive state of the field in input mode
$0$ associated with the detection of $n_0$, $n_1$ and $n_2$ photons. We can see from Eqn~(\ref{4.15}) that the
retrodictive fields $|n_0\rangle_0$ and $|n_2\rangle_2$ associated with the measurements in the output mode $0$ and
output mode $2$ evolve backwards in time and are entangled by means of beam-splitter $BS2$.  This entangled state is
projected onto $|\alpha\rangle_2$ to yield an unnormalised retrodictive probe state
\begin{equation}\label{4.16}
  |q\rangle_0={}_2\langle\alpha| \hat S_2^\dag |n_0\rangle_0|n_2\rangle_2
\end{equation}
in the intermediate mode $0$, that is, between the two beam-splitters.  As we shall see later this is a
retrodictive two-component state which performs a similar function to the predictive two-component probe of
\cite{Pegg99}. The state $|q\rangle_0$ is entangled by beam-splitter $BS1$ with the retrodictive state from
the measurement outcome of the detector $D_1$. This state in turn is projected onto the vacuum in input mode
$1$ to give the unnormalised retrodictive state $|\tilde\psi\rangle_0$ for projection onto the state to be
measured. We remark here that if the state $\hat\rho_0$ is a coherent state then in the predictive picture
there is no entanglement at all because all input states are coherent. The entanglement mentioned above
occurs only in retrodiction. In addition to giving new insight, working in terms of the retrodictive probe
state has practical calculational advantages. In our case, the fields that evolve backwards originate from
single photon number states associated with the measurement outcomes in contrast to the fields that evolve
forwards which contain multi-photon superpositions states.

The simplest possible retrodictive probe is associated with the detection event $n_0=n_2=0$.  In this case we find that
the retrodictive probe state $|q\rangle_0$ is just the vacuum and so, with the detector $D_1$ detecting $n_1=N$
photons, $|\tilde{\psi}\rangle{}_0$ is just proportional to $|N\rangle{}_0$. Thus only the diagonal matrix elements of
$\hat\rho_0$ are obtainable from the measured probabilities. The next simplest retrodictive probes are associated with
the measurement result $n_2=0$, $n_0=1$ and $n_2=1$, $n_0=0$. For a symmetric beam-splitter the output mode operators
are related to the input mode operators by the unitary transformation of Eqn~(\ref{3.9}). For a 50/50 beam-splitter
$\theta=\pi/4$. Using this value for $BS2$ we easily find, by writing $|1\rangle_0=\hat b^\dag_0|0\rangle_0$, the
unnormalised retrodictive probe state for $n_2=0$, $n_0=1$ to be
\begin{equation}\label{4.17}
  \sqrt{2}|q\rangle_0=a_0^*|1\rangle_0+ia_1^*|0\rangle_0
\end{equation}
When detector $D_1$ registers $n_1=N$ photons, this two-component
retrodictive probe evolves backwards through $BS1$ and becomes
entangled. In retrodiction, the detection of $N$ photon in the
output mode $1$ of $BS1$ acts like a photon source to the
retrodictive probe state. The result of the input mode being in a
vacuum state is to raise the photon occupation number in the state
$|q\rangle_0$ by $N$. This can be seen explicitly from
Eqn~(\ref{4.15}), where it is shown in the
Appendix~\ref{appendixB} that the operator ${}_1\langle 0|\hat
S_1^\dag|N\rangle{}_1$ is proportional to the raising operator to
the $N^{\textrm{th}}$ power. Using this expression it is trivial
to derive the retrodictive probe state at the input to mode $0$ as
\begin{equation}\label{4.18}
  \sqrt 2|\tilde\psi\rangle_0=(ir)^N\left(ia_1^*|N\rangle_0+t a_0^*\sqrt{N\!+\!1}\;|N+1\rangle_0\right)
\end{equation}
where $t=\cos\gamma$ and $r=\sin\gamma$ are the transmission and reflection coefficients of $BS1$ respectively.

This in turn allows the MDO $\hat\Gamma_0(1,N,0)$ to be calculated from Eqn~(\ref{4.14}) and hence, from
Eqn~(\ref{4.13}), the measurable probability
\begin{equation}\label{4.19}
  \pr(1,N,0)=\mbox{$\frac{1}{2}$}r^{2N}\left[|a_1|^2\rho_{N,N}+t^2|a_0|^2(N\!+\!1)\rho_{N\!+\!1\!,N\!+\!1}
  +\left(ia_0a_1^*t\sqrt{N\!+\!1}\,\rho_{N\!+\!1\!,N}+c.c\right)\right]
\end{equation}
From this we see that that the measured probability depends upon the first off-diagonal matrix elements $\rho_{N+1,N}$,
and its complex conjugate, of the input field $\hat\rho_0$. Unfortunately, is also depends upon the diagonal elements
$\rho_{n,n}$. To remove this dependence and obtain a measurable value for $\langle\sin\theta\rangle$ in Eqn~(\ref{4.1})
we also measure the probability in which $n_2=1$, $n_0=0$ and $n_1=N$ photon are detected. We find this occurs with a
probability given by
\begin{equation}\label{4.20}
  \pr(0,N,1)=\mbox{$\frac{1}{2}$}r^{2N}\left[|a_1|^2\rho_{N,N}+t^2|a_0|^2(N+1)\rho_{N\!+\!1\!,N\!+\!1}
  -\left(ia_0a_1^*t\sqrt{N\!+\!1}\,\rho_{N\!+\!1\!,N}+c.c\right)\right].
\end{equation}
For the first experiment we choose the phase of coherent reference state $|\alpha\rangle_2$ so that $a_n$ are real and
positive. From Eqns~(\ref{4.19}), (\ref{4.20}) and (\ref{4.2}) an expression for the mean of the sine in terms of the
measurable probabilities can then be written as
\begin{equation}\label{4.21}
  \langle\sin\theta\rangle=\sum_N\frac{\pr_0(0,N,1)-\pr_0(1,N,0)}{2tr^{2N}|a_0a_1|\sqrt{N+1}}
\end{equation}
where the subscript on the probability refers to the first experiment.

\subsection{Variance of the phase cosine and sine}

The next simplest possible retrodictive probe originates from the measurement event $n_0=n_2=1$. Again for the 50/50
beam-splitter BS1 we have $\theta=\pi/4$.  Writing $|1\rangle_0=\hat b^\dag_0|0\rangle_0$ and $|1\rangle_2=\hat
b^\dag_2|0\rangle_2$ we obtain from Eqn~(\ref{3.9}), the two-component retrodictive probe state
\begin{equation}\label{4.22}
  -i\sqrt{2}\,|q\rangle_0=a_0^*|2\rangle_0+a_2^*|0\rangle_0
\end{equation}
leading to
\begin{equation}\label{4.23}
  -i\sqrt{2}\,|\tilde\psi\rangle_0=(ir)^N\left(a_2^*|N\rangle_0+a_0^*t^2\sqrt{(N+1)(N+2)/2}\,|N+2\rangle_0\right).
\end{equation}
From Eqns~(\ref{4.13}) and (\ref{4.14}), this gives the probability of detecting a single photon at detectors $D_0$ and
$D_2$ and $N$ photons at $D_1$ as
\begin{eqnarray}\label{4.24}
  \pr(1,N,1)=&\mbox{$\frac{1}{2}$}r^{2N}\Big[|a_2|^2\rho_{N,N}+t^4(N+2)(N+1)/2|a_0|^2\rho_{N+2,N+2}\nonumber\\
  &+\left.\left(a_0a_2^*t^2\sqrt{(N+2)(N+1)/2}\,\rho_{N+2,N}+c.c\right)\right].
\end{eqnarray}
To allow the experiment with this probe to be conducted simultaneously with the experiment to find
$\langle\sin\theta\rangle$, we take $a_0$ and $a_2$ to be real and positive and find $\pr_0(1,N,1)$ by simply replacing
$a_0a_2^*$ in Eqn~(\ref{4.24}) by $|a_0a_2|$.

After measuring the probabilities $\pr_0(1,N,0)$, $\pr_0(0,N,1)$ and $\pr_0(1,N,1)$ the experiment is repeated with a
phase shift of $\pi/2$ in the reference state $|\alpha\rangle_2$, which has the effect of changing $a_n$ to
$a_n\exp(in\pi/2)$. Thus now $a_0=|a_0|$, $a_1=i|a_1|$, and $a_2=-|a_2|$ in Eqns~(\ref{4.19}), (\ref{4.20}) and
(\ref{4.24}), yielding $\pr_1(1,N,0)$, $\pr_1(0,N,1)$ and $\pr_1(1,N,1)$. From Eqn~(\ref{4.1}) we can then obtain the
mean phase cosine from the measured results as
\begin{equation}\label{4.25}
  \langle\cos\theta\rangle=\sum_N\frac{\pr_1(1,N,0)-\pr_1(0,N,1)}{2tr^{2N}|a_0a_1|\sqrt{N+1}}.
\end{equation}
We also find from (\ref{4.1}) that
\begin{equation}\label{4.26}
  \langle\cos(2\theta)\rangle=\sum_N\frac{\pr_0(1,N,1)-\pr_1(1,N,1)}{2t^2r^{2N}|a_0a_2|\sqrt{(N+1)(N+2)/2}}.
\end{equation}
After these values are obtained from the measured probabilities, the mean square phase cosine can be found from
Eqn~(\ref{4.5}) and finally the phase cosine variance calculated as
$\langle\cos^2\theta\rangle-\langle\cos\theta\rangle^2$. Further, we can also write from the phase formalism of
\cite{Pegg88,Barnett89,Pegg89}
\begin{equation}\label{4.27}
    \langle\sin^2\theta\rangle =\mbox{$\frac{1}{2}$}\left[1-\langle\cos(2\theta)\rangle\right],
\end{equation}
which allows us also to find the phase sine variance from the measured probabilities.

We have already assigned a value to the phase of the reference
state $|\alpha\rangle_2$ but still have freedom to choose its mean
photon number $|\alpha|^2$.  To avoid having to renormalise very
small probabilities, it is worth maximising the denominator, and
hence the numerator, in Eqns~(\ref{4.21}) and (\ref{4.25}).  Thus
we should choose a reference state to maximise $|a_0a_1|$ and for
Eqn~(\ref{4.26}) we should maximise $|a_0a_2|$. The former and
latter are maximised for mean photon numbers of 0.5 and 1
respectively. The experiment could in principle be run for both
these values but in practice it would be simpler just to use a
compromise value between 0.5 and 1. We have yet to choose the
reflection to transmission ratio of the beam-splitter $BS2$. Again
it is useful to choose a ratio which maximises the denominators of
the terms in Eqn~(\ref{4.21}). The optimum value of the reflection
coefficient $\sin\gamma$ for each term is $(1+2N)^{-1/2}$. For
Eqn~(\ref{4.26}) the optimum value of $\sin\gamma$ for each term
is $(1+N)^{-1/2}$. If necessary the experiment could be repeated
for different values of $N$ but, given we are mainly interested in
weak fields, the spread in values of $N$ should not be huge. Thus
a compromise value of around $\langle n\rangle$ should be adequate
for determining both Eqns~(\ref{4.21}) and (\ref{4.26}), where
$\langle n\rangle$ is the mean photon number of the field to be
measured. Thus for fields with a mean photon number around unity a
50/50 beam-splitter would be quite suitable. For stronger fields
an increase in the transmission of $BS2$ would be desirable.

The above measurable quantities can also be obtained by means of a
two-component probe field technique suggested in \cite{Pegg99}.
There are very important differences however.  In \cite{Pegg99},
where states are assigned to the probe fields according to the
usual predictive quantum formalism, the required probe states are,
as acknowledged in that paper, very difficult to prepare. The
preparation method suggested for \cite{Pegg99} was optical
truncation using quantum scissors \cite{Pegg98,Barnett99}, so the
measurement would require three beam-splitters in all, with
separate experiments being run with each different probe. More
seriously, the preparation of the probe in a one-photon and vacuum
superposition in \cite{Pegg99} requires the injection of a single
photon state into one input port and the probe in a two-photon and
vacuum superposition requires the simultaneous injection of a
single photon into two input ports. In contrast to the technique
suggested in \cite{Pegg99}, the method proposed here has real
practical advantages. Only two beam-splitters are used and, apart
from the state to be measured, the only states injected into other
input ports are vacuum and coherent states. These are not only
considerably easier to prepare, their coherence lengths can allow
longer gating times, reducing the effect of dead times. The
retrodictive one-photon states needed to construct the
retrodictive probe originate from photon detection events and are
thus more readily available than their predictive counterparts
which originate from preparation events. Further, because the
retrodictive probe states are produced by the detection events,
all three probe states, including the retrodictive vacuum state
used for measuring the diagonal density matrix elements, are
produced in the one experiment.  There is no need to run separate
experiments for different probe states.

It is interesting to compare this approach with the retrodictive
analysis of the quantum scissors device \cite{Pegg98}.  In the retrodictive
picture, the state to be truncated is in one input mode of a
beam-splitter with detectors in the two output modes.  When one of
these detects one photon and the other detects zero photons, the
retrodictive state in the other input mode is a superposition of
the vacuum and one photon states, so the actual cutting out of the
higher photon state components is done at this beam-splitter.  The
other beam-splitter of the quantum scissors creates a predictive
entangled state. Projection of the retrodictive state onto this
state effectively converts the retrodictive state into a
predictive state with the coefficients of the vacuum and one
photon components interchanged. The beam-splitter $BS2$ in Fig.~\ref{fig4.1} can be regarded as the part of the quantum scissors
that creates the retrodictive two-component state. As we wish to
use this retrodictive probe directly, there is no need to employ
another beam-splitter to convert it to a predictive probe. This
also dispenses with the necessity to produce and inject
single-photon fields. Finally, the insight provided by the
retrodictive formalism of quantum mechanics has enabled us to
propose a relatively simple experiment to measure some of the
canonical phase properties of light. There is now less need to
define separate operational phase properties based on easily
performed experiments.

\section{Higher order phase moments}\label{sec4.2}

For any given distribution the variance only provides a single measure of the spread of that distribution. For states of light such as $c_0|0\rangle+c_1|1\rangle+c_2|2\rangle$ which contain only diagonal, off-diagonal and next off-diagonal elements in the density matrix, when expressed in the photon number basis, we see from Eqns~(\ref{4.1}) and
(\ref{4.2}) that the first two moments are the only non-zero measures which can be acquired about the
distribution. In general, however, knowledge of higher order moments is necessary if a more complete
understanding is required of the distribution. Indeed, the set of all such moments provides enough
information to reconstruct the complete probability distribution \cite{Pregnell02a}, which, for a continuous
distribution, is generally an infinite set. Although such a reconstruction could be done in principle
provided all such moments could be measured, practically, it would be a very tedious way to acquire the
probability distribution. Instead, it would be more practical if these moments were used as a measure to
compare directly between different states of light, particularly if there were a simple procedure to measure
these moments in practice. Of course if it were found from the photon statistics during the experiment that
the field is truncated or sufficiently weak then only a small number of moments would be needed to construct
the phase probability.

In this section we show how the higher order moments
$\langle\cos(\lambda\theta)\rangle$ and
$\langle\sin(\lambda\theta)\rangle$ can be measured using the same
apparatus as introduced in the previous section. The only
modification necessary to the experiment is to replace the
coherent reference state in input mode $2$ with a mixed reference
state $\hat\varrho_2(\lambda)$. For now we do not specify the form
of $\hat\varrho_2(\lambda)$ but we do remark that such a state can
be derived easily from the original coherent reference state with
the preparation technique dependent upon the particular moment
which is being measured. We write the modified PDO for the new
experiment as,
\begin{equation}\label{4.28}
  \hat\Lambda=\hat\rho_0\otimes|0\rangle_{1}{}_1\langle0|\otimes\hat\varrho_2(\lambda).
\end{equation}
Since the remainder of the experiment is unchanged, the probability in which we observe the detection event
$(n_0,n_1,n_2)$ at detectors $D_0$, $D_1$ and $D_2$ respectively is given by Eqn~(\ref{4.13}), with the obvious
replacement of $\hat\Lambda$ above. After substituting Eqns~(\ref{4.11}) and (\ref{4.28}) for the MDO and PDO
respectively, we can derive the same simplified expression for the probability as given in Eqn~(\ref{4.14}), where now
the MDO, $\hat\Gamma_0$, which, when normalised represents the retrodictive state propagating backwards in time out of
input mode $0$, is mixed
\begin{equation}\label{4.29}
  \hat\Gamma_0(n_0,n_1,n_2)={}_1\langle 0|\hat S_1^\dag|n_1\rangle{}_1\,\Trs{2}{\hat\varrho_2(\lambda)
  |z\rangle\langle z|}\,{}_1\langle n_1|\hat S_1|0\rangle{}_1,
\end{equation}
where $|z\rangle=\hat S_2^\dag|n_0\rangle{}_0|n_2\rangle{}_2$ is defined for notational convenience. It can
be seen from Eqn~(\ref{4.29}) that the mixing arises because the coherent reference state $|\alpha\rangle_2$
has been replaced by a mixed state $\hat\varrho_2(\lambda)$. Interestingly, it turns out that the mixing is a
necessary feature of this measurement procedure as opposed to most measurement techniques which go to great
lengths to preserve the purity in their systems. With $|z\rangle$ defined above as the two-mode retrodictive
state at the input of $BS2$, we can interpret the single mode operator
$\Trs{2}{\hat\varrho_2(\lambda)|z\rangle\langle z|}$ which is the projection of $|z\rangle$ onto the mixed
reference state in input mode $2$ as the unnormalised retrodictive state $\hat\rho^{\textrm{ret}}_q$ in the
intermediate mode $0$. That is the state which propagates backwards in time away from the input mode $0$ of
$BS2$ and is incident on output mode $0$ of $BS1$. The subscript $q$ is attached as a reminder that this
state is a mixed generalisation of the retrodictive probe state $|q\rangle_{0}$ in Eqn~(\ref{4.16}). With the
remainder of the experiment unchanged, the retrodictive state $\hat\rho^{\textrm{ret}}_q$ undergoes the same
non-unitary evolution at $BS1$ as did the state $|q\rangle_{0}$ when the reference state was pure.

Previously in measuring $\langle\cos\theta\rangle$, the retrodictive probe state $|q\rangle_{0}{}_0\langle
q|$ was a two-component probe state engineered to contain only diagonal and first off-diagonal term in the
density matrix, while when measuring the second cosine moment $\langle\cos(2\theta)\rangle$, only the
diagonal and second off-diagonal terms in $|q\rangle_{0}{}_0\langle q|$ where non-zero. In keeping with this
trend we are going to require a retrodictive probe state $\hat\rho^{\textrm{ret}}_q$ which contains only
diagonal and $\lambda$ off-diagonal matrix elements if we are to generalise this technique to measure
$\langle\cos(\lambda\theta)\rangle$. We find that there are two necessary steps required to generate such a
retrodictive state. First, a total of $\lambda$ photons need be detected across both detectors $D_0$ and
$D_2$ such that $n_0+n_2=\lambda$. The specific sequence in which this happens is, for now, not important.
What is important is that a total of $\lambda$ photons be incident on the output of $BS2$ and evolve
backwards through the beam-splitter. From energy conservation the retrodictive state $|z\rangle$ at the entry
of $BS2$ will be, in general, some linear combination of all two mode photon number states that sum to
$\lambda$,
\begin{equation}\label{4.30}
  |z\rangle=\hat S_2^\dag|n_0\rangle{}_0|\lambda-n_0\rangle{}_2=\sum_{m=0}^\lambda z_m|m\rangle{}_0|\lambda-m\rangle{}_2.
\end{equation}
In addition to this we require the probe field $\hat\varrho_2(\lambda)$ to be of the form
\begin{equation}\label{4.31}
  \hat\varrho_2(\lambda)=\sum_n\varrho_{n,n}|n\rangle_2{}_2\langle n|+\left(\sum_n\sum_{m\ge\lambda}
  \varrho_{n,n+m}|n\rangle_2{}_2\langle n+m|+H.c.\right)
\end{equation}
which is a state with the first $(\lambda-1)$ off-diagonal matrix elements as zero. A specific example of
such a state would be $|0\rangle+|\lambda\rangle$ which is an extremely non-classical state and therefore
very difficult to produce. Alternatively, we could produce such a state as an equal mixture of coherent
states $|\alpha_j\rangle$ with $\alpha_j=|\alpha|\exp(i2\pi j/\lambda)$ as
\begin{eqnarray}\label{4.32}
  \hat\varrho(\lambda) &=&\mbox{$\frac{1}{\lambda}$}\sum_{j=0}^{\lambda-1}|\alpha_j\rangle\langle \alpha_j|\nonumber\\
  &=&\sum_{n,m}\bar\delta_{\lambda,n-m} \varrho_{n,m}|n\rangle\langle m|,
\end{eqnarray}
where $\bar\delta_{\lambda,n-m}=\mbox{$\frac{1}{\lambda}$}\sum_{j=0}^{\lambda-1}\exp[i(n-m)2\pi j/\lambda]$
is a periodic Kronecker delta function and $\varrho_{n,m}=\langle n|\alpha_0\rangle\langle
\alpha_0|m\rangle$. The last line in Eqn~(\ref{4.32}) can be derived by expressing $|\alpha_j\rangle$ in the
number state basis and summing over $j$. The periodic Kronecker delta function is zero unless $(n-m)$ is an
integer multiple of $\lambda$, in which case it is one. So, by mixing coherent states of selected values of
$\alpha_j$ all matrix elements except the leading diagonal, the $\lambda^{\mathrm{th}}$ off-diagonal, the
$(2\lambda)^\textrm{th}$ off-diagonal and so on in Eqn~(\ref{4.32}) average to zero. This is precisely the
form required of the reference state. To achieve this state in practice, one would initially run the
experiment with a coherent state of zero phase, $|\alpha_0\rangle$. To obtain reliable statistics the
experiment needs to be run many times. Each time the experiment is run the phase of the coherent reference
state is adjusted by $2\pi/\lambda$. To ensure even sampling, the experiment would need to be run an equal
number of times for each coherent reference state $|\alpha_j\rangle$. To ensure the reference state is then
mixed, the statistics of the experiments are compiled without
discriminating between the different values of $\alpha_j$ of the coherent reference states. 
It should be mentioned that the state $\hat\varrho(\lambda)$ in Eqn~(\ref{4.32}) need not be derived from a pure
coherent state. In general, this mixing procedure will remove selected off-diagonal elements in any general state. So
if the initial coherent state is mixed to begin with, as may often be the case in experiments, then this procedure will
still produce a valid reference state.

With these two requirements satisfied we can evaluate the projection of reference state
$\hat\varrho_2(\lambda)$ onto the two-mode retrodictive state $|z\rangle$ as
\begin{equation}\label{4.33}
  \hat\Gamma_q=\Trs{2}{\hat\varrho_2(\lambda)
  |z\rangle\langle z|}=\sum_{n=0}^{\lambda}|z_n|^2\varrho_{\lambda\!-\!n,\lambda\!-\!n}
  |n\rangle_0{}_0\langle n|+\left(z_0z_\lambda^*\varrho_{0\lambda}
  |0\rangle_0{}_0\langle \lambda|+H.c.\right)
\end{equation}
which is the unnormalised expression for the mixed retrodictive probe state
$\hat\rho_q^{\textrm{ret}}=\hat\Gamma_q/\textrm{Tr}[\hat\Gamma_q]$. From this expression we see that the
necessary off-diagonal element $|0\rangle_0{}_0\langle \lambda|$ and the conjugate are apparent in the
retrodictive probe state, while all other off-diagonal elements are not. To evaluate the final retrodictive
state at the input mode $0$ of $BS1$ we follow the evolution of the intermediate retrodictive probe state
backwards through $BS1$. The evolution, which is conditioned on detector $D_1$ detecting $n_1=N$ photons
while zero photon are in the input, can be viewed as a non-unitary transformation of the intermediate
retrodictive probe state. The result, identical to the that presented in the preceding section, is to raise
the photon occupation number of the intermediate retrodictive probe state by $N$ photons. This can be
explicitly derived from Eqn~(\ref{4.29}), using the expression for ${}_1\langle 0|\hat
S_1^\dag|n_1\rangle{}_1$ derived in Appendix~\ref{appendixB}, to be
\begin{equation}\label{4.34}
  \hat\Gamma_0=\sum_{n=0}^{\lambda}\sigma_{n,n}|N\!+\!n\rangle_0{}_0\langle N\!+\!n|
  +\left(\sigma_{0,\lambda}|N\rangle_0{}_0\langle N+\lambda|+H.c.\right)
\end{equation}
where the constant
\begin{equation}\label{4.35}
  \sigma_{n,m}=r^{2N}t^{n+m}z_nz_m^*\varrho_{\lambda\!-\!m,\lambda\!-\!n}\binom{N\!+\!n}{n}^{1/2}\binom{N\!+\!m}{m}^{1/2}
\end{equation}
is introduced for notational convenience and $t=\cos\gamma$ and $r=\sin\gamma$ are the transmission and reflection
coefficients of $BS1$. The probability of detecting $n_0$, $N$ and $(\lambda-n_0)$ photons at detectors $D_0$, $D_1$
and $D_2$ is then given by the overlap of the total MDO, $\hat\Gamma_0$, and the state to be measured,
\begin{equation}\label{4.36}
  \pr_0(n_0,N,\lambda-n_0)=\sum_{n=0}^{\lambda}\sigma_{n,n}\,\rho_{N\!+\!n,N\!+\!n}
  +\left(\sigma_{0,\lambda}\,\rho_{N+\lambda,N}+c.c.\right),
\end{equation}
where $\rho_{n,m}={}_0\langle n|\hat\rho_0|m\rangle_0$ is the matrix coefficient of the state to be measured.
A label is attached to the probability to indicate that this is the first experiment. To extract the
off-diagonal terms from the probability we need to remove the diagonal terms in Eqn~(\ref{4.36}). This is
done by repeating the experiment with a $\pi/\lambda$ phase shift applied to the input mode $2$ of $BS2$. The
effect is to alter the phase of the mixed reference state $\hat\varrho_2(\lambda)$ such that the element
$\varrho_{n,m}$ transforms to $\exp[i(n-m)\pi/\lambda]\varrho_{n,m}$, thereby changing $\sigma_{0,\lambda}$
to $-\sigma_{0,\lambda}$. The probability of such an outcome with this phase shift is then
\begin{equation}\label{4.37}
  \pr_1(n_0,N,\lambda-n_0)=\sum_{n=0}^{\lambda}\sigma_{n,n}\,\rho_{N\!+\!n,N\!+\!n}
  -\left(\sigma_{0,\lambda}\,\rho_{N+\lambda,N}+c.c.\right)
\end{equation}
where the subscript denotes this as the second experiment. If we set the phase of the mixed reference field to offset
any phase shift induced by $BS1$ such that $z_0z_\lambda^*\varrho_{0,\lambda}$ is real and positive, then by comparison
to Eqn~(\ref{4.1}), we find that there is sufficient information to obtain a measured value for the
$\lambda^{\mathrm{th}}$ cosine moment from the statistics of these two experiments as
\begin{equation}\label{4.38}
  \langle\cos(\lambda\theta)\rangle=\sum_N\frac{\pr_0(n_0,N,\lambda-n_0)-\pr_1(n_0,N,\lambda-n_0)}{4\,\sigma_{0,\lambda}}.
\end{equation}
To obtain a measured value for the $\lambda^{\mathrm{th}}$ sine moment we need to repeat the procedure a
further two times, once with a phase shift of $\pi/(2\lambda)$ applied to the reference field in mode $2$ and
then again with a phase shift of $3\pi/(2\lambda)$. Writing the probabilities for the outcomes of such
experiments as $\pr_{1/2}(n_0,N,\lambda-n_0)$ and $\pr_{3/2}(n_0,N,\lambda-n_0)$ respectively, we can obtain
an expression for $\langle\sin(\lambda\theta)\rangle$ in Eqn~(\ref{4.2}) in terms of measured probabilities
as
\begin{equation}\label{4.39}
  \langle\sin(\lambda\theta)\rangle=\sum_N\frac{\pr_{1/2}(n_0,N,\lambda-n_0)-\pr_{3/2}(n_0,N,\lambda-n_0)}
  {4\,\sigma_{0,\lambda}}.
\end{equation}

It can be seen from the two equations above that the observed probabilities need to be rescaled before a measured value
of the sine and cosine moments can be obtained. If the scaling factor is large then the measured probabilities will be
small in which case a large number of experiments will need to be performed before reliable statistics can be obtained
from the data. To avoid this the denominator in Eqn~(\ref{4.38}) and (\ref{4.39}), $\sigma_{0,\lambda}$, should be
optimised over all free parameters thereby minimising the number of experiments needed for a given level of accuracy.
This can be done in two ways. To begin with we consider the optimum detection sequence at detectors $D_0$ and $D_2$.

The scaling factor $\sigma_{0,\lambda}$ is proportional to $z_0z_\lambda^*$ which are the coefficients of the multimode
retrodictive state $|z\rangle$. Although an explicit form of $z_m$ in Eqn~(\ref{4.30}) is difficult to evaluate for a
general detection sequence, it is not so difficult to calculate the case when $m=0,\lambda$. By writing
$|n_0\rangle_0=(n_0!)^{-1/2}(\hat b^\dag_0)^{n_0}|0\rangle$ in Eqn~(\ref{4.30}) and similarly for
$|\lambda-n_0\rangle_2$ we can obtain from Eqn~(\ref{3.9}) the expression
\begin{equation}\label{4.40}
  z_0z_\lambda^*=(-itr)^\lambda(-1)^{n_0}\binom{\lambda}{n_0}.
\end{equation}
It is straightforward to see that this is optimised when $t=r=2^{-1/2}$ and $n_0$, the number of photons
detected at $D_0$, is $\lambda/2$ when $\lambda$ is even and $(n\pm 1)/2$ when $\lambda$ is odd.
Unfortunately, even with the optimised variables the scaling factor still scales exponentially with $\lambda$
as can be seen from Eqn~(\ref{4.40}) with $(tr)^\lambda=2^{-\lambda}$. One way of avoiding this is to include
all detection events at detectors $D_0$ and $D_2$ that sum to $\lambda$, not just the single optimum case
mentioned above. This is a more efficient process since we are not discarding data that can potentially be
used to measure the moments. To see just how much of an improvement can be obtained we rewrite the
coefficient $\sigma_{n,m}$ in Eqn~(\ref{4.35}) as
\begin{equation}\label{4.41}
  \sigma_{n,m}=g_{n,m}(2i)^{-\lambda}(-1)^{n_0}\binom{\lambda}{n_0}.
\end{equation}
Substituting this into the expression for the total MDO, Eqn~(\ref{4.34}), we find that summing $\hat\Gamma_0$ over
$n_0$ with a weighting factor of $(-1)^{n_0}$ has the effect of replacing $\sigma_{0,\lambda}$ with
$i^{-\lambda}g_{0,\lambda}$ since $2^{-\lambda}\sum_{n_0}\binom{\lambda}{n_0}=1$. Taking the trace of this expression
with the state to be measured leads to a modified expression for the moments as
\begin{eqnarray}\label{4.42}
  \langle\cos(\lambda\theta)\rangle&=&\sum_N\frac{\widetilde{\pr}_0(N,\lambda)-\widetilde{\pr}_1(N,\lambda)}
  {4\,g_{0,\lambda}}\\
  \langle\sin(\lambda\theta)\rangle&=&\sum_N\frac{\widetilde{\pr}_{1/2}(N,\lambda)-\widetilde{\pr}_{3/2}(N,\lambda)}
  {4\,g_{0,\lambda}}\label{4.43}
\end{eqnarray}
where we now have taken the phase of $\hat\varrho(\lambda)$ such that $i^{-\lambda}\varrho_{0\lambda}$ is real and
positive and defined
\begin{equation}\label{4.44}
  \widetilde{\pr}_0(N,\lambda)=\sum_{n_0}(-1)^{n_0}\pr_0(n_0,N,\lambda-n_0)
\end{equation}
as the corresponding weighted sum of measured probabilities. So we find by including all measurement outcomes that sum
to $\lambda$ at the output of $BS2$ we gain a modest reduction in the number of experiments that need to be performed
since the scaling factor increases to
\begin{equation}\label{4.45}
  g_{0,\lambda}=r^{2N}(-it)^{\lambda}\varrho_{0,\lambda}\binom{N\!+\!\lambda}{N}^{1/2}.
\end{equation}
Unfortunately there is still an exponential dependence on the measurement outcome $N$ and $\lambda$ in this
scaling factor. Although this cannot be removed it can be minimised. To do this we replace $t=\cos\gamma$ and
$r=\sin\gamma$ in Eqn~(\ref{4.45}) and find the minimum over $\cos^2\gamma$. After some simple algebra we
find the optimum $r:t$ ratio of $BS1$, $\tan\gamma$, to be $\sqrt{2N/\lambda}$. Since we are mainly
interested in weak fields, the spread of values that $N$ and $\lambda$ take should not be large. As such a
fixed value of $\tan\gamma$ around $\sqrt{2\langle n\rangle}$ should generally suffice when measuring lower
order moments, where $\langle n\rangle$ is the mean photon number of the state to be measured. For higher
order moments, a decrease in this ratio could be desirable. To complete the optimisation protocol we could
adjust the strength of the coherent reference state to maximise $\varrho_{0,\lambda}$ in the expression for
$g_{0,\lambda}$. This is perhaps the easiest of the three optimizations to implement physically. We find, by
writing $\varrho_{0,\lambda}$ as $\exp(-|\alpha|^2)\alpha^\lambda/\sqrt{\lambda!}$ and differentiating the
modulus with respect to the mean photon number $|\alpha|^2$, the optimum strength of the coherent reference
state to be $|\alpha|^2=\lambda/2$. This is independent of $N$, the detection outcome at $D_1$. This is
advantageous as the coherent strength need only be adjusted each time a different moment is to be measured,
not for each value of $N$ within a measurement.

A point of interest is the non-classical retrodictive probe state $\rho^{\textrm{ret}}_q$ in the intermediate
mode $0$. It was the motivation of this work to utilise easily prepared retrodictive probe states in lieu of
the more difficult to prepare predictive counterparts to design more practical experiments. What has been
proposed utilised a mixed retrodictive probe that was similar in function to the highly non-classical state
$|0\rangle+|\lambda\rangle$. What is interesting about this work is the way in which the non-classical
retrodictive probe was generated from a mixture of classical coherent states. Remarkably, it is the mixing of
the coherent state that provided the retrodictive probe with the necessary off-diagonal terms to measure the
$\lambda^{\mathrm{th}}$ sin and cosine phase moments. This is in contrast to most experiments which strive to
avoid classically induced uncertainty in their design.

\section{Measuring the density matrix elements of light}

It is now well established that the quantum state of light can be measured. The first experimental evidence of this
\cite{Smithey93a,Smithey93b} followed the work of Vogel and Risken \cite{Vogel89}, where it was shown that the Wigner
function could be reconstructed from a complete ensemble of measured quadrature amplitude distributions. The authors of
\cite{Smithey93a,Smithey93b} measured the quadrature distributions using balanced homodyne techniques. In the case of
inefficient homodyne detectors, a more general s-parameterized quasiprobability distribution is obtained resulting in a
smoothed Wigner function. In either case, to obtain the quasiprobability phase space distribution from the measured
data a rather complicated inverse transformation is required.

Novel techniques which avoid this transformation are aimed at measuring the quasiprobability distribution
more directly. This can be achieved, for example, in unbalanced homodyne counting experiments
\cite{Banaszek96,Wallentowitz96}, where a weighted sum of photocount statistics are combined to obtain a
single point in the phase space distribution. The entire distribution is then obtained by scanning the
magnitude and phase of the local oscillator over the region of interest while repeating the photon counting
at each point. Perhaps the most direct method of obtaining a quasiprobability distribution is to use
heterodyne \cite{Shapiro79} or double homodyne \cite{Walker84} detection techniques where the $Q$ function is
measured. The $Q$ function is related to the Wigner function through a convolution with a Gaussian
distribution which effectively washes out many of the interesting quantum features. It is possible to recover
these features by deconvoluting the $Q$ function, however this requires multiplying by an exponentially
increasing function thereby introducing a crucial dependence on sampling noise \cite{Vogel93}.

\begin{figure}
\begin{center}
\includegraphics{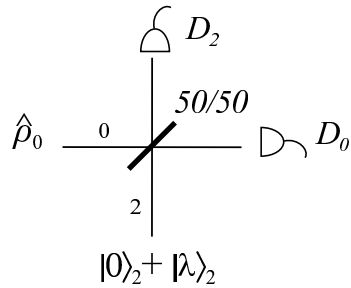}
\caption[Experimental proposal of Steuernagel and Vaccaro to measure the density matrix elements
$\rho_{N,N+\lambda}$ of an optical field]{Experimental proposal of Steuernagel and Vaccaro to measure the
density matrix elements $\rho_{N,N+\lambda}$ of an optical field. At the output of the 50/50 beam-splitter
are photodetectors $D_0$ and $D_2$. The field to be measured is $\hat\rho_0$ while
$|0\rangle+|\lambda\rangle$ is a reference state.} \label{fig4.2}
\end{center}
\end{figure}

A different approach has been suggested by Steuernagel and Vaccaro \cite{Steuernagel95}, who have proposed an
interesting nonrecursive scheme to measure not the quasiprobability distribution, but rather the density operator in
the photon number basis. The scheme is relatively direct in that only a finite number of different measurements are
required to determine each matrix element. The experimental arrangement of the proposal is illustrated in
Figure~\ref{fig4.2}.
It consists of a single 50/50 beam-splitter with input modes $0$ and $2$. At each of the two output modes
there is a photodetector label $D_0$ and $D_2$ respectively. The state to be measured is input mode $0$ of
the beam-splitter while the field in input mode $2$ is a reference field. The most practical arrangement of
this scheme requires the reference field to be in the state proportional to
$|0\rangle_2+e^{-i\lambda\theta}|\lambda\rangle_2$. With such a reference field Steuernagel and Vaccaro then
considered the case when the the two photodetectors detected a total of $N+\lambda$ photons. For a detailed
mathematical analysis of this proposal the reader is referred to the original paper \cite{Steuernagel95},
however, useful insight into the problem can be gained by considering the dynamics in the retrodictive
fromalism.

Following the arguments introduced in Chapter~\ref{chap3} we assign, conditioned on detector $D_0$ detecting
$n_0$ photons and detector $D_2$ detecting $N+\lambda-n_0$ photons, the retrodictive state
$|n_0\rangle{}_0|N+\lambda-n_0\rangle{}_2$ just prior to the detection event. Denoting the unitary action of
the beam-splitter by the operator $\hat S$ which acts in the joint Hilbert space of mode 0 and mode 2, we can
follow the evolution of the state backwards in time and find the retrodictive state at the input of the
beam-splitter as
\begin{equation}\label{4.46}
  |z\rangle=\hat S^\dag|n_0\rangle{}_0|N+\lambda-n_0\rangle{}_2=\sum_{m=0}^{N+\lambda} z_m
  |m\rangle{}_0|N+\lambda-m\rangle{}_2
\end{equation}
which, from conservation of energy, is a general superposition of all two-mode photon number states that sum
to $N+\lambda$. Projecting the predictive reference state in input mode 2, which is an equal superposition of
the vacuum state and a $\lambda$-photon state, onto $|z\rangle$ we find the single mode unnormalised
retrodictive state at the input mode 0 of the beam-splitter is
\begin{equation}\label{4.47}
  \sqrt{2}|\tilde\psi\rangle_0=z_N|N \rangle{}_0+z_{N+\lambda}e^{i\lambda\theta}|N+\lambda\rangle{}_0.
\end{equation}
The joint probability for such a detection event is then given by the overlap of the unnormalised retrodictive state
$|\tilde\psi\rangle_0$ and the state to be measured, $\hat\rho_0$, which from Eqn~(\ref{4.47}) above is
\begin{equation}\label{4.48}
  \textrm{Pr}_\theta(n_0,N+\lambda-n_0)=\mbox{$\frac{1}{2}$}\left[|z_N|^2\rho_{N,N}+|z_{N\!+\!\lambda}|^2
  \rho_{N\!+\!\lambda,N\!+\!\lambda}+\left(z_{N}^*z_{N\!+\!\lambda}e^{i\lambda\theta}\rho_{N,N\!+\!\lambda}+c.c\right)\right],
\end{equation}
where $\rho_{n,m}={}_0\langle n|\hat\rho_0| m\rangle_0$. To determine the value of the off-diagonal element
$\rho_{N,N\!+\!\lambda}$ the entire experiment need to be repeated four times in total, each time with a
different phase shift $\theta$ applied to the reference field. By selecting the four phase shift settings as
$\theta=j\pi/\lambda$ where $j=0,1/2,1,3/2$ a value of $\rho_{N,N\!+\!\lambda}$ can be extracted from the
measured probabilities as
\begin{equation}\label{4.49}
  \rho_{N,N\!+\!\lambda}=\frac{\Pr_0(N\!+\!\lambda)-i\Pr_{1/2}(N\!+\!\lambda)-\Pr_1(N\!+\!\lambda)
  +i\Pr_{3/2}(N\!+\!\lambda)}{2z_{N}^*z_{N\!+\!\lambda}}
\end{equation}
where $\Pr_j(N\!+\!\lambda)$ is shorthand for $\Pr_\theta(n_0,N+\lambda-n_0)$. So what was originally
proposed by Steuernagel and Vaccaro was an insightful way to relate the probability in which particular
detection events occur in an experiment to specific off-diagonal elements in the density matrix description
of an optical field. This might be viewed as a generalisation of a direct photon-counting experiment where
the probabilities of the detection events give the diagonal elements of the density matrix. The problem
however is that the reference state $|0\rangle+|\lambda\rangle$ is extremely difficult to produce in
practice. Even to use the quantum scissors device to truncate a coherent state for the case when $\lambda=1$
is by no means trivial. Remarkably, by introducing a second beam-splitter with the vacuum in one of the
inputs, it is possible to replace the non-classical reference state with a mixture of coherent states and
achieve the same results \cite{Pregnell02b,Pregnell03b}. The experiment from which the measured probabilities
are observed is exactly that which was proposed in the preceding section. The difference between the previous
sections work and this is in the way in which the probabilities are combined to ascertain something different
about the state we are observing. To see how the off-diagonal matrix element $\rho_{N,N\!+\!\lambda}$ can be
measured using the double beam-splitter device introduced in the previous section, consider the probability
of the detection event $(n_0,N,\lambda-n_0)$ at detectors $D_0$, $D_1$ and $D_2$ respectively. From
Eqn~(\ref{4.36}), we rewrite this here as
\begin{equation}\label{4.50}
  \pr_0(n_0,N,\lambda-n_0)=\sum_{n=0}^{\lambda}\sigma_{n,n}\,\rho_{N\!+\!n,N\!+\!n}
  +\left(\sigma_{0,\lambda}\,\rho_{N+\lambda,N}+c.c.\right).
\end{equation}
If instead of summing over $N$, we combine the four measured probabilities $\pr_j(n_0,N,\lambda-n_0)$,
$j=0,1/2,1,3/2$, relating to the four different phase shift settings and normalise we can extract the matrix
element $\rho_{N,N\!+\!\lambda}$ akin to expression (\ref{4.49}) we used to describe the approach of
Steuernagel and Vaccaro as
\begin{equation}\label{4.51}
  \rho_{N,N\!+\!\lambda}=\frac{\Pr_0(N,\lambda)-i\Pr_{1/2}(N,\lambda)-\Pr_1(N,\lambda)+i\Pr_{3/2}(N,\lambda)}
  {4\sigma_{0,\lambda}^*},
\end{equation}
where $\Pr_j(N,\lambda)$ is shorthand for $\Pr_j(n_0,N,\lambda-n_0)$. Since we are not modifying the double
beam-splitter experiment, merely changing the way in which we combine the measured probabilities, all the
optimisation protocols suggested in the previous section apply. As such we may write Eqn~(\ref{4.51}) in
terms of the weighted sum of probabilities $\widetilde{\Pr}_j(N,\lambda)$ introduced in Eqn~(\ref{4.44}) as
\begin{equation}\label{4.52}
  \rho_{N,N\!+\!\lambda}=\frac{\widetilde{\Pr}_0(N,\lambda)-i\widetilde{\Pr}_{1/2}(N,\lambda)
  -\widetilde{\Pr}_1(N,\lambda)+i\widetilde{\Pr}_{3/2}(N,\lambda)}
  {4g_{0,\lambda}^*},
\end{equation}
which includes all detection events at detectors $D_0$ and $D_2$ that sum to $\lambda$.

It is interesting to note that Steuernagel and Vaccaro proposal would still work if the reference state were
replaced by a general truncated state $\hat\rho^{tr}=\sum_{n,m=0}^\lambda\rho_{n,m}|n\rangle\langle m|$ in
$\lambda+1$ dimensions. Using the mixing technique presented in the previous section all but the diagonal and
$\lambda^{\mathrm{th}}$ off-diagonal elements would need to be removed to make the reference state of the
form $\hat\rho^{tr}_{mix}=\sum_{n=0}^\lambda\rho_{n,n}|n\rangle\langle
n|+\rho_{0,\lambda}|0\rangle\langle\lambda| +\rho_{\lambda,0}|\lambda\rangle\langle0|$. As the contribution
from the diagonal elements are removed when the probabilities are subtracted, this state is, for all purposes
considered here, equivalent to the state $|0\rangle+|\lambda\rangle$. Unfortunately, generating a predictive
state in a finite number of dimensions, or equivalently, truncating a state in a finite number of dimensions
is still a difficult process.

So we find, from Eqn~(\ref{4.47}), that the proposal of Steuernagel and Vaccaro takes the truncated
predictive state $|0\rangle+|\lambda\rangle$ and simultaneously turns it into a retrodictive state while
raising the photon occupation number by $N$. The double beam-splitter proposal presented here does these two
operations separately. At the first beam-splitter $BS1$ (in the retrodictive picture), the mixed retrodictive
state is truncated and turned into a retrodictive state $\rho_q^{\textrm{ret}}$, the retrodictive equivalent
of $\hat\rho^{tr}_{mix}$. The second beam-splitter performs the second of these operations which is to raise
the photon occupation number of the retrodictive state by $N$. The result is a \emph{retrodictive} probe
equivalent to that in the Steuernagel and Vaccaro scheme which can be used to measure individual elements of
the optical density matrix in the photon number basis. Both proposals detect a total of $N+\lambda$ photons.
The advantage of the the double beam-splitter arrangement is that it naturally truncates the reference state
in a $\lambda+1$ dimensional sub-space when turning it into a retrodictive state. This, in conjunction with
the mixing technique, allows an ordinary coherent state to be used as a reference state in lieu of the
non-classical truncated state, of which $|0\rangle+|\lambda\rangle$ is a specific example. So again we find
an asymmetry in the ease at which a retrodictive state can be produced in practice as opposed to the
predictive counterpart.

In summary, by utilising the more readily prepared retrodictive quantum states we were able to take two
experimental proposals that were proving difficult to implement physically and redesign them in a such a way
to make them more implementable with current technology. Both proposals are non-recursive in that the
measured quantity is extracted directly from observed probabilities of selected measurement events, in a
similar fashion to the original projection synthesis technique of \cite{Barnett96,Pegg97,Phillips98}.
Interestingly, one of these methods allows the density matrix elements of an unknown field to be obtained
quite simply from the density matrix of a mixed local oscillator state, even when the unknown field is in a
pure state.


%% file: Chap5.tex

\chapter{Quantum optical phase and its measurement}\label{chap5}
The quantum mechanical nature of the phase of light has been
studied since the beginnings of quantum electrodynamic theory
\cite{Dirac27} and with renewed interest recently. The study of
quantum phase is distinguished from the study of many other
quantum observables by the difficulties inherent not only in
finding a theoretical description but also in finding methods for
measuring the phase observable so described \cite{Pegg97b}.
Despite the method proposed in Chapter~\ref{chap3}, and others
like it \cite{Barnett96, Phillips98}, to engineer any general
retrodictive state expressible in a finite number of dimensions, a
``single-shot'' measure of quantum optical phase has been
illusive. In this chapter I present the first proposed method
capable, at least in principle, of providing a single shot measure
of canonical phase. This work has been published in our paper
\cite{Pregnell02c}. The technique is simple, involving only
beam-splitters, phase-shifters and photodetectors which can
discriminate between zero, one and many photons. Following this I
show that the eight-port interferometer used by Noh, Foug\`{e}res
and Mandel \cite{Noh91,Noh92b,Fougeres94} to measure their
operational phase distribution of light
\cite{Noh91,Noh92a,Noh92b,Noh93a,Barnett93,Noh93b} can,
remarkably, also be used to measure the canonical phase
distribution for weak optical fields \cite{Pregnell03a}. A
binomial reference state is required for this purpose which we
show can be obtained, to an excellent degree of approximation,
from a suitably squeezed state.

\section{Single-shot measure of quantum optical phase}\label{sec5.1}

\subsection{Canonical phase}\label{sec5.1.1}

Quantum-limited phase measurements of the optical field have important applications in precision measurements
of small distances in interferometry and in the emerging field of quantum communication, where there is the
possibility of encoding information in the phase of light pulses. Much work has been done in attempting to
understand the quantum nature of phase. Continuing our discussion of phase in Section~\ref{4.1}, we note that
some approaches have been motivated by the aim of expressing phase as the complement of photon number
\cite{Leonhardt95}. Examples of these approaches include the probability operator measure approach
\cite{Helstrom76,Shapiro91}, a formalism in which the Hilbert space is doubled \cite{Newton80}, a limiting
approach based on a finite Hilbert space \cite{Pegg88,Pegg89,Barnett89} and a more general axiomatic approach
\cite{Leonhardt95}. Although these approaches are quite distinct, they all lead to the same phase probability
distribution for a field in state $|\psi\rangle$ as a function of the phase angle $\theta$
\cite{Leonhardt95}:
\begin{equation}\label{5.1}
P(\theta)=\frac{1}{2\pi}\left|\sum_{n=0}^\infty \langle\psi|n\rangle \exp (in\theta)\right|^{2}
\end{equation}
where $|n\rangle$ is a photon number state and $\theta$ can take on any value within a $2\pi$ window which we
arbitrarily take as $0\le\theta\le 2\pi$. Leonhardt \emph{et al.} \cite{Leonhardt95} have called this common
distribution the ``canonical'' phase distribution to indicate a quantity that is the canonical conjugate, or
complement, to photon number. This distribution is shifted uniformly when a phase-shifter is applied to the
field and is not changed by a photon number shift. In accord with our previous discussion of phase, we can
continue to adopt this definition here and use the term canonical phase to denote the quantity whose
distribution is given by (\ref{5.1}). We can write the definition in Eqn~(\ref{5.1}) in a more compact
notation as $|\langle\psi|\theta\rangle|^2$, where we use the (improper) state vector
\begin{equation}\label{5.2}
  |\theta\rangle=\frac{1}{\sqrt{2\pi}}\sum_{n=0}^\infty \exp (in\theta)|n\rangle.
\end{equation}
We note that this state is not orthogonal to a
state of different phase $|\theta'\rangle$, even in the sense of a Dirac delta function. Interestingly, many of the difficulties associated with finding
an Hermitian phase operator in the infinite-dimensional Hilbert space can be attributed to this fact which is due to the semi-bounded property of the associated Hilbert space. Howbeit, we refer to the state defined
by Eqn~(\ref{5.2}) as a phase state. In the case of a mixed state $\hat\rho$, we can generalise
Eqn~(\ref{5.1}) to
\begin{equation}\label{5.3}
  P(\theta)=\textrm{Tr}\left[\hat\rho|\theta\rangle\langle\theta|\right].
\end{equation}
Using the definition of the phase state it can be shown that the set of all such operators
$|\theta\rangle\langle\theta|$ form a resolution of the identity operator in the infinite dimensional Hilbert space,
\begin{equation}\label{5.4}
  \int_{0}^{2\pi}\!\!\!\!|\theta\rangle\langle\theta|\,d\theta=\hat 1.
\end{equation}
As it is obvious from Eqn~(\ref{5.3}) that
$|\theta\rangle\langle\theta|$ is a non-negative operator, the set
of all such operators constitutes a valid POM, with elements
$\hat\Pi_\theta=|\theta\rangle\langle\theta|$. Since the set of
POM elements $\hat\Pi$ is sufficient to derive the canonical phase
distribution of Eqn~(\ref{5.1}), we see that the phase POM
provides a way of representing the phase observable without
defining a phase operator. By redefining phase in terms of a POM,
we can associate the observable phase with outcomes of a
measurement apparatus while maintaining consistency with other
descriptions of canonical phase\footnote{The consistency of
quantum descriptions of phase has been studied in
\cite{Vaccaro93}. There it is shown how the POM description can be
derived from the limiting approach of
\cite{Pegg88,Pegg89,Barnett89}}.

Despite these theoretical advances, much less progress has been
made on ways to measure canonical phase. Homodyne techniques can
be used to measure phase-like properties of light but are not
measurements of canonical phase. For very weak states of light an
adaptive technique can improve the homodyne methods to provide a
quite good approximate measurement of canonical phase
\cite{Wiseman95,Wiseman98}. These have been implemented
experimentally recently \cite{Armen02}. Using the apparatus
described in Chapter~\ref{chap2}, or even the original projection
synthesis technique described in \cite{Barnett96}, it is possible
in principle to measure the canonical phase \emph{distribution} by
a series of experiments on a reproducible state of light but there
has been no known way of performing a single-shot measurement.
Even leaving aside the practical issues, the concept that a
particular fundamental quantum observable may not be measurable,
even in principle, has interesting general conceptual
ramifications for quantum mechanics. A different approach to the
phase problem, which avoids difficulties in finding a way to
measure canonical phase, is to define phase operationally in terms
of observables that can be measured \cite{Leonhardt95}. The best
known of these operational phase approaches is that of Noh
\emph{et al.} \cite{Noh91,Noh92a,Noh92b,Noh93a,Barnett93,Noh93b,
Fougeres94}. Although the experiments to measure this operational
phase produce excellent results, they were not designed to measure
canonical phase as defined here and, as shown by the the measured
phase distribution \cite{Noh91,Noh92b,Noh93a,Fougeres94}, they do
not measure canonical phase. In this chapter I show how, despite
these past difficulties, it is indeed possible, at least in
principle, to perform a single-shot measurement of canonical phase
in the same sense that the experiments of Noh \emph{et al.} are
single-shot measurements of their operational phase.

A single-shot measurement of a quantum observable must not only yield one of the eigenvalues of the observable, but
repeating the measurement many times on systems in identical states should result in a probability distribution
appropriate to that state. If the spectrum of eigenvalues is discrete, the probabilities of the results can be easily
obtained from the experimental statistics. Where the spectrum is continuous, the probability density is obtainable by
dividing the eigenvalue range into a number of small bins and finding the number of results in each bin. As the number
of experiments needed to obtain measurable probabilities increases as the reciprocal of the bin size, a practical
experiment will require a non-zero bin size and will produce a histogram rather than a smooth curve.

Although the experiments of Noh \emph{et al.} were not designed to
measure canonical phase, it is helpful to be guided by their
approach. In addition to their results being measured and plotted
as a histogram, some of the experimental data are discarded,
specifically photon count outcomes that lead to an indeterminacy
of the type zero divided by zero in their definitions of the
cosine and sine of the phase
\cite{Noh91,Noh92b,Noh93a,Fougeres94}. The particular experiment
that yields such an outcome is ignored and its results are not
included in the statistics. Such a measurement procedure is a
specific example of a biased measurement procedure introduced in
Chapter~\ref{chap2}, the statistics of which are governed by the
general expression of Eqn~(\ref{2.12}).

We seek now to approximate the continuous distribution (\ref{5.3})
by a histogram representing the probability distribution for a
discrete observable $\theta_{m}$ such that when the separation
$\delta\theta$ of consecutive values of $\theta_{m}$ tends to zero
the continuous distribution is regained. A way to do this is first
to define a (proper) state vector
\begin{equation}\label{5.5}
|\theta_m\rangle =\frac 1{(N+1)^{1/2}}\sum_{n=0}^N\exp (in\theta_m)|n\rangle.
\end{equation}
There are $N+1$ orthogonal states $|\theta_{m}\rangle$
corresponding to $N+1$ values $\theta_{m}=m\delta\theta$ with
$\delta\theta = 2\pi/(N+1)$ and $m=0,1,\dots N$. This range for
$m$ ensures that $\theta_{m}$ takes values between $0$ and $2\pi$.
Then, if we can find a measurement technique that yields the
result $\theta_{m}$ with a probability of
$\tr{\hat\rho|\theta_m\rangle \langle \theta _m|}$, the resulting
histogram will approximate a continuous distribution with a
probability density of
$\delta\theta^{-1}\tr{\hat\rho|\theta_m\rangle \langle \theta
_m|}$. It follows that, as we let $N$ tend to infinity, there will
exist a value of $\theta_{m}$ as close as we like to any given
value of $\theta$ with a probability density approaching
$P(\theta)$ given by (\ref{5.3}). If we keep $N$ finite so that we
can perform an experiment with a finite number of outcomes, then
the value of $N$ must to be sufficiently large to give the
resolution $\delta\theta$ of phase angle required and also for
$|\psi \rangle$ to be well approximated by $\sum_{n} \langle
n|\psi \rangle |n\rangle$ where the sum is from $n = 0$ to $N$.
The latter condition ensures that the terms with coefficients
$\langle n|\psi\rangle$ for $n > N$ have little effect on the
probability $\tr{\hat\rho|\theta_m\rangle \langle \theta _m|}$. As
we shall be interested mainly in weak optical fields in the
quantum regime with mean photon numbers of the order of unity, the
phase resolution $\delta\theta$ desired will usually be the
determining factor in the choice of $N$.

When $N$ is finite, the states $|\theta _m\rangle $ do not span
the whole Hilbert space, so the projectors $|\theta _m\rangle
\langle \theta _m|$ will not sum to the unit operator $\hat{1}$ in
this space. Thus these projectors by themselves do not form the
elements of a POM in this space. Conveniently however, they do sum
to the identity operator in an $N+1$ dimensional Hilbert space,
$\hat 1_{N}$,
\begin{equation}\label{5.6}
  \sum_{m=0}^N |\theta_m\rangle \langle \theta _m|=\hat 1_{N}.
\end{equation}
To complete the POM acting in the whole Hilbert space we need to include an element $\hat{1}-\hat 1_{N}$, such that the
sum of all elements is the identity in the whole space. If we are to discard the outcome associated with this element,
that is, treat an experiment with this outcome as an unsuccessful attempt at a measurement in a similar way that Noh
\emph{et al.} \cite{Noh91,Noh92b,Noh93a,Fougeres94} treated experiments with indeterminate outcomes, then we should
associate each of the non-discarded events with a MDO $\hat\Gamma_m$ proportional to $|\theta_m\rangle \langle \theta
_m|$. From Eqn~(\ref{2.12}), with Eqn~(\ref{5.6}) above, the probability that the recorded outcome of a measurement is
the phase angle $\theta _m$ is given by
\begin{equation}\label{5.7}
\pr(\theta _m)=\frac{\text {Tr}[\hat{\rho }|\theta _m\rangle \langle \theta _m|]} {\text {Tr}[\hat{\rho }\hat 1_{N}]}.
\end{equation}
We now require a single-shot measuring device that will reproduce this probability in repeated experiments.

\subsection{Experimental proposal}\label{sec5.1.2}

\begin{figure}
\begin{center}
\includegraphics{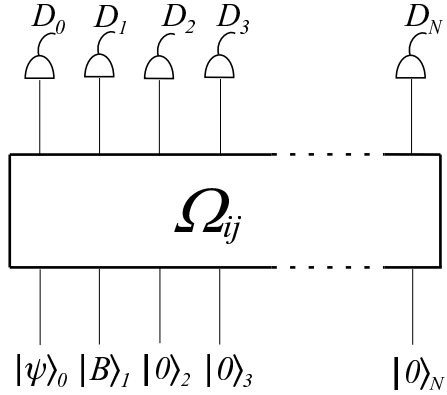}
\caption[Multiport device for single-shot phase
measurement]{Multiport device for measuring phase. The input and
output modes are labelled $0, 1, \ldots N$ from the left. In input
mode $0$ is the field in state $|\psi\rangle_{0}$ to be measured
and in input mode $1$ is the reference field in state
$|B\rangle_{1}$. Vacuum states form the other inputs. There is a
photodetector in each output mode. If all the photodetectors
register one count except the detector $D_{m}$ in output mode $m$,
which registers no counts, then the detector array acts as a
digital pointer mechanism indicating a measured phase angle of
$\theta_{m}$.}\label{fig5.1}
\end{center}
\end{figure}
We demonstrated in Section~\ref{sec3.3} that a multiport device is
capable of producing a \emph{single} retrodictive truncated phase
state of the form given by Eqn~(\ref{5.5}). The probability of
generating such a state is proportional to $\text {Tr}[\hat{\rho
}|\theta _m\rangle \langle \theta _m|]$ and was conditioned on
observing the photodetection sequence $(0,1,1,\dots,1)$ at the
$N+1$ photodetectors $D_i$, $i=0,1,\dots,N$. In all input modes,
except the one containing the state to be measured, a coherent
field was present. To construct a device capable of producing a
single shot measure of phase in the sense outlined above, we need
a device which is capable of generating \emph{all} of the $N+1$
retrodictive states of the form in Eqn~(\ref{5.5}), with
$m=0,1,\dots,N$. We ask the question: can we modify this apparatus
such that each of the $N+1$ permutations of the detection sequence
$(0,1,1,\dots,1)$ generates one of the $N+1$ truncated phase
states? If we can, then we have an apparatus capable of performing
a single shot measurement of phase.

Retaining the linear optical multiport and the photodetectors at
each of the output modes means that the only modification we
can make to the apparatus is to allow for more general reference
states than coherent states. We consider the simplest case of
replacing only one of the coherent reference state
$|\alpha\rangle_1$ in input mode $1$ with a general reference
state $|B\rangle_1=\sum_n b_n|n\rangle_1$, and replace all others
with a vacuum field. Such a multi-mode input state is written as
\begin{equation}\label{5.8}
  \left(\prod_{i=2}^{N}|0\rangle_i\right)|B\rangle_1
\end{equation}
Following on from Section~\ref{sec3.3}, we now consider the retrodictive state that is generated with this multi-mode
reference state when each photodetector registers a single photon, except one, which registers zero photons. There are
$N+1$ different ways in which this can happen resulting from the the $N+1$ different detectors, we consider all of
them. If we label the detector which does not detect any photons as the $m^{\textrm{th}}$ detector, then we can assign
a POM element to the multi-photon detection event as
\begin{equation}\label{5.9}
  \hat\Pi(m)=|0\rangle_m{}_m\langle 0|\prod_{j\neq m}^{N}|1\rangle_j{}_j\langle 1|,
\end{equation}
where the index on the product is taken over all mode labels,
$j=0$ to $N$, but does not include $j=m$. We can write this
operator as a projector $\hat\Pi(m)=|\Psi_m\rangle\langle\Psi_m|$,
where
\begin{equation}\label{5.10}
  |\Psi_m\rangle=|0\rangle_m\prod_{j\neq m}^{N}|1\rangle_j.
\end{equation}
It follows then from Section~\ref{sec3.3}, with the multi-mode
reference state given by Eqn~(\ref{5.8}), that the single-mode
retrodictive state at the input of mode $0$ conditioned on
detecting one photon in each output mode, except the
$m^{\textrm{th}}$ mode, is
\begin{equation}\label{5.11}
  |\tilde{\psi}_m\rangle_0={}_1\langle B|\left(\prod_{i=2}^{N}{}_i\langle 0|\right)\hat S^\dag|\Psi_m\rangle.
\end{equation}
The operator $\hat S^\dag$ again represents the unitary evolution of the multi-mode state, backwards in time, and is
characterised by the mode transformation matrix elements $U_{ij}^*$ in Eqn~(\ref{3.20}). We require such a device for
which the associated unitary matrix is
\begin{equation}\label{5.12}
  U_{ij}^*=\frac{\omega ^{ij}}{\sqrt{N+1}}
\end{equation}
where $\omega =\exp[-i2\pi /(N+1)]$ that is, a $(N+1)^{\textrm{th}}$ root of unity. This is precisely the
transformation introduced in Section~\ref{sec3.2.1} as the discrete Fourier transformation in $N+1$ dimensions.
Substituting Eqn~(\ref{5.10}) into the expression for the retrodictive state (\ref{5.11}) and writing $|1\rangle_j$ as
$\hat a^\dag_j |0\rangle$ gives, after some algebra,
\begin{equation}\label{5.13}
  |\tilde{\psi}_m\rangle_0={}_1\langle B|\left(\prod_{i=2}^{N}{}_i\langle 0|\right)\left(\prod_{j\neq m}^{N}\hat
  S^\dag\hat a_j^\dag\hat S\right)|0\rangle.
\end{equation}
where $|0\rangle$ is the multi-mode vacuum which is invariant under the linear transformation of $\hat S^\dag$.
Substituting Eqn~(\ref{3.20}) for the mode transformation and using the matrix elements of (\ref{5.12}) simplifies this
expression to
\begin{eqnarray}\label{5.14}
  |\tilde{\psi}_m\rangle_0=\kappa_{1}\,{}_1\langle B|\left [\prod_{j\neq m}^N(\hat{a}_0^{\dagger }
  +\omega ^j\hat{a} _1^{\dagger })\right ] |0\rangle _0|0\rangle _1
\end{eqnarray}
where $\kappa_{1} =(N+1)^{-N/2}$.

To evaluate Eqn~(\ref{5.14}) we divide both sides of the identity
\begin{equation}\label{5.15}
  X^{N+1}+(-1)^N=(X+1)(X+\omega )(X+\omega ^2)\ldots (X+\omega ^N)
\end{equation}
by $X+\omega ^m$ to give, after some rearrangement and application of the relation $\omega ^{m(N+1)}=1$,
\begin{equation}\label{5.16}
  \prod_{j\neq m}^N(X+\omega ^j)=(-1)^N\omega ^{mN}\frac{1-(-X\omega ^{-m})^{N+1} }{1-(-X\omega ^{-m})}\text{ .}
\end{equation}
The last factor is the sum of a geometric progression. Expanding this and substituting $X=x/y$ gives eventually the
identity
\begin{equation}\label{5.17}
  \prod_{j\neq m}^N(x+\omega ^jy)=\sum_{n=0}^N x^n(-\omega ^my)^{N-n}\text{ .}
\end{equation}

We now expand $|B\rangle _1$ in terms of photon number states as $|B\rangle _1=\sum_{n=0}^Nb_n|n\rangle_{1}$ and put
$x=a_0^{\dagger }$ and $y=a_1^{\dagger }$ in Eqn~(\ref{5.17}). With this we find that the retrodictive state of
Eqn~(\ref{5.14}) becomes
\begin{equation}\label{5.18}
  |\tilde\psi_m\rangle_0=\kappa_{2}\sum_{n=0}^{N}(-1)^{N-n}{N\choose n}^{-1/2}\omega^{-nm}b_{N-n}^* |n\rangle_{0}
\end{equation}
where $\kappa_{2}= \kappa_{1}\omega^{-m}(N!)^{1/2}$. We see then that, if we let $|B\rangle_{1}$ be the binomial state
\begin{equation}\label{5.19}
   |B\rangle_{1}=2^{-N/2}\sum_{n=0}^{N}(-1)^{n}{N\choose n}^{1/2}|n\rangle_{1},
\end{equation}
then Eqn~(\ref{5.18}) is proportional to $\sum_{n}\omega^{-nm}|n\rangle_{0}$, that is, to $|\theta_{m}\rangle_{0}$.

The unnormalised retrodictive state at the input of mode $0$ of
the device, $|\tilde{\psi}_m\rangle_0$, can be associated with a
MDO $\hat\Gamma_0(m)$ for the measuring device consisting of
everything in Figure~\ref{fig5.1} except the state to be measured.
We found above that this MDO is proportional to the projector
$|\theta_m\rangle_0{}_0\langle\theta_m|$. With this we see from
Eqn~(\ref{2.12}) that the probability that zero photons are
detected in output mode $m$ and one photon is detected in all the
other output modes, given that only outcomes associated with the
$(N+1)$ events of this type are recorded in the statistics, is
consistent with Eqn~(\ref{5.7}), where we note that the
proportionality constant $\kappa_{2}$ will cancel from this
expression. Thus \emph{the measurement event that zero photons are
detected in output mode {\em m} and one photon is detected in all
the other output modes can be taken as the event that the result
of the measurement of the phase angle is $\theta_{m}$}. Thus the
photodetector with zero photocounts, when all other photodetectors
have registered one photocount, can be regarded as a digital
pointer to the value of the measured phase angle.

We have shown, therefore, that it is indeed possible in principle to conduct a single-shot measurement of
canonical phase to within any given non-zero error, however small. This error is of the order $2\pi/(N+1)$
and will determine the value of $N$ chosen.

While the aim of this section is to establish how canonical phase can be measured in principle, it is worth
briefly considering some practical issues. Although we have specified that the photodetectors need only be
capable of distinguishing among zero, one and more than one photons, reflecting the realistic case, there are
other imperfections such as inefficiency. These will give rise to errors in the phase measurement, just as
they will cause errors in a single-shot photon number measurement. In practice, there is no point in choosing
the phase resolution $\delta\theta$ much smaller than the expected error due to photodetector inefficiencies,
thus there is nothing lost in practice in keeping $N$ finite. A requirement for the measuring procedure is
the availability of a binomial state. Such states have been studied for some time \cite{Stoler85,Dodonov02}
but their generation has not yet been achieved. In practice, however, we are usually interested in measuring
weak fields in the quantum regime with mean photon numbers around unity \cite{Noh91,Noh93a} and even
substantially less \cite{Torgerson96}. Only the first few coefficients of $|n\rangle_{0}$ in Eqn~(\ref{5.18})
will be important for such weak fields. Also, it is not difficult to show that the reference state need not
be truncated at $n = N$, as indicated in Eqn~(\ref{5.19}), as coefficients $b_{n}$ with $n > N$ will not
appear in Eqn~(\ref{5.18}). Thus we need only prepare a reference state with a small number of its photon
number state coefficients proportional to the appropriate binomial coefficients. Additionally, of course, in
a practical experiment we are forced to tolerate some inaccuracy due to photodetector errors, so it will not
be necessary for the reference state coefficients to be exactly proportional to the corresponding binomial
state coefficients. These factors give some latitude in the preparation of the reference state. The multiport
depicted in Figure~\ref{fig5.1} can be constructed in a variety of ways. As mentioned in
Section~\ref{sec3.2.1}, Reck \emph{et al.} provide an algorithm for constructing a triangular array to
realise such a transformation. In addition, it was also discussed how the plate beam-splitter design of
T\"{o}rm\"{a} and Jex \cite{Torma95} will provide the same transformation  with a quadratic reduction in the
number of optical elements necessary.

\section{Canonical phase distribution for weak optical fields}\label{sec5.2}

In this section I show that the eight-port interferometer used by
Noh, Foug\`{e}res, and Mandel \cite{Noh91,Noh92b,Fougeres94} to
measure their operational phase distribution of light can also be
used to measure the canonical phase distribution of weak optical
fields. Such a result was originally published in Ref.~
\cite{Pregnell03a} by D. T. Pegg and myself. Here I show that both
the multimode projection synthesis technique introduced in
Chapter~\ref{chap3} and a method for measuring the canonical phase
distribution based on the single-shot method presented above can
be implemented with the eight-port interferometer. The obvious
difference between these two methods is the reference states
required at the input of the interferometer. In the case of the
multi-mode projection synthesis technique, it is necessary to
combine the state being measured with three coherent fields at the
input of the interferometer, while for the single-shot technique
the state being measured is combined with a single binomial state
and two vacuum states. We find, remarkably, that the binomial
reference state can be obtained to an excellent degree of
approximation from a suitably squeezed state. Given that the
operational phase measurements have already been conducted using
the eight-port interferometer and single photon detectors provides
encouraging support that there should be no insurmountable
physical challenges preventing the measurement of the canonical
phase distribution for weak optical fields.

\subsection{Reconstructing and measuring the phase distribution}\label{sec5.2.1}

As phase is an intrinsically continuous observable, akin to
position, it is necessary to represent the phase distribution by a
continuous probability density as opposed to a discrete
probability distribution. In a practical experiment, however, only
discrete probability distributions can be observed. Accordingly,
the continuous probability density can be approximated by
observing the probability that the outcome falls in a particular
range, called a bin. The probability divided by the associated bin
size, can be plotted as a discrete probability density. It follows
then that in the limit as the bin size tends to zero, the
continuous probability density can be obtained from the discrete
density. In practice to obtain a continuous probability
experimentally for an unknown state would take an infinite time
so, in general, a continuous probability density can only be
measured to within some nonzero error. We show in this section by
decomposing the canonical phase distribution of (\ref{5.1}) into
the Fourier coefficients, that truncated states, for example, with
at most $N$ photons cannot produce a phase distribution
$P(\theta)$ with oscillations more rapid than $\exp(i\theta N)$.
It follows then that for such a state $2N+2$ points
are the minimum needed to determine the phase distribution.

Making use of the Dirac delta function we can express the
canonical phase distribution as
\begin{equation}\label{5.20}
  P(\theta)=\int P(\phi)\delta(\theta-\phi)\,d\phi.
\end{equation}
After writing the delta function as an infinite summation of orthonormal functions $\exp(in\theta)$, we see that the
phase distribution can be expressed as a weighted sum of the orthonormal functions
\begin{equation}\label{5.21}
  P(\theta)=\frac{1}{2\pi}\sum_{n=-\infty}^\infty\alpha_n\exp(in\theta),
\end{equation}
where the weighting coefficients
\begin{equation}\label{5.22}
  \alpha_n=\int P(\phi)\exp(-in\phi)\,d\phi
\end{equation}
are the $n^{\textrm{th}}$ order exponential phase moments. From
(\ref{5.21}), the infinite set of coefficients $\alpha_n$ then
provides an equivalent representation to the continuous
distribution $P(\theta)$ of (\ref{5.1}). We refer to
Eqn~(\ref{5.21}) as the Fourier decomposition of the canonical
phase distribution, where the Fourier coefficients are just the
weighting coefficients of Eqn~(\ref{5.22}). In general each
Fourier coefficient can take on any independent value within a
unit circle on the complex number plane, provided however
$\alpha_n^*=\alpha_{-n}$. This condition can be derived in a
straightforward manner by taking the complex conjugate of
Eqn~(\ref{5.22}).

The Fourier coefficients are a set of values general enough to
represent the probability distribution for any state
$|\psi\rangle=\sum_{n=0}^\infty\psi_n|n\rangle$ contained within
the infinite dimensional Hilbert state. In practice however, all
states possess only a finite amount of energy. As such, most
states can be sufficiently well represented by truncating
$|\psi\rangle$ at some finite energy value $N$, however large,
\begin{equation}\label{5.23}
  |\psi\rangle=\sum_{n=0}^N\psi_n|n\rangle.
\end{equation}
It is shown in Appendix~\ref{appendixC} that the Fourier coefficients $\alpha_n$ associated with such states are zero
for all $|n|>N$. In such case the probability distribution can be written as a \emph{finite} summation of oscillating
functions
\begin{equation}\label{5.24}
  P(\theta)=\frac{1}{2\pi}\sum_{n=-N}^N\alpha_n\exp(in\theta).
\end{equation}
So for states represented by Eqn~(\ref{5.23}), this introduces a
maximum limit of $2\pi/N$ to the period in which the probability
distribution can oscillate. Thus we see, for states with a finite
number of energy terms, that the continuous phase probability
density can be represented by a finite set of $N+1$ unique complex
numbers\footnote{In general there are $2N+1$ nonzero Fourier
coefficients in Equation~(\ref{5.24}), however $N$ of them are
just the complex conjugate, leaving $N+1$ unique coefficients.}.
So, provided there is a way of obtaining the $N+1$ coefficients,
we can reconstruct the continuous phase probability distribution.

One way of obtaining the nonzero coefficients is to sample the continuous distribution (\ref{5.24}) at evenly spaced
angles $\gamma_m=\theta_m/2=2\pi m/(2N+2)$,
\begin{equation}\label{5.25}
  P(\gamma_m)=\frac{1}{2\pi}\sum_{n=-N}^N\alpha_n\exp(in\gamma_m),
\end{equation}
where $m=0,1,\dots,2N+2$ to ensure the entire distribution is sampled completely. We can invert this expression by
taking the discrete Fourier transform of these $2N+2$ values to give the Fourier coefficients in terms of the measured
probability densities as
\begin{equation}\label{5.26}
  \alpha_n=\frac{\pi}{N+1}\sum_{m=0}^{2N+1}\exp(-in\gamma_m)P(\gamma_m).
\end{equation}
Thus we need at least $2N+2$ points to reproduce the distribution.
For a state that is not truncated, we will need an infinite number
of such points. However if such a field is sufficiently weak for
the phase distribution to be obtained to a good approximation by
projection onto a phase state truncated after the $N$ photon
component then a correspondingly smaller number of points is
needed. We now consider how to obtain such points by experimental
measurement. Such a measurement, as previously mentioned, would be
associated with the projection of the state $|\psi\rangle$ onto
the phase state of Eqn~(\ref{5.2}). Since the state $|\psi\rangle$
is truncated after $n=N$, there would be no observable difference
if instead we projected onto the truncated phase state of
Eqn~(\ref{5.5}), with $\gamma_m=\theta_{m}/2$. Since the outcome
of this would be represented by a discrete observable, it can be
associated with a probability $\pr(\gamma_m)$, as opposed to a
density. Indeed, projecting $|\psi\rangle$ onto both phase states
we find that
\begin{equation}\label{5.27}
  2\pi\,P(\gamma_m)=(N+1)\,\pr(\gamma_m)
\end{equation}
showing the probability is proportional to the probability density at the point $\theta=\gamma_m$. So, after
substituting Eqns~(\ref{5.26}) and (\ref{5.27}) into (\ref{5.21}), we arrive at an expression for the continuous
canonical phase distribution for a truncated state $|\psi\rangle$ as
\begin{equation}\label{5.28}
  P(\theta)=\frac{1}{4\pi}\sum_{n=-N}^{N}\sum_{m=0}^{2N+1}\exp[in(\theta-\gamma_m)]\,\pr(\gamma_m).
\end{equation}
We note that this is already normalised \cite{Pregnell03a}. This expression
depends only on the $2N+2$ measurable probabilities $\pr(\gamma_m)$, as opposed
to the infinite amount necessary to measure the phase distribution of a
non-truncated state. So for states definitely containing at most $N$ photons,
it is possible, at least in principle, to obtain the continuous canonical phase
distribution perfectly.

In practice although $P(\theta)$ can be reconstructed from just
$2N+2$ experimental points, obtaining just this minimum number of
points leaves no room for error and there must always be
statistical errors when determining probabilities. Also there will
in general be experimental errors. For this reason it may be
preferable to make direct measurements of probabilities at a
sufficiently large number of phase settings to identify points
severely affected by errors. The possibility of directly plotting the histogram from the measurements gives this approach an advantage over the technique
of Section~\ref{sec4.2}, which we could use only for
reconstructing the distribution using a similar procedure to the one outlined above.

\subsection{Eight-port interferometer}\label{sec5.2.2}

In general, any method which can measure the minimum $2N+2$
probabilities $\pr(\gamma_m)$ described in the preceding section
can produce the continuous canonical phase distribution associated
with the truncated state $|\psi\rangle$ of Eqn~(\ref{5.23}). The
original projection synthesis of \cite{Barnett96} was the first
proposed technique capable, at least in principle, of measuring
such probabilities. The technique, although theoretically simple,
is very difficult to implement practically as it requires the use
of an exotic reciprocal binomial state as a reference field,
which, to date, has yet to be produced in the laboratory. An
extension of the original projection synthesis proposal is the
recently proposed multi-mode technique \cite{Pregnell04},
presented here in Chapter~\ref{chap3}. As discussed, the advantage
of this technique is that it would need only coherent references
states to obtain the probability $\pr(\gamma_m)$ necessary to
produce the canonical phase distribution. By replacing one of the
coherent references states with a binomial state, we saw in the
preceding section how $N+1$ probabilities $\pr(\theta_m)$
associated with the phase angles $\theta_m$ can be obtained from a
single measurement apparatus. For such a measurement the input
mode operators need to be related to the output mode operators by
a discrete Fourier transformation. In general, this is not
necessary for the case when the reference states are coherent
states as the amplitudes and phases can be adjusted to compensate.
We saw in Section~\ref{sec3.4}, however, that such a
transformation of mode operators did produce near optimal results
for the specific case when $N+1=4$. So, with such a multiport, in
conjunction with $N+1$ photodetectors that can discriminate
between zero, one, and many photons, it should be possible to
obtain experimentally, for the first time, a direct measurement of
the canonical phase distribution for a weak optical field. This is
in contrast to reconstructing the distribution from its moments,
as we have suggested earlier, or by reconstructing the complete
state first and then calculating the distribution (see references
given in \cite{Pegg97}).

\begin{figure}
\begin{center}
\includegraphics[width=0.40\textwidth]{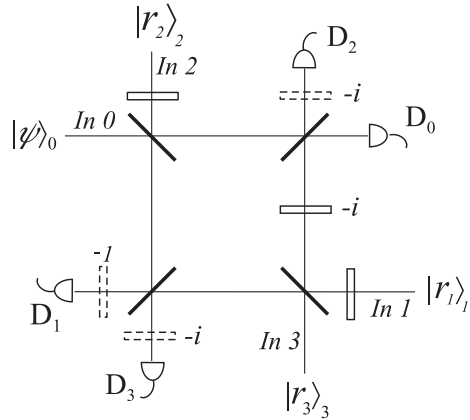}
\caption[Eight-port interferometer for measuring the canonical phase distribution of weak fields.]{Eight-port
interferometer for measuring the canonical phase distribution of weak fields. The field in state $|\psi\rangle_0$ to be
measured is in the input mode labelled \emph{In 0}, while the reference field in state $|r_j\rangle_j$, $j=1,2,3$ is in
input mode \emph{In~j}. A photodetector is in each output port. The dotted phase-shifters are for mathematical
convenience only, and do not affect the results.}\label{fig5.2}
\end{center}
\end{figure}

Given that operational phase was introduced to circumvent the
problems associated with measuring canonical phase, it is perhaps
a surprising coincidence that the experiment of Noh \emph{et al.}
\cite{Noh91,Noh93a} and Torgerson and Mandel
\cite{Torgerson96,Torgerson97} to measure operation phase does, in
fact, use a linear optical multiport of the form mentioned above.
To see that the linear multiport does in fact relate the input
mode operators to the output mode operators by a discrete Fourier
transformation consider the eight-port interferometer used by Noh
\emph{et al.} and Torgerson and Mandel in their experiments. Such
an apparatus, illustrated in Figure~\ref{fig5.2},
consists of four 50:50 symmetric beam-splitters at the corners of a square. The phase-shifter labelled $-i$ between the
two beam-splitters on the right shifts the phase by $\pi /2$. The field state $|\psi\rangle_0$ to be measured is in
input mode 0. The phase-shifter in input mode $1$ allows the phase of the reference field state $|r_1\rangle_1$ to be
changed, similarly for the phase-shifter in input mode 2. In the experiment to measure operational phase, the reference
states $|r_2\rangle_2$ and $|r_3\rangle_3$ in input mode 2 and 3 were in the vacuum states. The dotted phase-shifters
before detectors D$_{1}$, D$_2$ and D$_3$, which are not present in the original interferometer, are merely inserted
here for mathematical convenience. As the detectors detect photons, their operation will not be affected by
phase-shifters in front of them.

As defined by Eqn~(\ref{3.27}), a single 50:50 symmetric beam-splitter transforms the input creation operators
$\widehat{b}^{\dagger}$ and $\widehat{c}^{\dagger }$ in accord with
\begin{equation}\label{5.29}
  \widehat{S}_1\widehat{b}^{\dagger }\widehat{S}_1^{\dagger }=2^{-1/2}(\widehat{b}^{\dagger }+i\widehat{c}^{\dagger})
\end{equation}
\begin{equation}\label{5.30}
  \widehat{S}_1\widehat{c}^{\dagger }\widehat{S}_1^{\dagger }=2^{-1/2}(i\widehat{b}^{\dagger }+\widehat{c}^{\dagger })
\end{equation}
where $\widehat{S}_1$ is the unitary operator for the action of the single beam-splitter. By using this relation
successively, it is not difficult to show that the input creation operators for the eight-port interferometer,
including the dotted phase-shifters, are transformed as
\begin{equation}\label{5.31}
  \widehat{S}\widehat{a}_i^{\dagger }\widehat{S}^{\dagger }=\exp (i\gamma )\sum_{i=0}^3U_{ij}\widehat{a}_j^{\dagger}
\end{equation}
where $\hat S$ represents the total unitary transformation of all the optical elements and
\begin{equation}\label{5.32}
  U_{ij}^*=\frac{\omega ^{ij}}2
\end{equation}
with $\omega=\exp(-i\pi/2)$, provided we set the phase-shifter in input mode $1$ and $2$ to shift the phase by $\pi
/2$, that is to attach a value $-i$ to them. Expressions (\ref{5.31}) and (\ref{5.32}) are in agreement with
(\ref{5.12}) for $N=3$ apart from the phase factor $\exp (i\gamma )$, which depends on the difference between the
distance between beam-splitters and an integer number of wavelengths. This phase factor does not affect the photocount
probabilities and can be ignored. So we see that the eight-port interferometer and the associated photodetectors used
by Noh \emph{et al.} to measure operation phase is precisely equivalent to the apparatus needed to measure the
canonical phase distribution for the weak state $|\psi\rangle$ given by Eqn~(\ref{5.23}), with $N+1=4$. The difference
between the operational phase measurement and the proposed canonical phase measurement is in the reference states
present at the input of the interferometer.

\subsubsection{Binomial reference state}

For the probabilities associated with the photo-detection events
to correspond to canonical phase measurements, it is necessary for
the reference state in input mode~1 of the interferometer to have
the first four terms in its number state expansion proportional to
the square root of the binomial coefficients,
\begin{equation}\label{5.33}
  |0\rangle+\sqrt 2|1\rangle+\sqrt 2|2\rangle+|3\rangle,
\end{equation}
as well as there being vacuum states present in input modes 2 and
3. Such an arrangement is then consistent with the single-shot
apparatus introduced in Section~\ref{sec5.1.2} with $N+1=4$. Thus
we see that the eight-port interferometer, without modification,
can be used to synthesize the projection of the state to be
measured onto one of four truncated phase states
$|\theta_n\rangle$, with $\theta_n=0,\pi/2,\pi,3\pi/2$.
Specifically the probability of the measurement event $(0,1,1,1)$,
that is the detection of zero photocounts at detector D$_0$ and
one at each of D$_1$, D$_2$ and D$_3$, is given by Eqn~(\ref{5.7})
and is proportional to the square of the modulus of the projection
of the measured state onto the truncated phase state
\begin{equation}\label{5.34}
  \left| \theta_0\right\rangle =2^{-1}(\left| 0\right\rangle +\left| 1\right\rangle +\left| 2\right\rangle +\left|
  3\right\rangle )
\end{equation}
while the probability of the event $(1,0,1,1)$ is proportional the square of the modulus of the projection of the
measured state onto the truncated phase state
\begin{equation}\label{5.35}
  \left| \theta_1\right\rangle =2^{-1}(\left| 0\right\rangle +i\left| 1\right\rangle -\left| 2\right\rangle -i\left|
  3\right\rangle )
\end{equation}
and so on, in accord with (\ref{5.5}) with $N+1=4$. Repeating the
experiment a number of times with a reproducible state will allow
a probability $\mathrm{Pr}(\theta_n)$ with $n=0,1,2,3$ to be
measured for each of the four events (0,1,1,1), (1,0,1,1),
(1,1,0,1) and (1,1,1,0) respectively. To plot the
canonical phase distribution, it is necessary to obtain more probabilities associated with different phase angles. This can be achieved by
repeating the experiment with a $\varphi$ phase shift applied to the
binomial reference state. We see from Eqn (\ref{5.18}) that such a
phase shift will shift the phase of all four retrodictive phase
states from $\theta_n$ to $\theta_{n}+\varphi$. Figure~\ref{fig5.3} shows a simulated
measured distribution containing sixteen points obtained by shifting the reference phase three times.

So we find that the canonical phase distribution can be obtained
from the existing apparatus used by Noh \emph{et al.} to measure
their operational phase. The only changes that need to be made to
the experiment is to replace the low intensity coherent reference
state present in input mode 1 of the operation phase measurement
device with a binomial reference state.

\subsubsection{Squeezed reference state}

\begin{figure}
\begin{center}
\includegraphics[width=0.40\textwidth]{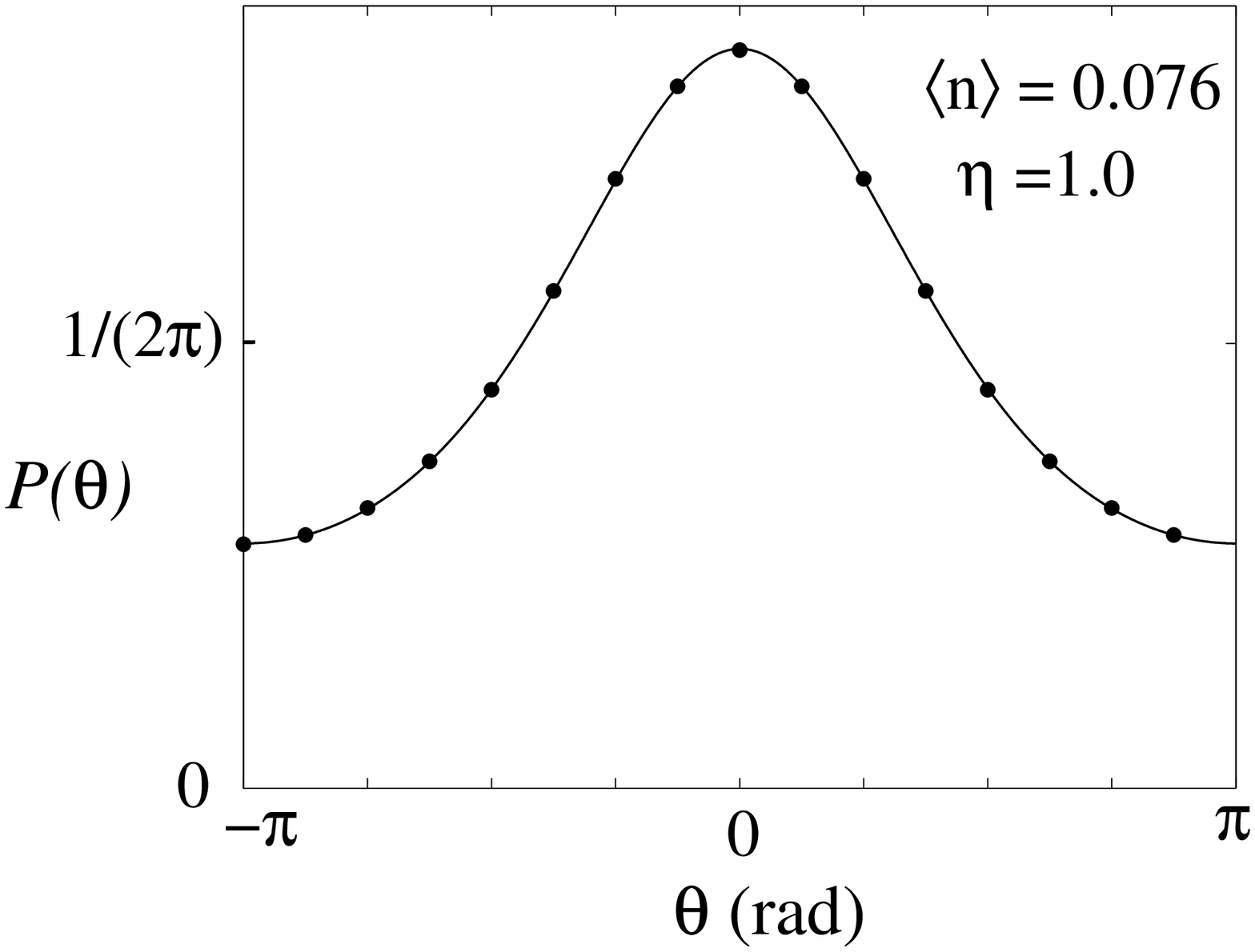}
\includegraphics[width=0.40\textwidth]{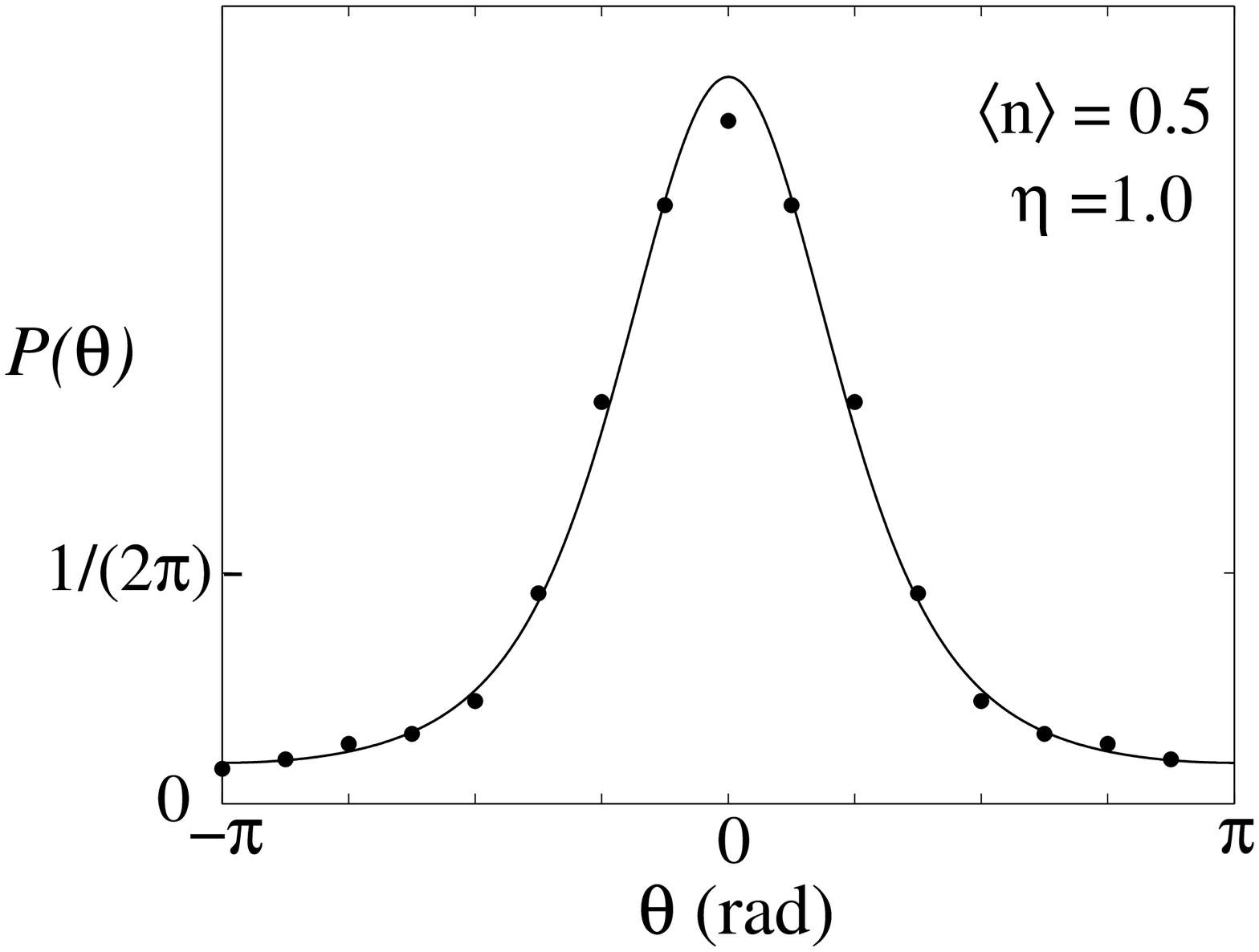}
\caption[Simulated canonical phase distribution for low intensity
coherent states]{Canonical phase probability distribution
$P(\theta)$ for a coherent state field with, on the left, a mean
photon number of 0.076 and, on the right, a mean photon number of
0.5. The full line is the theoretical result and the dots are
simulated measured results with ideal detectors and a squeezed
reference state.}\label{fig5.3}
\end{center}
\end{figure}

The above analysis and suggested procedure assumes that the reference field is in a perfect binomial state. If,
instead, we use the squeezed state approximation to the binomial state as derived in Appendix~\ref{appendixD}, then the
vacuum state coefficient differs from the ideal value and the measured state is no longer projected onto the truncated
phase state $\left| \theta _m\right\rangle $ but is instead projected onto a state proportional to
\begin{equation}\label{5.36}
| 0\rangle +\exp (i\theta _m)| 1\rangle +\exp (2i\theta _m)| 2\rangle +1.0146\exp (3i\theta _m)| 3\rangle.
\end{equation}
We would expect that this would lead to some small errors when the
procedure suggested above is applied. In practice, if we are
measuring a state, such as a coherent or squeezed state, that does
not have a truncated photon number distribution, the error caused
by the modulus of the three-photon coefficient in expression
(\ref{5.36}) differing from unity may in general be smaller that
the error caused by assuming the input state can be sufficiently
well approximated by the truncated expression of Eqn~(\ref{5.23})
with $N+1=4$. In Figure~\ref{fig5.3}
we show the canonical phase distribution histogram, with points
rather than bars for clarity, obtained from a simulated experiment
for a coherent state with a mean photon number of 0.076, which is
comparable to the field strength of interest in Ref.
\cite{Torgerson96}, using a squeezed reference state. The close
agreement with the canonical distribution is apparent. For weaker
fields, for example the other field of interest in Ref.
\cite{Torgerson96} with a mean photon number of 0.047, the
agreement is even closer. Agreement is still good for stronger
coherent state fields with mean photon numbers of 0.139 and 0.23,
as used in Ref. \cite{Torgerson97}, with divergence from the
canonical distribution becoming apparent for mean photon numbers
of around 0.4. The histogram on the right in Figure~\ref{fig5.3}
shows simulated results for a coherent state with a mean photon
number of 0.5. The error here is almost entirely due to the
truncation of the coherent state after the three photon component rather than to
the non-unit coefficient of the fourth term in Eqn~(\ref{5.36}). A
mean photon number of 0.5 represents the approximate limit to the
field strength for a coherent state for which this measurement
technique is suitable.

\subsubsection{Coherent reference states}

The alternative method to the above approach for
measuring the phase distribution is the projection synthesis
method outlined in Chapter~\ref{chap3}. This technique, as
discussed, requires three coherent reference states
$|\alpha_1\rangle_1$, $|\alpha_2\rangle_2$ and
$|\alpha_3\rangle_3$ in the input modes 1, 2 and 3, respectively,
of the interferometer. By appropriately choosing the coherent
amplitudes $\alpha_1$, $\alpha_2$ and $\alpha_3$, we can engineer
the retrodictive state conditioned on observing the
photo-detection sequence $(0,1,1,1)$ at photo-detectors D$_{0}$,
D$_{1}$, D$_2$ and D$_3$ to be proportional to the truncated phase
state
\begin{equation}\label{5.37}
  \left| \theta_0\right\rangle =2^{-1}(\left| 0\right\rangle +\left| 1\right\rangle +\left| 2\right\rangle +\left|
  3\right\rangle).
\end{equation}
To find the amplitudes of the coherent reference states, we solve the characteristic polynomial (\ref{3.4}) for $N=3$
with $\psi_n=1$ to give the three complex roots
\begin{eqnarray*}
  \beta_1&=&-0.2168+i1.3563\\
  \beta_2&=&-1.2984\\
  \beta_3&=&-0.2168-i1.3563.
\end{eqnarray*}
Then from Eqn~(\ref{3.32}), with $U_{ij}^*$ defined by (\ref{5.32}), we find the necessary amplitudes of the coherent
reference states to be
\begin{eqnarray*}
  \alpha_1&=&1.4358\\
  \alpha_2&=&0.2168\\
  \alpha_3&=&0.0795.
\end{eqnarray*}
The probability of observing the specific photocount sequence
$(0,1,1,1)$ over repeated trials is then found from
Eqns~(\ref{3.30}) and (\ref{2.12}) with $\hat\Gamma=\hat 1$ to be
$\mathrm{Pr}(0,1,1,1)=0.0425|\langle\theta_0|\psi\rangle|^2$. So the value of the projection of the input state onto the truncated phase state of Eqn~(\ref{5.37}) is
$(0.0425)^{-1}$, or approximately 23.5, times greater than the
probability of observing the photocount sequence $(0,1,1,1)$. Measuring the occurrence frequency for this specific observation gives one value which can be plotted on the phase probability histogram. To obtain additional points corresponding to different values of phase the experiment needs to be repeated with a phase-shift applied to the state to be measured in input mode 0. The distribution histogram can then be built up in this way by directly measuring the value of the phase distribution for different values of phase.

\section{Some practical considerations}

In addition to the statistical error inherent in obtaining
probabilities from measured relative frequencies, in a practical
experimental situation errors can arise from collection
inefficiencies, non-unit quantum efficiencies for one, two and
multiple photon detection, dead times and accidental counts
arising from dark counts and background light. The fact, however,
that Noh \emph{et al.} \cite{Noh91,Noh93b} have performed
successful experiments involving the measurement of joint
detection probabilities with an eight-port interferometer, by
means of photon counting, for states with field strengths similar
to those of interest in this paper is an encouraging indication
that there should be no insurmountable difficulties for the method
proposed here arising from such errors. It is still however worth
considering some specific aspects of the sources of error. In the
experiments of Noh \emph{et al.} \cite{Noh91,Noh93b} photon count
rates were of the order of $10^4$ per second with a counting
interval of about 5 $\mu$s, to give the required small mean photon
number, and dead-time effects were negligible. In the present
proposed experiment dead times are even less important because it
is only necessary to discriminate among zero, one and many counts
rather than among general numbers of counts\footnote{In the case
where dead times are significant their effect can be substantially
reduced by use of beam-splitters. See, for example,
\cite{Paul96}.}. Dark counts can be reduced to about 200 per
second \cite{Noh91,Noh93a} or even to 20 per second
\cite{Trifonov00} by cooling the detectors and background light
can be reduced by appropriate shielding.  In the event that the
residual dark and background counts are not negligible, the
measured joint probabilities of the four photocount events can be
corrected by a deconvolution procedure using the data obtained by
blocking the input signals \cite{Noh91,Noh93a}.

Concerning detector efficiencies, even if collection efficiencies
are made to approach unity by, for example, suitable geometry and
reflection control, there will still be some detector inefficiency
due to non-unit quantum efficiency, so some correction for
detector inefficiency may be needed. Conventional single-photon
counting module detectors can have an efficiency of around 0.7
\cite{Kim99,Takeuchi99,Tsujino02}, while visible light photon
counters that distinguish between single-photon and two-photon
incidence can have quantum efficiencies of about 0.9 with some
sacrifice of smallness of dark count rate
\cite{Kim99,Takeuchi99,Tsujino02}. We denote the one-photon
detection efficiency, that is the probability of recording one
photocount if one photon is present, by $\eta$. Then, as dead
times are not important, the general multiple detection efficiency
is such that the probability of recording $n$ photocounts if $N$
photons are present is ${N\choose n}\eta^{n}(1-\eta)^{N-n} $ where
the first factor is the binomial coefficient
\cite{Lee93,Barnett98}. If $\eta$ is the same for all four
detectors the probability for the joint four-count detection event
$(m,n,p,q)$ is given by
\begin{eqnarray}\label{5.38}
    P_{c}(m,n,p,q)=\sum_{s=m}^{\infty}\sum_{t=n}^{\infty}
    \sum_{u=p}^{\infty}\sum_{v=q}^{\infty} {s\choose m}{t\choose n}
    {u\choose p}{v\choose q}\nonumber
    \\ \times
    \eta^{m+n+p+q}(1-\eta)^{s+t+u+v-m-n-p-q}P_{I}(s,t,u,v)
\end{eqnarray}
where $P_{I}(s,t,u,v)$ is the probability that an ideal detector would have detected the joint four-count event
$(s,t,u,v)$. The relation (\ref{5.38}) can be inverted by use of the four-function Bernoulli transform, which is
straightforward to derive in a similar manner to that of the one-function transform stated in the
Appendix~\ref{appendixA}, or the two-function transform derived in Ref. \cite{Pegg99c}. This allows us to calculate the
ideal probabilities from the measured probabilities and thus correct for non-unit efficiencies.

\begin{figure}
\begin{center}
\includegraphics[width=0.40\textwidth]{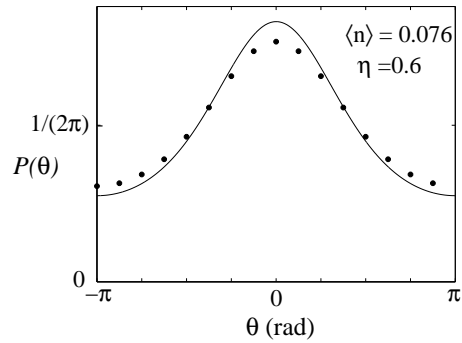}
\caption[Simulated canonical phase distribution for low intensity coherent state with imperfect
photo-detection]{Uncorrected simulated measurements (dots) and the theoretical canonical phase distribution (full line)
for a coherent state field with a mean photon number of 0.076 where the photodetectors have an efficiency $\eta =
0.6$.}\label{fig5.5}
\end{center}
\end{figure}

Although we can correct for non-unit efficiencies an analysis shows that the effect of not correcting for them is not
as serious as it may first appear. Essentially this is because the four measured probabilities are always normalized so
their sum is unity. The major effect of $\eta $ not being unity is, as can be seen from (\ref{5.38}), that the
probabilities for the four events (0,1,1,1), (1,0,1,1), (1,1,0,1) and (1,1,1,0) to be actually recorded are reduced by
a factor $\eta ^3$. As this affects the event probabilities uniformly, however, the effect vanishes upon normalization.
The next order effect is that some ideal four-count events, such as (1,1,1,1) and (0,2,1,1), are registered, for
example, as (0,1,1,1) because of the inefficiency. The effect of this is only partially removed by the normalization.
Ideal higher-count events also contribute to the error, but for weak fields the probability of ideal high-count events
is not large. A numerical calculation of the total effect of non-unit efficiency, including the effect of
normalization, shows that the proposed procedure is not highly sensitive to detector inefficiency, provided the
efficiency is reasonable, for the weak states of interest. More precisely, for coherent states with a mean photon
number up to 0.5 photons, as discussed above, the error in the final normalized probabilities is less than 2\% for
$\eta \geq 0.9$. For a mean photon number of 0.076, the error is less that 0.5\% for such efficiencies. In Figure~
\ref{fig5.5} we show the effect of a poorer efficiency of $\eta =0.6$ for a mean photon number of 0.076.

To produce the squeezed state required as an approximate binomial state, we note that squeezed vacuum states can be
transformed into various types of squeezed states, the squeezing axis can be rotated, coherent amplitude can be added
and the squeezing can be controlled independently of the coherent amplitude. The degree of squeezing needed here is of
a magnitude that is a realistic expectation either now or in the near future \cite{Bachor98}.

\section{Summary}

We have shown in this chapter how a single-shot measurement of
canonical phase can be performed. We emphasis that this is not a
deterministic measurement, that is, only some of the experiments
provide successful measurements. The results of other measurements
are discarded. This happens in a significant number of measurement
techniques, for example, that used by Noh \emph{et al.}
\cite{Noh91,Noh92b,Noh93b} to measure their operational phase. We
have also shown how the eight-port interferometer used by Noh
\emph{et al.} can also be used to measure the canonical phase
distribution given by Eqn~(\ref{5.1}), where the canonical phase
is defined as the complement of photon number. The procedure is
applicable for weak fields in the quantum regime, by which we mean
explicitly states for which number state components for photon
numbers greater than three are negligible. For coherent states,
this requirement translates to a mean photon number of a half a
photon or less. This is precisely the quantum regime in which
large differences between the operational phase and the canonical
phase distributions are most apparent. For example fields of
interest in Refs \cite{Torgerson96,Torgerson97} are coherent
states with mean photon numbers of 0.23, 0.139, 0.076 and 0.047.
The success of the experiments in the foregoing references
indicates that the procedure proposed in this paper should be
viable, given a reliable source of the required reference state.

The procedures in this chapter have advantages over the original projection synthesis method proposed for measuring the
canonical phase distribution. The most significant of these is that the procedures require reference states that are
either coherent states, or states that can be derived from coherent states by reliable procedures such as squeezing.
Another advantage is that we require only photodetectors that can distinguish among zero, one and more than one
photocounts. The measurements are not particularly sensitive to photodetector inefficiency and, for reasonably good
detector efficiencies, no corrections should be needed. Overall, we feel that the proposal in this thesis brings the
measurement of the canonical phase distribution for weak optical fields closer to reality.

%% file: Conc.tex
\chapter{Conclusion}

The generation of retrodictive quantum optical states was the
focus of this thesis. For this purpose we investigated in
particular the use of the lossless optical multiport. The
motivation for this was to try to introduce simple experimental
techniques capable of extending the current range of measuring
apparatuses in the field of quantum optics. In Chapter~\ref{chap3}
I showed that a lossless optical multiport, constructed from an
array of beam-splitters and phase-shifters, a coherent reference
state and photodetection is all that is needed to generate any
retrodictive state with a finite number of number-state
components. Such an apparatus has applications in measurement,
particularly projection synthesis, and quantum state preparation.
We showed in Chapter~\ref{chap4} how some measurement techniques
involving predictive optical probe states can be converted to
experiments involving retrodictive probe states which are far
easier to generate in practice. Another finding of this thesis was
the single-shot measuring device for canonical phase presented in
Chapter~\ref{chap5}. With the theoretical description of canonical
phase already well formulated, the introduction of such an
apparatus removes the final distinction separating canonical phase
from all other observables in quantum optics, that is, canonical
phase can now be measured, at least in principle.

Just as there exists an intimate relationship between the process
of preparation and a predictive state in the conventional
predictive formalism of quantum mechanics, there exists a
symmetric relationship between the process of measurement and a
retrodictive state in the retrodictive formalism. One is said to
be the cause of the other. In Chapter~\ref{chap2}, I review a
formalism of quantum mechanics which did not discriminate against
either process. From such a formalism, both the predictive and
retrodictive formalisms were derived. We found, consistent with
the original conclusion of Aharonov \emph{et al.}
\cite{Aharonov64}, that the difference between preparation and
measurement is not an intrinsic property of quantum mechanics. It
is reasonable to assume, therefore, that it results from the
macroscopic laws of the universe as a whole and must be introduced
into quantum mechanics as an additional postulate. By introducing
causality in the form of the postulate that messages can be sent
only forwards in time, we regained the usual normalisation
conditions which do not apply symmetrically to the predictive
density operator and the measurement POM elements. Provided the
asymmetric \emph{normalisation conditions} are maintained,
causality allows a symmetry in the use of predictive and
retrodictive states. That is causality does not lie in the time
direction of propagation of the states as may have been thought at
first glance as mentioned in the Introduction. Using this
symmetric formalism, we found that any physical ensemble of
preparation and measurement devices can always be reduced to
either a single measurement device or a single preparation device.
We used this result to investigate linear optical multiport
devices as a general measurement device.

This investigation, although similar to the original projection
synthesis proposal, had two important differences. First, we
generalised the optical element, a single beam-splitter, to
include an arbitrary array of beam-splitters and phase-shifters.
In doing so we found that we can synthesise the same arbitrary set
of projections with classical, that is coherent, reference states
as achieved by the original projection synthesis with
non-classical reference states. So by limiting ourselves to
coherent reference states, linear optics and photodetection we can
construct a general apparatus that is capable of producing any
retrodictive quantum state with a finite number of photon number
state coefficients. That is, a measuring apparatus that can
project an initial predictive state onto any state vector of a
finite dimensional Hilbert space. We found, quite surprisingly,
that it is less demanding practically to generate a wide class of
retrodictive states that it is to produce the predictive
counterparts. This asymmetry originates from the ease in which
simultaneous single photons can be observed as opposed to created.

A procedure for the optimum measurement technique, defined in the
text as the one which will produce the desired retrodictive state
with maximum likelihood, was derived. We found that, although the
amplitudes and phases of the coherent reference states are
constrained, the elements of the transformation matrix are not.
This corresponds to adjusting the transmission to reflection ratio
of the beam-splitters, the basic elements of the multiport. To
illustrate such a protocol, we considered three apparatuses which
are capable of projecting onto a truncated phase state of $N+1=3$
dimensions, one of which is optimal.

In Chapter~\ref{chap4}, we exploited the fact that non-classical
retrodictive states are simpler to produce than the predictive
equivalent. This was done by redesigning two experimental
proposals with a need for such non-classical probe states, so that
the necessary probe state could be a \emph{retrodictive} state. On
the whole, both experiments are simpler, involving only a single
coherent reference state, two beam-splitters and photodetection.
This example serves to demonstrate the advantage in viewing
measurement in the retrodictive formalism as it is computationally
much easier to evolve backwards the single photon number states
conditioned on detection, than it is to evolve forward all
possible superposition states that may be at the input of such an
apparatus.

Another notable result of this thesis was presented in
Chapter~\ref{chap5}. We found that by replacing just one of the
coherent inputs in the general measuring apparatus of
Chapter~\ref{chap3}, it was possible to produce a single-shot
measurement of canonical phase. The price paid for obtaining a
single-shot measuring apparatus is a tradeoff with ease of
construction, since the necessary reference state is a binomial
state. However, we were able to show that such a state can be
sufficiently well approximated by a suitably squeezed state. Such
a measurement scheme has long been sought after as it was the last
difficulty associated with the challenging concept of phase. Quite
ironically, the apparatus used to measure ``measured phase'', a
concept introduced because of the difficulties in measuring
canonical phase, can in fact be used to measure the canonical
phase distribution, without alteration. By expressing the
canonical phase distribution in terms of the Fourier coefficients,
it was demonstrated that only a \emph{finite} number of points
need be sampled from the continuous distribution in order to
reconstruct the whole canonical phase distribution for a state
possessing a finite number of photons. Alternatively, more
probabilities can be measured to produce the canonical phase
distribution histogram directly. With the introduction of such
phase measurement schemes, both the single-shot method and the
linear optical apparatus, there seems little need to define phase
dependent observables based upon a measurement scheme which is
simple to implement.

In addition to the application to measurement discussed in this
thesis, it has already been recognized that the retrodictive
formalism of quantum mechanics should have applications such as
quantum communication \cite{Barnett00b}. A potential application
of retrodictive quantum state engineering that we have mentioned
but not explored in this thesis is for predictive state
engineering. If a two-mode entangled state can be prepared with a
significant number of non-negligible photon-number-state
coefficients then sending a suitably engineered retrodictive state
into one of these modes will result in an associated predictive
state in the other mode. As it is easier to generate exotic
retrodictive states than exotic predictive states, such a
technique could prove useful.

Overall, we can conclude that the retrodictive formalism of
quantum mechanics is not just a curiosity of philosophical value
only. Instead, it has the potential to be of real practical value.

%% file: AppendixA.tex
\chapter{Corrections for imperfect photon detection}\label{appendixA}

There are many physical processes which manifest themselves as an
imperfection in realistic photodetectors. Practical photodetectors
suffer from non-unit quantum efficiencies that reduce the number
of counts, the presence of dark counts not associated with the
absorption of a photon and a non-zero dead time following a count
during which no other counts are registered. Using weak fields, in
which we are particularly interested, and sufficiently long gating
times reduces the effect of the dead time. Dark counts are related
to thermal excitations within the detector and are independent of
the number of photons incident upon the detector. Generally,
sufficient cooling of the detectors can minimise this effect. If
the remaining dark counts are not negligible, then a deconvolution
of the measured data with the counts obtained when the detector is
blocked from the light source can remove the majority of the
remaining counts \cite{Noh92b}. Detector inefficiencies is a
collective term including the effects of such processes as quantum
efficiency, mode mismatching and coupling efficiency. All these
inefficiencies can be represented by one parameter, $\eta$, the
detector efficiency.

In this appendix I discuss the most dominate processes relevant to
the proposals presented in this thesis; detector inefficiency.
Most importantly, I review how the measured probabilities obtained
from the photocount distribution of a single photodetector can be
corrected to obtain the photon number probability distribution
that would be observed if the detector had unit efficiency.

\section{Detector inefficiency}

Real photodetectors are imperfect. The probability that $m$ photocounts are registered, $P(m)$, is related to the
probability that $n$ photos where present, $Q(n)$, by
\begin{equation}\label{A1}
  P(m)=\sum_n p(m|n)Q(n)
\end{equation}
where $p(m|n)$ is the conditional probability that there are $m$ photocounts if there was $n$ photons present. For a
photodetector with finite efficiency $\eta$, it has been shown \cite{Kelly64,Loudon83} that the conditional probability
can be expressed as,
\begin{equation}\label{A2}
  p(m|n)={n\choose m}(1-\eta)^{m-n}\eta^n.
\end{equation}
In general, a distribution of the form
\begin{equation}\label{A3}
  P(m)=\sum_{n=m}^\infty Q(n){n\choose m}(1-\eta)^{n-m}\eta^m
\end{equation}
is known as a Bernoulli distribution, and can be inverted to give \cite{Pegg99c}
\begin{equation}\label{A4}
  Q(n)=\sum_{m=n}^\infty  P(m){m\choose n}(\eta-1)^{m-n}\eta^{-m}.
\end{equation}
The two distributions $P(m)$ and $Q(n)$ form a Bernoulli transform pair. For the case considered here,
Equation~(\ref{A4}) gives a way in which the true photon number distribution $Q(n)$ can be extracted from the observed
photocount distribution $P(m)$. For situations where there is a joint probability distribution corresponding to the
photocount of two photodetectors, then a two-function Bernoulli transform pair can be derived \cite{Pegg99c}. Such a
derivation can readily be extended to a multi-function Bernoulli transform pair which is necessary to extract the
photon number distribution from the photocount distribution obtained from the multiport devices considered in this
thesis.

%% file: AppendixB.tex
\chapter{Raising operator}\label{appendixB}

It is the purpose of this appendix to derive an expression for the
non-unitary operator ${}_1\langle 0 |\hat S^{\dag} | N
\rangle_{1}$, governing the evolution of a retrodictive state
backwards in time through a beam-splitter. The evolution is
conditioned on measuring $N$ photons in the output mode 1 of the
beam-splitter, while a vacuum field is present in the input mode
1. We represent the (backward time) unitary evolution operator of
the beam-splitter as $\hat S^\dag $

Writing the identity operator for the field mode 0 as $\sum_m |m\rangle_{0} {}_0\langle m|$, where $|m\rangle_0$ is a
photon number state, and expressing the kets as raising operators acting on the vacuum as $(\hat
b^\dag)^n/\sqrt{n!}\,|0\rangle$, we obtain
\begin{equation}\label{B1}
  {}_1\langle 0 |\hat S^{\dag} | N \rangle_{1}=\sum_m{}_1\langle 0 |\frac{(\hat S^{\dag}\hat b_0^\dag\hat S)^m
  (\hat S^{\dag}\hat b_1^\dag\hat S)^N}{\sqrt{m! N!}}\;\hat S^\dag|0\rangle_{1} |0\rangle_{0} {}_0\langle m|,
\end{equation}
where the identity, $\hat S \hat S^\dag=\hat 1$, has been used
repeatedly. The backward-time evolution is now represented by a
mode transformation of the output mode operators to the input mode
operators. The set of operators are related by the unitary
transformation of Eqn~(\ref{3.9}). After substituting this
relation into (\ref{B1}) and taking the inner product with the
vacuum state it follows that
\begin{equation}\label{B2}
  {}_1\langle 0 |\hat S^{\dag} | N \rangle_{1}=\sum_m\frac{(t\hat a_0^\dag)^m
  (ir\hat a_0^\dag)^N}{\sqrt{m! N!}}\;|0\rangle_{0} {}_0\langle m|,
\end{equation}
where $t=\cos\theta$ and $r=\sin\theta$ are the transmission and reflection coefficients. After some simple algebra,
this can be written as
\begin{equation}\label{B3}
  {}_1\langle 0 |\hat S^{\dag} | N \rangle_{1}=\frac{(ir\hat a_0^\dag)^N t^{\hat n}}{\sqrt{ N!}}  ,
\end{equation}
where
\begin{equation}\label{B4}
  t^{\hat n}=\sum_m t^m|m\rangle_{0} {}_0\langle m|
\end{equation}
is, in general, a non-unitary operator since $t=\cos\theta$ can
take on non-unit values. It in interesting to consider briefly the
effect of the transformation (\ref{B3}) when the transmission
coefficient takes on the two extreme values 0 and 1. In the limit
as $t\rightarrow 1$, it is straightforward to see that (\ref{B4})
approaches the identity operator. Since in the same limit the
reflectance goes to zero, (\ref{B3}) is only non-zero for $N=0$,
in which case it is just the identity, implying that the field
propagates unchanged. This is consistent with what one would
expect of the transformation operator for a beam-splitter which is
totally transmitting as there is no coupling between the two
fields.

In the other limit as $t\rightarrow 0$, the beam-splitter is
totally reflecting acting like a double-sided mirror. In such case
(\ref{B4}) goes to the vacuum state projector
$|0\rangle_0{}_0\langle 0|$, giving, as the non-unitary
transformation operator (\ref{B3}),
\begin{equation}\label{B5}
{}_1\langle 0 |\hat S^{\dag} | N \rangle_{1}=i^N |N\rangle_0{}_0\langle 0|.
\end{equation}
Remembering that this is a backward-time-evolving operator, it is
most natural to analyse this in the retrodictive picture. In this
picture, the photodetector in output mode 1 acts like a photon
source. Because the beam-splitter acts as a double-sided mirror,
the photons are reflected into mode 0. Considering just the
dynamics of mode 0, this would look as though the photon number of
the state originally in that mode, that is the vacuum, suddenly
was raised by $N$ photons. This is consistent with the operator of
(\ref{B5}). So we find that this operator does conform with our
expectations in these simple limits.

%% file: AppendixC.tex
\chapter{Fourier coefficients of the phase distribution}\label{appendixC}
It is the aim of this appendix to show that the Fourier coefficients introduced in Chapter~\ref{chap5},
\begin{equation}\label{AB.1}
  \alpha_q=\int P(\theta)\exp(-iq\theta)\,d\theta
\end{equation}
associated with the state $|\psi\rangle=\sum_{n=0}^N\psi_n|n\rangle$ are zero for $|q|>N$.

Following on from Eqn~(\ref{5.3}), the probability density $P(\theta)$ can be expressed as the trace of the product of
the state to be measured $\hat\rho$ and the POM element $|\theta\rangle\langle\theta|$ as
\begin{equation}\label{AB.2}
  P(\theta)=\textrm{Tr}\left[\hat\rho|\theta\rangle\langle\theta|\right],
\end{equation}
where
\begin{equation}\label{AB.3}
  |\theta\rangle=\frac{1}{2\pi}\sum_{n=0}^\infty\exp(in\theta)|n\rangle
\end{equation}
is the phase state introduced in Eqn~(\ref{5.2}). Substituting Eqn~(\ref{AB.2}) into the Eqn~(\ref{AB.1}) allows the
Fourier coefficient $\alpha_q$ to be associated with an operator $\hat\alpha_q$ as
\begin{equation}\label{AB.4}
  \alpha_q=\textrm{Tr}[\hat\rho\hat\alpha_q]
\end{equation}
where the operator
\begin{equation}\label{AB.5}
  \hat\alpha_q=\int |\theta\rangle\langle\theta|\exp(-iq\theta)\,d\theta
\end{equation}
is defined. Substituting the expression for the phase state, Eqn~(\ref{AB.3}), allows the operator of (\ref{AB.5}) to
be written in the photon number basis as
\begin{equation}\label{AB.6}
  \hat\alpha_q=\sum_{n=0}^\infty|n+q\rangle\langle n|
\end{equation}
for $q=0,1,\dots$, where the definition of the Kronecker delta
\begin{equation}\label{AB.7}
  \delta_{n,m}=\int \exp[i(n-m)\theta]\,d\theta
\end{equation}
has been used to simplify. By taking the conjugate transpose of
Eqn~(\ref{AB.5}) it is straightforward to derive the operator
relation $\hat\alpha_q^\dag=\hat\alpha_{-q}$, which gives, from
Eqn~(\ref{AB.6}), an expression for the operator $\hat\alpha_q$,
for $q=-1,-2,\dots$, in the photon number basis. We note as an
aside that the operator $\hat\alpha_q$ is equivalent to the
non-unitary Susskind-Glogower ``exponential'' phase operator
\cite{Susskind64} introduced as an attempt to represent phase as a
non-Hermitian operator. As this derivation does not depend on
phase being represented as an operator, the problems associated with the
Susskind-Glogower formalism are not of concern here.

The most general description for the state of a system is to assign a density operator,
\begin{equation}\label{AB.8}
  \hat\rho=\sum_{n,m=0}^\infty\rho_{n,m}|n\rangle\langle m|
\end{equation}
expressed here in the photon number basis. For such a state, we see from Eqns~(\ref{AB.4}) and (\ref{AB.6}) that the
Fourier coefficients $\alpha_q$ can be expressed in term of the off-diagonal matrix elements $\rho_{n,n+q}$ as
\begin{equation}\label{AB.9}
  \alpha_q=\sum_{n=0}^\infty\rho_{n,n+q},
\end{equation}
with $\alpha_{-q}=\alpha_q^*$. It is straightforward then to see that for a system containing at most $N$
photons, that is the density matrix can be represented by a $(N+1)\times(N+1)$ dimensional matrix, that the
coefficients $\alpha_q$ are zero for $q>N$. Since $\alpha_{-q}=\alpha_q^*$, then is must follow that the
coefficient are also zero for $q<N$.

So we find, since $|\psi\rangle\langle\psi|$ is a specific example of such a density matrix, that $\alpha_q=0$ for all
integer $|q|>N$, as required.

%% file: AppendixD.tex
\chapter{Squeezed states as approximate binomial states}\label{appendixD}

In this Appendix we show how the required binomial reference of Section~\ref{sec5.2.2} state can be approximated by a
suitably squeezed state. The particular binomial state of interest to us is given by
\begin{equation}\label{AD.1}
\left| B\right\rangle =\sum_{n=0}^N\,b_n\left| n\right\rangle =2^{-N/2}\sum_{n=0}^N\,{N\choose n}^{1/2}\left|
n\right\rangle
\end{equation}
where ${N\choose n}$ is the binomial coefficient. The binomial state derived in Ref.~\cite{Pregnell03a} with
alternating signs for the number state coefficients can be obtained by phase shifting this state by $\pi$.

The general form for a squeezed state is \cite{Loudon83}
\begin{eqnarray}
\left| \alpha ,\zeta \right\rangle &=&\sum_{n=0}^\infty \,\alpha _n\left| n\right\rangle \nonumber
\\&=&(\cosh |\zeta |)^{-1/2}\exp \{-\frac 12[|\alpha |^2+
t(\alpha ^{*})^2]\} \nonumber
\\
&&\times \sum_{n=0}^\infty \frac{(t/2)^{n/2}}{(n!)^{1/2}}H_n\left[\frac{\alpha +t\alpha ^{*}}{(2t)^{1/2}}\right]\left|
n\right\rangle \label{AD.2}
\end{eqnarray}
where $\zeta =|\zeta |\exp (i\phi )$ with $|\zeta |$ being the squeezing parameter, $t=\exp(i\phi )\tanh |\zeta |$ and
$H_n(x)$ is a Hermite polynomial of order $n$. $\alpha$ is the complex amplitude of the coherent state obtained in the
limit of zero squeezing.

The first case we study is where we are interested in finding a squeezed state whose coefficients $\alpha _n$ are
proportional to the coefficients $b_n$ of binomial state for the early terms, that is for $n<<N$. In this case we can
approximate the binomial coefficient by
\begin{eqnarray}
{N\choose n}^{1/2} &=&\frac{N^{n/2}}{\sqrt{n!}}\sqrt{(1-\frac{1}{N}) (1-\frac{2}{N})\ldots (1-\frac{n-1}{N})} \nonumber
\\
&\approx &\frac{N^{n/2}}{\sqrt{n!}}\left[ 1-\frac{n(n-1)}{4N}\right]\label{AD.3}
\end{eqnarray}
We can approximate the Hermite polynomial for large $x$ by its leading terms:
\begin{equation}\label{AD.4}
H_n(x)\approx (2x)^n-n(n-1)(2x)^{n-2}
\end{equation}
We find, remarkably, that choosing $t=0.5$ and $\alpha
=(2/3)N^{1/2}$ allows us to write
\begin{equation}\label{AD.5}
{N\choose n}^{1/2}\approx \frac{(t/2)^{n/2}}{(n!)^{1/2}}H_n\left[\frac{\alpha +t\alpha ^{*}}{(2t)^{1/2}}\right]
\end{equation}
for $n<<N$. Thus the first $n$ number state coefficients of a squeezed state with these values of $t$ and $\alpha$ will
be proportional to the required binomial coefficients to a good approximation. With this degree of squeezing, the
squeezed quadrature variance is 1/3 that of the vacuum level, that is, 4.77 dB below the standard quantum limit.

The opposite case to the above is where we require a small number of coefficients $\alpha _n$ for $n=N,N-1,N-2\ldots $
to be proportional to $b_n$. It is not as easy to obtain as general a relationship as the above so we look at each case
individually. In this paper we are interested in the particular case with four values of $b_n$, that is, $N=3$. By
using the explicit form of the Hermite polynomials in Eqn~(\ref{AD.2}) and setting $\alpha _2/\alpha _3=b_2/b_3$ and
$\alpha _1/\alpha _3=b_1/b_3$ we find that the values $t=0.5$ and $\alpha =(2+2^{1/2})/3$ satisfy the two simultaneous
equations obtained. We note that the required squeezing parameter $\tanh ^{-1}0.5$ is the same as for the first case
above but the value 1.138 of $\alpha $ varies slightly from 1.155, the value of $(2/3)N^{1/2}$ with $N=3$, which is
required to make the first few coefficients of $\left| \alpha ,\zeta \right\rangle $ proportional to binomial
coefficients. We also note that with perfect matching of the last three coefficients the ratio $\alpha _0/\alpha _3$
becomes 1.0146, a mismatch of only 1.5\% with the corresponding binomial coefficient.